\documentclass[twocolumn,tighten]{aastex631}

\usepackage{color,comment,multirow,booktabs,upgreek}
\usepackage{multirow,enumitem}
\usepackage[flushleft]{threeparttable}
\usepackage{amsmath}
\usepackage{CJK}
\graphicspath{{./}{images/}}

\newcommand{\pcc}{{\rm cm}^{-3}}
\newcommand {\kms}{\ifmmode{\rm km \, s^{-1}}\else{$\rm km \, s^{-1}$}\fi} 
\newcommand {\mo}{{\rm M}_\odot}

\newcommand{\multiref}[3]{\autoref{#1}-\ref{#2}-\ref{#3}}

\shorttitle{CRs in Varying
Galactic Environments}
\shortauthors{Armillotta et al.}

\begin{document}
\begin{CJK*}{UTF8}{gbsn}

\title{Cosmic-Ray Transport in Varying Galactic Environments}

\correspondingauthor{Lucia Armillotta}
\email{lucia.armillotta@princeton.edu}

\author[0000-0002-5708-1927]{Lucia Armillotta}
\affiliation{Department of Astrophysical Sciences, Princeton University, Princeton, NJ 08544, USA}

\author[0000-0002-0509-9113]{Eve C. Ostriker}
\affiliation{Department of Astrophysical Sciences, Princeton University, Princeton, NJ 08544, USA}

\author[0000-0002-2624-3399]{Yan-Fei Jiang(姜燕飞)}
\affiliation{Center for Computational Astrophysics, Flatiron Institute,
New York, NY 10010, USA}

\begin{abstract}
We study the propagation of mildly-relativistic cosmic rays (CRs) in multiphase interstellar medium environments with conditions typical of nearby disk 
galaxies. We employ the techniques developed in \citet{Armillotta+21} to post-process three high-resolution TIGRESS magnetohydrodynamic simulations modeling local patches of star-forming galactic disks. Together, the three simulations cover a wide range of gas surface density, gravitational potential, and star formation rate (SFR). Our prescription for CR propagation includes the effects of advection by the background gas, streaming along the magnetic field at the local ion Alfv\'en speed, and diffusion relative to the Alfv\'en waves, with the diffusion coefficient set by the balance between streaming-driven Alfv\'en wave excitation and damping mediated by local gas properties. We find that the combined transport processes are more effective in environments with higher SFR. These environments are characterized by higher-velocity hot outflows (created by clustered supernovae) that rapidly advect CRs away from the galactic plane. As a consequence, the ratio of midplane CR
pressure to midplane gas pressures decreases with increasing SFR. We also use the post-processed simulations to make predictions regarding potential dynamical impacts of CRs. The relatively flat CR pressure profiles near the midplane argue that they would not provide significant support against gravity for most of the ISM mass.  
However, the CR pressure gradients are larger than the other pressure gradients in the extra-planar region ($|z|>0.5$~kpc), suggesting that CRs may affect the dynamics of galactic fountains and/or winds. The degree of this impact is expected to increase in environments with lower SFR. 
\end{abstract}

\vspace*{1cm}

\section{Introduction}

Cosmic rays (CRs) are charged particles moving with relativistic speeds, directly detected in and near the solar system and observed from emission they create in the Milky Way and in other galactic disks. 
Except for the highest energy particles,
CRs are believed to be mostly produced in shocks created by supernovae, with $\sim 10\%$ of the injected supernova energy going into CR acceleration  \citep[e.g.][]{Bell04,Morlino&Caprioli12}. Direct observations of CRs at the Earth and in the heliosphere indicate that their kinetic energy spectrum extends from at least $ \sim 10^6$~eV up to $\sim 10^{20}$~eV, and for the protons that comprise most of the CR energy is well approximated by a broken power law that peaks at energies near $10^9$~eV \citep[see reviews by][]{Strong+07,Grenier+15}. The total CR energy density in the solar neighborhood is $\sim 1$~eV~cm$^{-3}$, a value comparable to the measured  thermal, turbulent and magnetic energy densities  \citep[e.g.][]{Boulares&Cox90, Beck01}. This evidence suggests that CRs can significantly contribute to the dynamics of the interstellar medium (ISM).
A fundamental question is whether the rough equipartition among different pressure components holds in other galactic environments.

Far from the solar system,
where CRs cannot be directly detected, indirect observations of hadronic CRs (protons and heavier nuclei) come from high-energy $\gamma$-ray emission. CRs with kinetic energies $\gtrsim 1$~GeV collide with thermal gas in the ISM producing pions, which decay into $\gamma$-rays. 
So far, $\gamma$-ray emission has been observed in the Milky Way, in star-forming galaxies in the Local Group, and in a few low-redshift starburst galaxies \citep{Abdo+10a, Abdo+10b, Abdo+10c, Ackermann+12, Acero+16,Ahronian2020,Peron2021}, revealing on large scales a tight correlation with the far-infrared luminosity of the galaxy that is emitted by dusty gas surrounding star-forming regions. Since pionic $\gamma$-ray luminosity is proportional to the CR energy density, the correlation between $\gamma$-ray and far-infrared luminosity hints at a connection between the CR energy density and the star formation rate (SFR). 
A relationship of this kind is understandable given that the majority of supernovae originate from recently formed massive stars, but CR transport 
as well as the CR production rate affects 
the CR energy density. 
Combining $\gamma$-ray observations with observations of synchrotron emission by CR electrons, several works have tried to constrain the energy density of CRs in external galaxies and suggested that while the energy equipartition observed in the solar neighborhood holds in local star-forming galaxies, starburst environments are characterized by CR energy densities lower than the other relevant energy densities \citep[e.g.][]{Lacki11,Yoast-Hull+13,Yoast-Hull+16}. The number of galaxies detected in $\gamma$-rays thus far however remains too limited to draw any robust conclusion about the relevance of CRs in different star-forming environments. 

From a theoretical point of view, the dynamical impact of CRs 
is of particular interest for their possible role in driving galactic winds.
This process been widely studied in both one-dimensional analytic models \citep[e.g.][]{Ipavich75, Breitschwerdt+91, Everett+08, Dorfi&Breitschwerdt12, Mao&Ostriker18, Crocker+21a, Quataert+21a, Quataert+21b, Recchia21} and numerical simulations of isolated galaxies or cosmological zoom-ins \citep[e.g.][]{Booth+13,Hanasz2013,Salem&Bryan13,Pakmor+16,Ruszkowski+17,Chan+19,Dashyan+20,Hopkins+20,Girichidis+21} and portions of ISM \citep[e.g.][]{Girichidis+16, Simpson+16, Farber+18, Girichidis+18}.
In addition to driving galactic outflows, CRs may also contribute to the internal support of disks against gravity (regulating their level of star formation), and contribute to heating and ionization of both the ISM and circumgalactic medium \citep[e.g.][]{Wiener+19,Butsky+20, Ji+20,Kempski+20,Bustard+21}. However, the degree to which CRs are able to affect these phenomena is strongly sensitive to 
CR propagation (both in models and in reality).

One of the main uncertainties in modeling the propagation of CRs is that the microphysical processes coupling CRs to the thermal gas are not completely understood \citep[see review by][]{Amato&Blasi18}. The interaction between CRs and thermal gas is mostly collisionless and mediated by the ambient magnetic field. As CRs stream along magnetic field lines, they scatter off small-scale (of order the CR gyroradius) magnetic fluctuations, reducing their effective propagation speed. It is still unclear to what extent these fluctuations are Alfv\'{e}n waves excited by the CRs themselves via resonant streaming instability \citep[the ``self-confinement'' scenario; e.g.][]{Kulsrud&Pearce69, Wentzel74,Bai2019}, or background turbulent fluctuations \citep[``extrinsic turbulence'' scenario; e.g.][]{Chandran00, Yan&Lazarian02}, although detailed spectral modeling supports self-confinement for the lower-energy CRs representing most of the total energy, and external turbulence for very high energy CRs \citep{Blasi2012,Evoli+18}. In the self-confinement scenario, scattering by resonant Alfv\'{e}n waves can in principle prevent CRs from streaming faster than the local Alfv\'{e}n speed if wave amplitudes are sufficiently large.
However, wave amplitudes and therefore scattering rates are reduced by wave damping, which is especially effective in the higher-density, lower-ionization portions of the ISM containing almost all of the mass and the majority of the volume near the midplane   
\citep[e.g.][]{Kulsrud05,Plotnikov2021,Bambic2021}. For 
self-excited waves, the transport of CRs relative to the background gas can be described as a combination of streaming down CR pressure gradients at the local Alfv\'{e}n speed and diffusion relative to the Alfv\'{e}n waves.
In the extrinsic turbulence scenario, CRs propagate relative to the gas through field-aligned diffusion only. In both scenarios, the magnetic field mediates exchange of energy and momentum between CRs and background gas.

In most studies of ISM dynamics and thermodynamics the CR kinetic scales are much smaller than the spatial scales of interest and CRs must be approximated as a fluid. The transport of the CR fluid is generally described in terms of advection along with the background thermal gas velocity and either streaming at the local Alfv\'en speed or diffusing  (primarily along the magnetic field) relative to the gas, or a combination of these \citep[see review by][]{Hanasz+21}. As explained above, the dichotomy between streaming and diffusion comes from the distinction between the self-confinement versus the extrinsic-turbulence picture for the formation of scattering waves. Another uncertainty in CR-fluid prescriptions concerns the dependence between the dominant scattering mechanism and the properties of the background gas (e.g. magnetic field structure, gas density, ionization fraction). The most common approach in previous magnetohydrodynamic (MHD) simulations has been to assume a CR scattering rate (or diffusion coefficient) that ignores the multiphase structure of the gas. In these works, the degree of scattering is generally parametrized by a constant coefficient, whose value is based on empirical estimates in the Milky Way \citep[e.g.][]{Trotta+11, Cummings+16, Johannesson+16}, although other approach to setting the scattering rate have recently been explored in galactic-scale MHD simulations by \citet{Hopkins+20}.

With the goal of studying the 
dependence of CR propagation on the properties of the underlying multiphase ISM, in \citet{Armillotta+21}, we post-processed the TIGRESS\footnote{ Three-phase Interstellar medium in Galaxies Resolving Evolution with Star formation and Supernova feedback} MHD simulation modeling a region of a galactic disk representative of our solar neighborhood \citep{Kim&Ostriker17} with a two-moment fluid algorithm for CR transport \citep{Jiang&Oh18}. The solar neighborhood TIGRESS simulation we used employs a tall box that intersects the galactic midplane in a square kpc patch and extends seven kpc vertically, with uniform resolution $\Delta x =8$~pc so that both hot and cool ISM phases are well resolved.  
In addition to simple propagation prescriptions with spatially-constant scattering, we explored the physically-motivated case in which the scattering coefficient varies spatially. We mostly focused on GeV~CRs as they contain most of the energy and momentum of the CR population and are therefore more relevant for the gas dynamics. Since estimates for the Galactic disk suggests that the waves that scatter GeV~CRs are mostly driven by the streaming instability 
\citep[e.g.][]{Zweibel13, Zweibel17, Evoli+18}, in our physically-motivated model we assumed that CRs are scattered by self-excited Alfv\'{e}n waves and that the wave amplitude is set by the balance of streaming-driven growth and damping \citep[considering both ion-neutral damping and non-linear Landau damping,][]{Kulsrud&Pearce69, Kulsrud05}. 
We also separately ran post-processing transport models of CRs with kinetic energy 30 MeV (representative of the population most important for producing ionization), which have a streaming instability growth rate and collisional loss terms that differ from those of the GeV CRs.  

In \citet{Armillotta+21}, we found that advection by thermal gas is the main CR transport mechanism in the fast-moving hot gas, while both diffusion and streaming are important in the cooler and denser gas, which moves at lower velocity. The analysis of our physically-motivated model showed that the scattering coefficient may vary over more than four orders of magnitude depending on properties of the background gas. 
The scattering rate in in the neutral gas is quite low due to strong ion-neutral wave damping, which makes the CR pressure nearly uniform in warm-cold gas at density $\gtrsim 0.1 \, \pcc $.
The propagation of CRs out of the neutral gas is however limited by the high scattering rate in the surrounding hotter and lower-density gas. As a consequence, CRs are strongly confined in the dense galactic disk, where most of the neutral mass resides.

In this work, we go beyond the solar-neighborhood environment and apply the physically-motivated transport prescription developed in \citet{Armillotta+21} to other galactic conditions. For this analysis, we compare the solar neighborhood model with two other TIGRESS simulations from the suite described in \citep{Kim+20}, which cover a range of input gas surface density and gravitational potential and output SFR surface density and thermal, turbulent, and magnetic pressures. Our overall goal is to understand how the propagation of CRs in star-forming galaxies is affected by the detailed interstellar properties. Key questions we address are: are there systematic variations across environment in (1) diffusion coefficients and effective transport speeds? (2) ratios of CR pressure to other pressures and ratios of CR pressure to the star formation rate? (3)  the potential for CRs to drive winds, based on the CR momentum flux? Related to the last question, we also explore the CR pressure gradient forces and the gas flow and Alfv\'en speeds at high altitudes, which may have implications for understanding cloud acceleration.

The layout of the paper is as follows. In \autoref{sec:methods}, we briefly describe the suite of TIGRESS simulations analyzed in this work and the algorithms we use to compute the transport of CRs. In \autoref{sec:results-Ipart}, we present the results of our post-processed simulations and characterize how and why CR propagation differs with environment. 
In \autoref{sec:results-IIpart}, we use the simulation outcomes to investigate 
the potential for CR momentum transfer to drive large-scale galactic winds.
Finally, in \autoref{sec:conclusions}, we summarize and discuss our main results.

\section{Methods}
\label{sec:methods}

We apply the methods used in \citet{Armillotta+21} to compute the propagation of CRs depending on the underlying distribution of thermal gas density, velocity, and magnetic field.  Here we briefly summarize our models and methods,  and refer readers to \citet{Armillotta+21} for further details. 

\subsection{TIGRESS models}
\label{sec:Tigress}

In the TIGRESS MHD simulations, local patches of galactic disks are self-consistently modeled including star formation and feedback in the form of far-UV (FUV) heating and resolved supernova remnant expansion   \citep{Kim&Ostriker17, Kim+20}. The TIGRESS framework is built on the grid-based MHD code \textit{Athena} \citep{Stone+08}. The ideal MHD equations are solved in a shearing-periodic box \citep{Stone&Gardiner10} representing a $\sim$kpc-sized patch of a differentially-rotating galactic disk. The physics treated includes gas self-gravity and gravitational forces from an old stellar disk and dark matter halo (treated via fixed external potentials), optically thin cooling, and FUV photoelectric heating. Sink particles are created to represent star cluster formation in cells undergoing unresolved gravitational collapse. Each sink/star particle is treated as a star cluster with coeval stellar population that fully samples the Kroupa initial mass function \citep{Kroupa01}. Young massive stars (star particle age $t_\mathrm{sp} \lesssim 40$~Myr) provide feedback to the  ISM representing effects of FUV radiation and core-collapse supernova. The instantaneous FUV luminosity and the rate of supernova explosions for each star cluster are adopted from the \textsc{STARBURST99} population synthesis model \citep{Leitherer+99}.

The TIGRESS simulations are run for a time long enough to cover several star-formation/feedback cycles. After the first star formation burst and feedback cycle, an overall self-regulated state -- with periods of enhanced star formation followed by periods of enhanced feedback -- is reached, and a realistic multiphase ISM is produced. Feedback drives turbulent motions and heats the ISM, thus providing the turbulent, thermal, and magnetic support needed to offset the vertical weight of the gas. Part of the gas heated and accelerated by supernova blast waves breaks out of the galactic plane, generating large-scale outflows in the coronal region. These outflows present a multiphase structure consisting of hot winds and warm fountains  \citep{Kim&Ostriker18, Vijayan+20}, with the dependence of outflow ``loading'' on SFR and other ISM properties characterized in \citet{Kim+20,Kim+20b}.

\begin{deluxetable*}{cccccccccccc}
\tablecaption{Model parameters and temporally-averaged ISM properties. \label{tab:tab1}}
\tablehead{
\colhead{Model} &
\colhead{$L_\mathrm{z}$} &
\colhead{$\Delta x$} &
\colhead{$\rho_\mathrm{DM}$} & 
\colhead{$\Sigma_\mathrm{star}$} & 
\colhead{$\Sigma_\mathrm{gas,ini}$} & 
\colhead{$t_\mathrm{orb}$} & 
\colhead{$\langle \Sigma_\mathrm{gas} \rangle$} & 
\colhead{$\langle \Sigma_\mathrm{SFR} \rangle$} & 
\colhead{$\langle n_\mathrm{mid} \rangle$} & 
\colhead{$\langle P_\mathrm{mid}/k_\mathrm{B}\rangle$} &
\colhead{$\langle H_\mathrm{gas}\rangle$}
\\
\colhead{} &
\colhead{\footnotesize(pc)} &
\colhead{\footnotesize(pc)} & 
\colhead{\footnotesize($\mo$/pc$^{3}$)} & 
\colhead{\footnotesize($\mo$/pc$^{2}$)} & 
\colhead{\footnotesize($\mo$/pc$^{2}$)} & 
\colhead{\footnotesize(Myr)} & 
\colhead{\footnotesize($\mo$/pc$^{2}$)} &
\colhead{\footnotesize(M$_\odot$/kpc$^{2}$/yr)} & 
\colhead{\footnotesize(cm$^{-1}$)} & 
\colhead{\footnotesize(K/cm$^{3}$)} & 
\colhead{\footnotesize(kpc)}
}
\colnumbers
\startdata
R2 & $\pm 1792$ & 4 & $8.0\times10^{-2}$ & 450 & 150 & 61 & 74 &1.1 &7.7 & $2.5\times10^6$& 3.5\\
R4 & $\pm 1792$ & 4 & $2.4\times10^{-2}$ & 208 & 50 & 110 & 30 &0.13 &1.4 & $4.1\times10^5$& 3.4\\
R8 & $\pm 3584$ & 8 & $6.4\times10^{-3}$ & 42  & 12 & 220 & 11 &$5.1\times10^{-3}$ &0.9 &$1.9\times10^4$ & 3.3\\
\enddata
\tablecomments{
Columns: (1) model name; (2) vertical box size; (3) spatial resolution; (4) dark matter volume density; (5) old-star surface density; (6) initial gas surface density; (7) orbital time; (8) time-averaged gas surface density; (9) time-averaged SFR surface density; (10) time-averaged midplane gas number density; (11) time-averaged midplane gas total pressure; (12) time-averaged gas scale height. The time-averaged quantities are averaged over the interval $0.5<t/t_\mathrm{orb}<1.5$.
}
\end{deluxetable*}

In this work, we extend our analysis to other two TIGRESS environments (hereafter denoted as R2 and R4), in addition to the solar neighborhood model (hereafter denoted as R8) already studied in \citet{Armillotta+21}. In \autoref{tab:tab1}, we summarize the key parameters and properties of the three models. These are meant to represent environments in a generic Milky Way-like star-forming galactic disk at radial distances of roughly 2, 4, and 8 kpc from the galactic center. From R8 to R2, the models are initialized with increasing old-star, dark-matter and gas surface densities. While the former are fixed in time, the latter decreases over time because gas turns into sink particles due to star formation and it is vertically lost as a wind. The scale height of the stellar disk is $z_* = 245$ pc in all cases, with midplane stellar volume density related to surface density by $\rho_* = \Sigma_* /(2z_* )$.  The R2 and R4 simulations have box size $L_\mathrm{x} = L_\mathrm{y} = 512$~pc and $L_\mathrm{z} = 3584$~pc with a uniform spatial resolution $\Delta x = 4$~pc, while the R8 simulation has box size $L_\mathrm{x} = L_\mathrm{y} = 1024$~pc and $L_\mathrm{z} = 7168$~pc with resolution $\Delta x = 8$~pc. The larger box size 
in R8 is needed because, due to the lower mean gas density, individual superbubbles created by correlated supernovae explosions can fill the whole midplane volume if the box size is too small.  The higher resolution in R2 and R4 is required to achieve robust convergence of several ISM and outflow properties \citep[see][]{Kim+20}. In \citet{Armillotta+21}, we found that a resolution $\Delta x \leq 16$~pc is sufficient to achieve convergence of CR properties.

For each TIGRESS model, we select and post-process about ten snapshots at equal intervals within the time range $0.5<t/t_\mathrm{orb}<1.5$, with $t_\mathrm{orb} = 2 \pi/ \Omega$ the orbital time (col. 7 in \autoref{tab:tab1}) and $\Omega$ the angular velocity of galactic rotation at the domain center. In this way, we exclude the initial transient state from our analysis. In \autoref{tab:tab1} (col. $5-7$), we list some relevant properties of the three models averaged over the time interval investigated here. As a consequence of the stronger gravitational potential and the higher gas surface density, the time-averaged gas density, pressure, and SFR  surface density increase from R8 to R4 to R2. The three models thus cover a wide range of environmental properties, in terms of gas surface density ($ \Sigma_\mathrm{gas} \sim 10-100 \, \mo$~pc$^{-2}$), SFR surface density ($\Sigma_\mathrm{SFR} \sim 0.005 - 1 \, \mo$~kpc$^{-2}$~yr$^{-1}$), and midplane total pressure ($P_\mathrm{mid}/k_\mathrm{B} \sim 10^4-10^6 $~cm$^{-3}$~K), computed as the sum of thermal, turbulent, and magnetic pressure averaged over two horizontal slices at $z = \pm \Delta x/2$. We note that, unlike the other properties, the value of the gas scale height $H_\mathrm{gas}$ happens to be quite similar in the three models.

\subsection{Algorithm for CR transport}
\label{sec:algorith}

We post-process the TIGRESS simulations with the two-moment algorithm for CR transport implemented in the \textit{Athena}++ code \citep {Stone+20} by \citet{Jiang&Oh18} and extended by \citet{Armillotta+21}. The two-moment equations governing the CR transport are:
\begin{equation}
\frac{\partial e_\mathrm{c}}{\partial t} + \mathbf{\nabla} \cdot \mathbf{F_\mathrm{c}} = - (\mathbf{v} + \mathbf{v_\mathrm{s}}) \cdot 
\tensor{\mathrm{\sigma}}_\mathrm{tot} \cdot   [  \mathbf{F_\mathrm{c}} - \mathbf{v} \cdot (\tensor{{\mathbf{P}}}_\mathrm{c} + e_\mathrm{c} \tensor{\mathbf{I}}) ]
\;,
\label{eq:CRenergy}
\end{equation}
\begin{equation}
\frac{1}{v_\mathrm{m}^2} \frac{\partial \mathbf{F_\mathrm{c}}}{\partial t} + \mathbf{\nabla} \cdot \tensor{\mathbf{P}}_\mathrm{c} = - \tensor{\mathrm{\sigma}}_\mathrm{tot} \cdot [  \mathbf{F_\mathrm{c}} - \mathbf{v} \cdot (\tensor{{\mathbf{P}}}_\mathrm{c} + e_\mathrm{c} \tensor{\mathbf{I}}) ] \;,
\label{eq:CRflux}
\end{equation}
where $e_\mathrm{c}$, $\mathbf{F_\mathrm{c}}$ and $\tensor{{\mathbf{P}}}_\mathrm{c}$ are the CR energy density, energy flux and pressure tensor, respectively. We assume approximately isotropic pressure, so that $\tensor{\mathbf{P}}_\mathrm{c} \equiv P_\mathrm{c}\tensor{\mathbf{I}}$, with $P_\mathrm{c} = (\gamma_\mathrm{c} -1) \,e_\mathrm{c} = e_\mathrm{c}/3$, where $\gamma_\mathrm{c} = 4/3$ is the adiabatic index of the relativistic fluid, and $\tensor{\mathbf{I}}$ is the identity tensor. The speed $v_\mathrm{m}$ represents the maximum velocity CRs can propagate. In principle, $v_\mathrm{m}$ is equal to the speed of light $c$ for relativistic CRs. However, here we adopt $v_\mathrm{m} = 10^4 \, \kms \ll c$ as it is demonstrated that the simulation outcomes are not sensitive to the exact value of $v_\mathrm{m}$ as long as $v_\mathrm{m}$ is much larger than any other speed in the simulation \citep[][]{Jiang&Oh18}. Adoption of a ``reduced speed of light'' enables larger numerical timesteps based on the CFL condition for this set of hyperbolic equations.  

In \autoref{eq:CRenergy} and \autoref{eq:CRflux}, $\mathbf{v}$ indicates the gas velocity which directly advects the CR fluid, 
while $\mathbf{v_\mathrm{s}}$ represents the CR streaming velocity,
\begin{equation}
\mathbf{v_\mathrm{s}} \equiv - \mathbf{v_\mathrm{A,i}}
\, \frac{\mathbf{B} \cdot (\nabla \cdot \tensor{\mathbf{P}}_\mathrm{c})}{\vert \mathbf{B} \cdot (\nabla \cdot \tensor{\mathbf{P}}_\mathrm{c})\vert}
= -\mathbf{v_\mathrm{A,i}} 
\frac{\hat{B} \cdot \nabla P_\mathrm{c}}{\vert \hat{B} \cdot \nabla P_\mathrm{c}\vert}
\;,
\label{eq:vs}
\end{equation}
defined to have the same magnitude as the local Alfv\'{e}n speed in the ions $\mathbf{v_{\rm A,i}} \equiv \mathbf{B}/\sqrt{4\pi\rho_i}$, oriented along the local magnetic field and pointing down the CR pressure gradient. Here, $\mathbf{B}$ is the magnetic field vector and $\rho_\mathrm{i} $
is the ion mass density (see Section 2.2.5 of \citealt{Armillotta+21} for the derivation of $\rho_\mathrm{i}$ in our simulations). 

Finally, the diagonal tensor $\tensor{\mathbf{\sigma}}_\mathrm{tot}$ is the wave-particle interaction coefficient, defined to allow for both scattering and streaming along the direction parallel to the magnetic field, 
\begin{equation}
    \sigma_{\rm tot,\parallel}^{-1}= \sigma_\parallel^{-1} + \frac{v_\mathrm{A,i}}{|\hat B \cdot \nabla P_\mathrm{c} |} (P_\mathrm{c} + e_\mathrm{c}) \, ,
\label{eq:sigmatotpar}    
\end{equation}
and only scattering in the directions perpendicular to the magnetic field, 
\begin{equation}
    \sigma_{\rm tot,\perp}= \sigma_{\perp}\, .
\label{eq:sigmatotperp}    
\end{equation}
For the relativistic case, $\sigma_\parallel=\nu_\parallel/c^2$ and $\sigma_\perp=\nu_\perp/c^2$, where $\nu_\parallel$ is the scattering rate parallel to the magnetic field direction due to Alfv\'en waves that are resonant with the CR gyro-motion and $\nu_\perp$ is an effective perpendicular scattering rate (see \autoref{sec:sigma}).

CRs transfer 
momentum to the ambient gas at a rate per unit volume given by the term $ - \tensor{\mathrm{\sigma}}_\mathrm{tot} \cdot   [ \mathbf{F_\mathrm{c}} - \mathbf{v} \cdot (P_\mathrm{c} + e_\mathrm{c})\tensor{\mathbf{I}}] = \tensor{\mathrm{\sigma}}_\mathrm{tot} \cdot   ( \mathbf{F_\mathrm{c}} - 4/3 \mathbf{v} e_\mathrm{c})$ (RHS of \autoref{eq:CRflux}), and transfer energy at a rate per unit volume given by
$ - (\mathbf{v} + \mathbf{v_\mathrm{s}}) \cdot \tensor{\mathrm{\sigma}}_\mathrm{tot} \cdot ( \mathbf{F_\mathrm{c}} - 4/3 \mathbf{v} e_\mathrm{c})$ (RHS of \autoref{eq:CRenergy}); with a sign change these would be applied as respective source terms in the gas momentum and energy equations (although in the current work we do not include an MHD ``back-reaction''). For the energy equation, $- \mathbf{v} \cdot  \tensor{\mathrm{\sigma}}_\mathrm{tot} \cdot   ( \mathbf{F_\mathrm{c}} - 4/3 \mathbf{v} e_\mathrm{c}) $ describes the direct CR pressure work done on or by the gas, while $- \mathbf{v_\mathrm{s}} \cdot  \tensor{\mathrm{\sigma}}_\mathrm{tot} \cdot ( \mathbf{F_\mathrm{c}} - 4/3 \mathbf{v} e_\mathrm{c}) $ represents the rate of energy transferred to the gas via wave damping. We note that the RHSs of \autoref{eq:CRenergy} and \autoref{eq:CRflux} reduce to zero in the absence of wave-particle interaction
(i.e., $\sigma_\mathrm{tot} \simeq 0$).
In this limit, CRs can freely stream at the ``reduced'' speed of light $v_\mathrm{m}$, as encoded in the time-dependent and divergence terms 
of 
\autoref{eq:CRflux} and \autoref{eq:CRenergy}.

In the limit of negligible 
time-dependent term in \autoref{eq:CRflux} (large $v_m$), we obtain the canonical expression for $\mathbf{F_\mathrm{c}}$,
\begin{equation}
\mathbf{F_\mathrm{c}} =   \frac{4}{3}\,e_\mathrm{c}\, (\mathbf{v} + \mathbf{v_\mathrm{s}})  
- \tensor{\mathbf{\sigma}}^{-1} \cdot \nabla P_\mathrm{c}\, ,
\label{eq:SteadyFlux}
\end{equation}
by combining \autoref{eq:CRflux} -- \autoref{eq:sigmatotpar}. 
\autoref{eq:SteadyFlux} shows that 
for quasi-steady state 
CR transport is given as a sum of advection ($4/3 e_\mathrm{c} \mathbf{v}$), streaming ($4/3 e_\mathrm{c} \mathbf{v_\mathrm{s}}$) and diffusion ($- \tensor{\mathbf{\sigma}}^{-1} \cdot \nabla P_\mathrm{c}$), where the diffusion term becomes small if wave amplitudes are large 
(large $\sigma$). In addition to the gas-advection and streaming velocity, we can define the diffusion velocity as
\begin{equation}
  \mathbf{v}_\mathrm{d} \equiv - \frac{3}{4} \, \tensor{\mathbf{\sigma}}^{-1} \cdot \frac{  \nabla P_\mathrm{c}}{e_\mathrm{c}} \, ,  
 \label{eq:vd_def}
\end{equation}
which indicates the CR propagation speed relative to the wave frame. 

In \citet{Armillotta+21}, we supplement \autoref{eq:CRenergy} and \autoref{eq:CRflux} with additional source and sink terms representing injection of CR energy from supernovae and collisional losses due to the interaction of CRs with the star-forming ISM. These additional terms are described in \autoref{sec:source terms}. In \autoref{sec:sigma}, we explain how the scattering coefficients $\sigma_\parallel$ and $\sigma_\perp$ are calculated in the code. 

\subsubsection{Source/sink terms}
\label{sec:source terms}

The injection of CR energy from supernovae enters in the RHS of \autoref{eq:CRenergy} through a source term $Q$, representing the injected CR energy density per unit time. We assume that the injected energy is distributed around each star cluster particle following a Gaussian profile, and, in each cell, we calculate $Q$ as 
\begin{equation}
Q = \frac{1}{2 \pi \sqrt{2 \pi} \, \sigma_\mathrm{inj}^3 }\,\sum_{\mathrm{sp} =1}^{N_\mathrm{sp}} \dot{E}_\mathrm{c, sp} \cdot \mathrm{exp} (-r_\mathrm{sp}^2/2 \sigma_\mathrm{inj}^2)\;,
\label{eq:injectedenergy}
\end{equation}
where the sum is taken over all the star cluster particles in the simulation box. In \autoref{eq:injectedenergy}, $r_\mathrm{sp}$ is the distance between the cell center and the star particle, $\sigma_\mathrm{inj} = 4 \, \Delta x$ is the standard deviation of the distribution\footnote{In \citet{Armillotta+21}, we explored a range of different $\sigma_\mathrm{inj}$, from $2 \, \Delta x$ to $10 \, \Delta x$, and we found that the simulation outcomes are independent of this choice.}, while $\dot{E}_\mathrm{c, sp}$ is the rate of injected CR energy. The latter is calculated as $\dot{E}_\mathrm{c, sp} = \epsilon_\mathrm{c} \, E_\mathrm{SN} \,\dot{N}_\mathrm{SN}$, where $\epsilon_\mathrm{c}$ is the fraction of supernova energy that goes into production of CRs, assumed to be equal to 0.1 \citep[e.g.][]{Morlino&Caprioli12,Ackermann+13}, $E_\mathrm{SN} = 10^{51}$~erg is the energy released by an individual supernova event, and $\dot{N}_\mathrm{SN} = m_\mathrm{sp} \,\xi_\mathrm{SN} (t_\mathrm{sp})$ is the number of supernovae per unit time, with $m_\mathrm{sp}$ the star particle mass and $t_\mathrm{sp}$ the mass-weighted age. $\xi_\mathrm{SN}$, defined as the number of supernovae per unit time per star cluster mass at a given time $t_\mathrm{sp}$, is determined from the \textsc{STARBURST99} code \citep[see][]{Kim&Ostriker17}. 

Sink terms, associated with the interaction of CRs with the surrounding gas, are included by adding the terms  
\begin{equation}
\Gamma_\mathrm{e_c} = - \Lambda_\mathrm{coll}(E)n_\mathrm{H} e_\mathrm{c}
\label{eq:lostenergy}
\end{equation}
and
\begin{equation}
\mathbf{\Gamma}_\mathrm{F_c} = - \frac{\Lambda_\mathrm{coll}(E) n_\mathrm{H}}{v_\mathrm{p}^2} \,\mathbf{F}_\mathrm{c} \;.
\label{eq:lostflux}
\end{equation}
to the RHS of \autoref{eq:CRenergy} and \autoref{eq:CRflux}, respectively. Here, $n_\mathrm{H}$ is the hydrogen number density, while $v_\mathrm{p} = \sqrt{1-(m_\mathrm{p} c^2/E)^2}$ is the CR proton velocity, where $m_\mathrm{p}$ is the proton mass and $E \equiv E_\mathrm{k} + m_\mathrm{p} c^2$ is the total relativistic energy, with $E_\mathrm{k}$ the kinetic energy. For CRs with $E_\mathrm{k} \simeq 1$~GeV, $v_\mathrm{p} \approx c$. Finally, $\Lambda(E)$ is defined as $\Lambda(E) = v_\mathrm{p} L(E)/E$, where $L(E)$ is the energy loss function for protons, defined as the product of the energy lost per ionization event and the cross section of the collisional interaction.

$L(E)$ is a function of the CR energy and its value at a given energy can depend on one or more collisional processes. GeV CRs, which are the focus of this study, primarily collide with the ambient gas through hadronic interactions leading to a decay of pions into $\gamma$-rays. We extract the value of $L(E)$ at $E_\mathrm{k} \simeq 1$~GeV from the gray line in Figure~2 of \citet{Padovani+20}, representing the loss function for a medium of pure atomic hydrogen, and multiply it by a factor 1.21, to account for elements heavier than hydrogen. We adopt $L(E) = 3 \times 10^{-17}  $~eV~cm$^{2}$, meaning that $\Lambda_\mathrm{coll}$ in \autoref{eq:lostenergy} and \autoref{eq:lostflux} is equal to $4 \times 10^{-16}$ cm$^{3}$~s$^{-1}$.

\subsubsection{Scattering coefficient}
\label{sec:sigma}

For GeV CRs -- which are the focus of this paper -- the dominant transport mode is self-confinement via streaming instability \citep{Zweibel13, Zweibel17}. In this picture, CRs with a bulk drift speed greater than the Alfv\`{e}n  speed can excite Alfv\`{e}n  waves through gyro-resonance and scatter off these waves as they propagate in the direction of decreasing CR density \citep{Kulsrud&Pearce69, Wentzel74}. We derive the scattering coefficient $\sigma_\parallel$ based on the predictions of the self-confinement picture and assuming that, in steady state, the conversion of CR energy to wave energy is balanced by some form of wave damping \citep{Kulsrud&Pearce69,Kulsrud&Cesarsky1971}. 

In \citet{Armillotta+21}, we demonstrate that in steady state the growth rate of streaming-driven Alfv\'{e}n waves \citep[from][]{Kulsrud05} can be written as 
\begin{equation}
\Gamma_\mathrm{stream} (p_1) = \frac{\pi^2}{4} \frac{\Omega_0  m_\mathrm{p} v_\mathrm{A,i}}{B^2} \frac{\vert \mathbf{{\hat{B}}} \cdot \nabla  P_\mathrm{c}\vert}{ \sigma_\parallel P_\mathrm{c}}\, n_\mathrm{1} \;,
\label{eq:GrowthRate}
\end{equation}
where $p_1=m_p \Omega_0/k$ is the resonant momentum for wavenumber $k$ and $\Omega_0 = e \vert \bf{B} \vert / (m_\mathrm{p} c)$ is the cyclotron frequency for $e$ the electron charge. $n_\mathrm{1}$ is defined as
\begin{equation}
n_1 \equiv 4 \pi p_1 \int_{p_1}^\infty p F(p) dp\;,
\label{eq:n1}
\end{equation}
where $F(p)$ is the CR distribution function in momentum space, normalized as $4\pi\int_0^\infty F(p)p^2 dp /n_\mathrm{c} =1$ with $n_\mathrm{c}$ the CR number density. The CR spectrum is well determined in the solar neighborhood for CRs with kinetic energies $E_\mathrm{k} \gtrsim 1$~GeV, although there are considerable uncertainties at lower energy \citep[see e.g.][and references therein]{Padovani+18,Padovani+20}. At $E_\mathrm{k} \gtrsim 1$~GeV, $F(p)$ can be parametrized with a power law distribution, whose slope is $-4.7$ \citep[e.g.][]{Aguilar+14, Aguilar+15}. In \citet{Armillotta+21}, we show that $n_1 = 1.1 \times 10^{-10} \, [e_\mathrm{c}(E_\mathrm{k} 
\ge
1\mathrm{GeV})/1\mathrm{eV}] $~cm$^{-3}$ for $p_1 = p_1(E_\mathrm{k}=1\mathrm{GeV})$ even allowing for a range of low-energy slopes.

We consider two mechanisms that can limit the amplitude of 
Alfv\`{e}n waves, namely ion-neutral damping and nonlinear Landau damping. The ion-neutral damping arises from friction between ions and neutrals in partially ionized gas, where the latter are not tied to magnetic fields.
The rate of ion-neutral damping is \citep{Kulsrud&Pearce69}
\begin{equation}
\Gamma_\mathrm{damp,in} = \frac{1}{2} \frac{n_\mathrm{n} m_\mathrm{n}}{m_\mathrm{n}+m_\mathrm{i}} \langle \sigma v \rangle_\mathrm{in}\;,
\label{eq:DampingRateIN}
\end{equation}
where $n_\mathrm{n}$ is the neutral number density, $m_\mathrm{n}$ is the mean mass of neutrals, $m_\mathrm{i}$ is the mean mass of ions (see Section 2.2.5 of \citealt{Armillotta+21} for the derivation of $n_\mathrm{n}$, $m_\mathrm{n}$ and $m_\mathrm{i}$) and $\langle \sigma v\rangle_\mathrm{in}$ is the rate coefficient, equal to $\sim 3\times10^{-9}$~cm$^3$~s$^{-1}$ for  ion-neutral collisions between H and H$^+$
\citep[][Table 2.1]{Draine11}. 

Nonlinear Landau damping occurs when thermal ions have a resonance with the beat wave formed by the interaction of two resonant Alfv\`{e}n waves. The rate of nonlinear Landau damping is \citep{Kulsrud05}
\begin{equation}
\Gamma_\mathrm{damp,nll} =0.3 \, \Omega \, \frac{v_\mathrm{t,i}}{c} \left(\frac{\delta B}{B}\right)^2 = 0.3 \,\frac{v_\mathrm{t,i} v_\mathrm{p}^2}{c}\sigma_\mathrm{\parallel}\;,
\label{eq:DampingRateNLL}
\end{equation}
where $\Omega = \Omega_0 / \gamma (p_1)$ is the relativistic cyclotron frequency, with $\gamma$ the Lorentz factor of CRs with momentum $p_1$, $v_\mathrm{t,i}$ is the ion thermal velocity (which we set equal to the gas sound speed $c_\mathrm{s}$), and ${\delta B}/{B}$ is the magnetic field fluctuation at the resonant scale. The quasi-linear theory predicts that the  scattering rate is $\nu_s \sim \Omega ({\delta B}/{B})^2$, while the scattering coefficient is $\sigma_\parallel  \sim \nu_s/v_\mathrm{p}^2 \sim \Omega ({\delta B}/{B})^2 /v_\mathrm{p}^2$. This relation explains the last equivalence in \autoref{eq:DampingRateNLL}.

Assuming wave growth and damping balance, we set 
\begin{equation}
\Gamma_\mathrm{stream} = \Gamma_\mathrm{damp,in} + \Gamma_\mathrm{damp,nll}
\label{eq:sigma}
\end{equation}
and solve\footnote{We note that in \citet{Armillotta+21}, we do not solve \autoref{eq:sigma}, but rather we set $\sigma_\parallel$ equal to the minimum of \autoref{eq:sigmaNLL} and \autoref{eq:sigmaIN}. The distribution of $\sigma_\parallel$ is however almost unaffected by this change.} this equation for $\sigma_\parallel$. The solution of \autoref{eq:sigma} reduces to 
\begin{equation}
\begin{split}
\sigma_\mathrm{\parallel,nll} =& \sqrt{\frac{\pi}{16} \, \frac{\vert \mathbf{{\hat{B}}} \cdot \nabla P_\mathrm{c}\vert}{v_\mathrm{A,i} P_\mathrm{c}} \frac{\Omega_0 c}{ 0.3 v_\mathrm{t,i} v_\mathrm{p}^2} \frac{m_\mathrm{p}}{m_\mathrm{i}} \frac{n_\mathrm{1}}{n_\mathrm{i}}}  \\
\sim 1.3 \times& 10^{-28} \mathrm{\frac{s}{cm ^{2}}} \, {\vert \mathbf{\hat{B}} \cdot  \nabla P_\mathrm{c}\vert}_\mathrm{-4}^{1/2} c_\mathrm{s,200}^{-1/2} (x_\mathrm{i} n_\mathrm{H,-3})^{-1/4}
\end{split}
\label{eq:sigmaNLL}
\end{equation}
in well ionized, low-density 
gas where nonlinear Landau dominates, and to \begin{equation}
\begin{split}
\sigma_\mathrm{\parallel,in} (p_1) = &\frac {\pi}{8} \, \frac{\vert \mathbf{\hat{B}} \cdot \nabla  P_\mathrm{c}\vert}{v_\mathrm{A,i} P_\mathrm{c}}  \frac{\Omega_0}{ \langle \sigma v \rangle_\mathrm{in}} \, \frac{m_\mathrm{p} (m_\mathrm{n} + m_\mathrm{i})}{ n_\mathrm{n} m_\mathrm{n}  m_\mathrm{i}}  \frac{n_1}{n_\mathrm{i}}  \\
 \sim 3.4& \times 10^{-31} \mathrm{\frac{s}{cm ^{2}}} \,  {\vert \mathbf{\hat{B}} \cdot  \nabla P_\mathrm{c}\vert}_\mathrm{-4} x_\mathrm{i,-2}^{-1/2} n_\mathrm{H,0}^{-3/2}
\end{split}
\label{eq:sigmaIN}
\end{equation}
in primarily-neutral, denser gas where ion-neutral damping dominates \citep{Armillotta+21}. In the above, $x_i=n_i/n_H$ is the ion fraction, with $n_\mathrm{i}$ the ion number density. For gas at $T>2 \times 10^4$~K, the ion faction is calculated from the values tabulated by \citet{Sutherland&Dopita93}, while for gas at $T \leqslant 2\times 10^4$~K, the ion fraction is calculated as in Equation 16.5 in \citet{Draine11}. In the latter, $x_i$ depends on the CR ionization rate, which we evaluate in each cell depending on the local value of the CR energy density (see \citealt{Armillotta+21} for more details).
We find $x_i\approx 1.099$ for collisionally ionized gas at $T\gg 2\times 10^4$~K, and $x_i\ll 1$, decreasing at higher density, in the atomic and molecular gas that is ionized by low-energy CRs. In the dimensional versions of \autoref{eq:sigmaNLL} and \autoref{eq:sigmaIN}, ${\vert \mathbf{\hat{B}} \cdot  \nabla P_\mathrm{c}\vert}_\mathrm{-4} = {\vert \mathbf{\hat{B}} \cdot  \nabla P_\mathrm{c}\vert}/(10^{-4} \rm{eV \, cm}^{-3} \, \rm{pc}^{-1})$, $c_\mathrm{s,200} = c_\mathrm{s}/(200 \, \kms)$, $n_\mathrm{H,-3} = n_\mathrm{H}/(10^{-3} \rm{cm}^{-3})$, $n_\mathrm{H,0} = n_\mathrm{H}/(1\, \rm{cm}^{-3})$, $x_\mathrm{i,-2} = x_\mathrm{i}/10^{-2}$.
We note that two different normalizations are used in the dimensional versions of \autoref{eq:sigmaNLL} and \autoref{eq:sigmaIN} based on typical values in model R8 in the regions where NLL and IN damping are important; pressure gradients and densities are overall higher in R4 and R2.

We note that, from MHD-PIC simulations of CRs in which the theoretical quasi-linear prediction is compared to an effective fluid scattering rate and to measured pitch angle diffusion of individual particles \citep[][]{Bambic2021}, the value of \autoref{eq:sigmaNLL} or \autoref{eq:sigmaIN} may be reduced by a factor $\sim 2$.  
While $\sigma_\parallel$ represents the gyro-resonant scattering coefficient along the local magnetic field direction, $\sigma_\perp$ can be understood as scattering along unresolved fluctuations of the mean magnetic field. Even though in our simulations we directly follow the CR transport along the magnetic field, we cannot resolve this all the way down to the gyroradius scales ($\sim 10^{-6}$~pc $\ll \Delta x$), and there would be an effective perpendicular scattering along unresolved magnetic-field perturbations. The scattering coefficient in the direction perpendicular to the mean magnetic field can be expressed as $\sigma_\perp \sim \sigma_\parallel \, (B/ \delta B)^2 $, with $(\delta B/B)$ the fractional magnetic field perturbation \citep[see][]{Shalchi2019,Shalchi2020}. If we assume order-unity perturbations at the scale height of the disk ($\sim 300$~pc) and extrapolate the large-scale power down to the resolution of our simulations ($4-8$~pc), we obtain $(\delta B/B)^2 \approx 0.1$. Although this argument is 
only heuristic  (and should be replaced by direct numerical measurements of the effective perpendicular diffusion with realistic ISM turbulence), for current purposes we simply   
set $\sigma_\perp = 10 \, \sigma_\parallel$ for our post-processing. In \citet{Armillotta+21}, we explored the transport of CRs in the absence of perpendicular scattering ($\sigma_\perp \gg \sigma_\parallel$) as well as the case $\sigma_\perp = 10 \, \sigma_\parallel$, and did not found any substantial difference in the CR distribution.

\section{Cosmic-ray transport in different environments}
\label{sec:results-Ipart}

\begin{figure*}
\centering
\includegraphics[width=\textwidth]{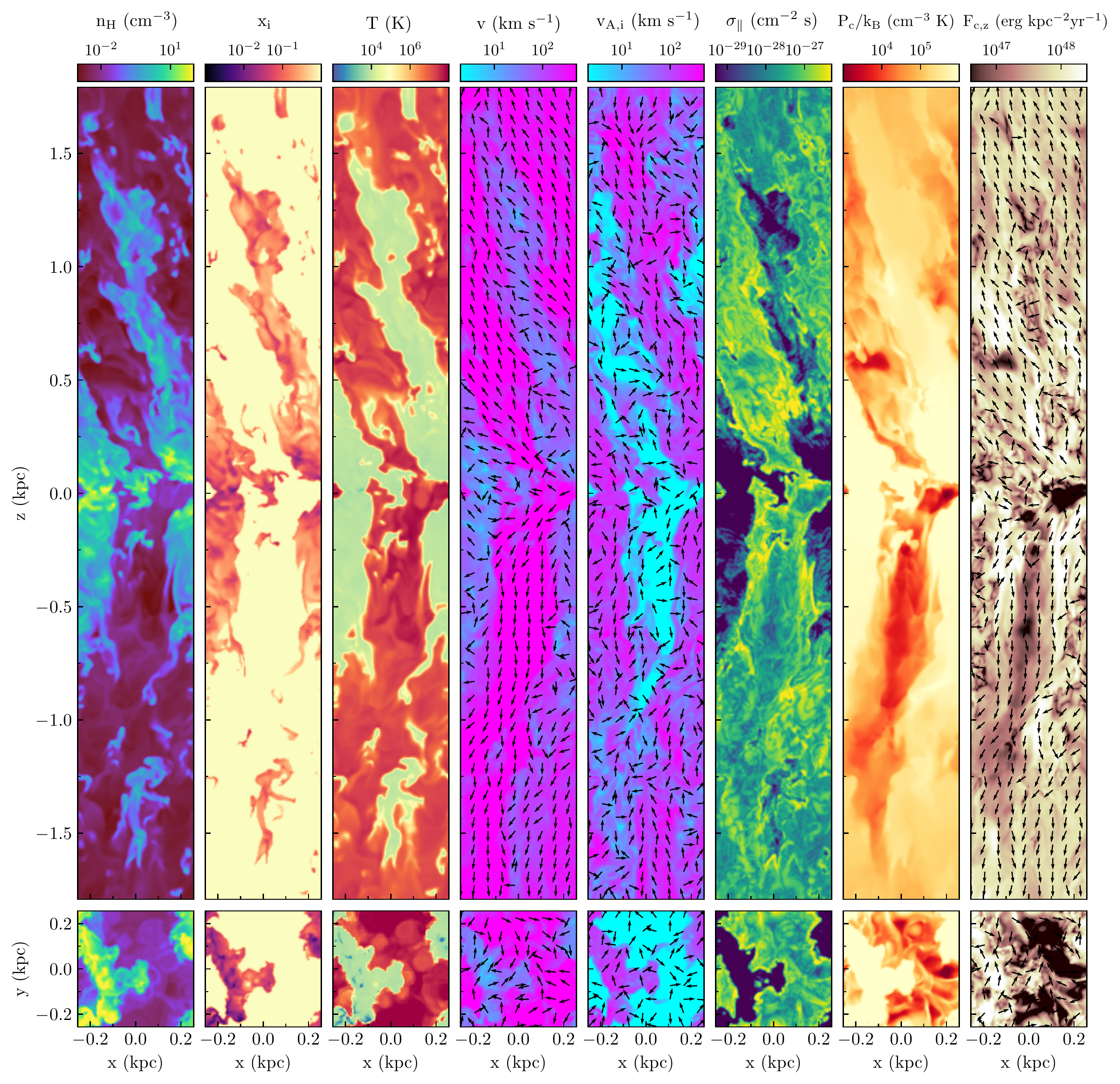}
\caption{Sample snapshot from the R2 simulation. The upper (lower) row of panels shows $x$-$z$ ($x$-$y$) slices through the center of the simulation 
box, where $x$, $y$, and $z$ are the local radial, azimuthal, and vertical directions.  
From left to right, columns show hydrogen number density $n_\mathrm{H}$, 
ion fraction $x_\mathrm{i}$, gas temperature $T$, gas speed $v$, ion Alfv\'{e}n speed $v_\mathrm{A,i}$, scattering coefficient $\sigma_\parallel$, cosmic ray pressure $P_\mathrm{c}$, and vertical cosmic ray flux $F_{\rm c,z}$. The arrows overlaid on the gas velocity, Alfv\'{e}n speed and vertical CR flux slices indicate the projected directions of the gas velocity, Alfv\'{e}n speed and CR flux, respectively, in each slice. 
}
\label{fig:R2snapshot}
\end{figure*}

\begin{figure*}
\centering
\includegraphics[width=\textwidth]{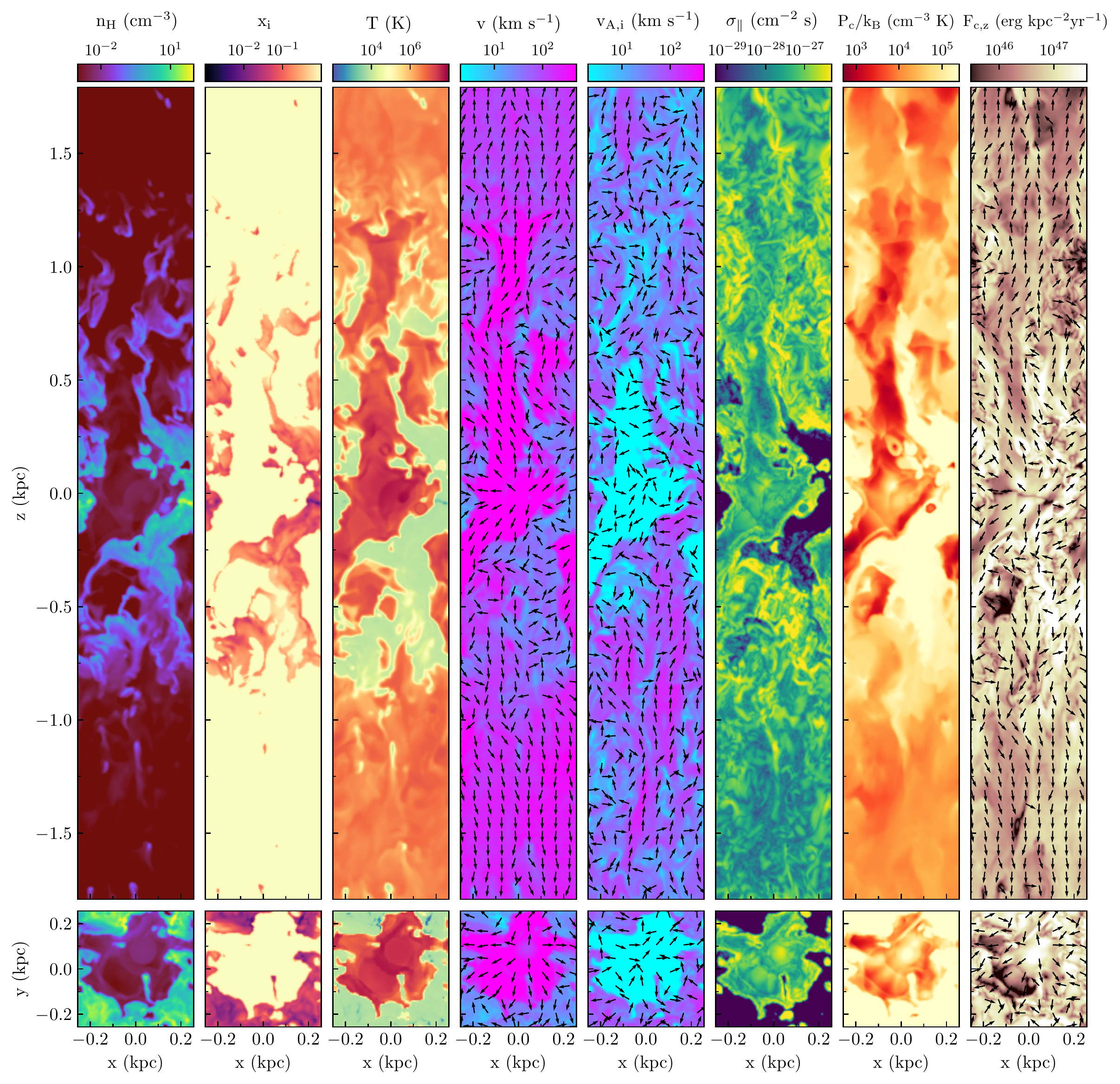}
\caption{Same as \autoref{fig:R2snapshot}, but for a sample snapshot from the R4 simulation.}
\label{fig:R4snapshot}
\end{figure*}

\begin{figure*}
\centering
\includegraphics[width=\textwidth]{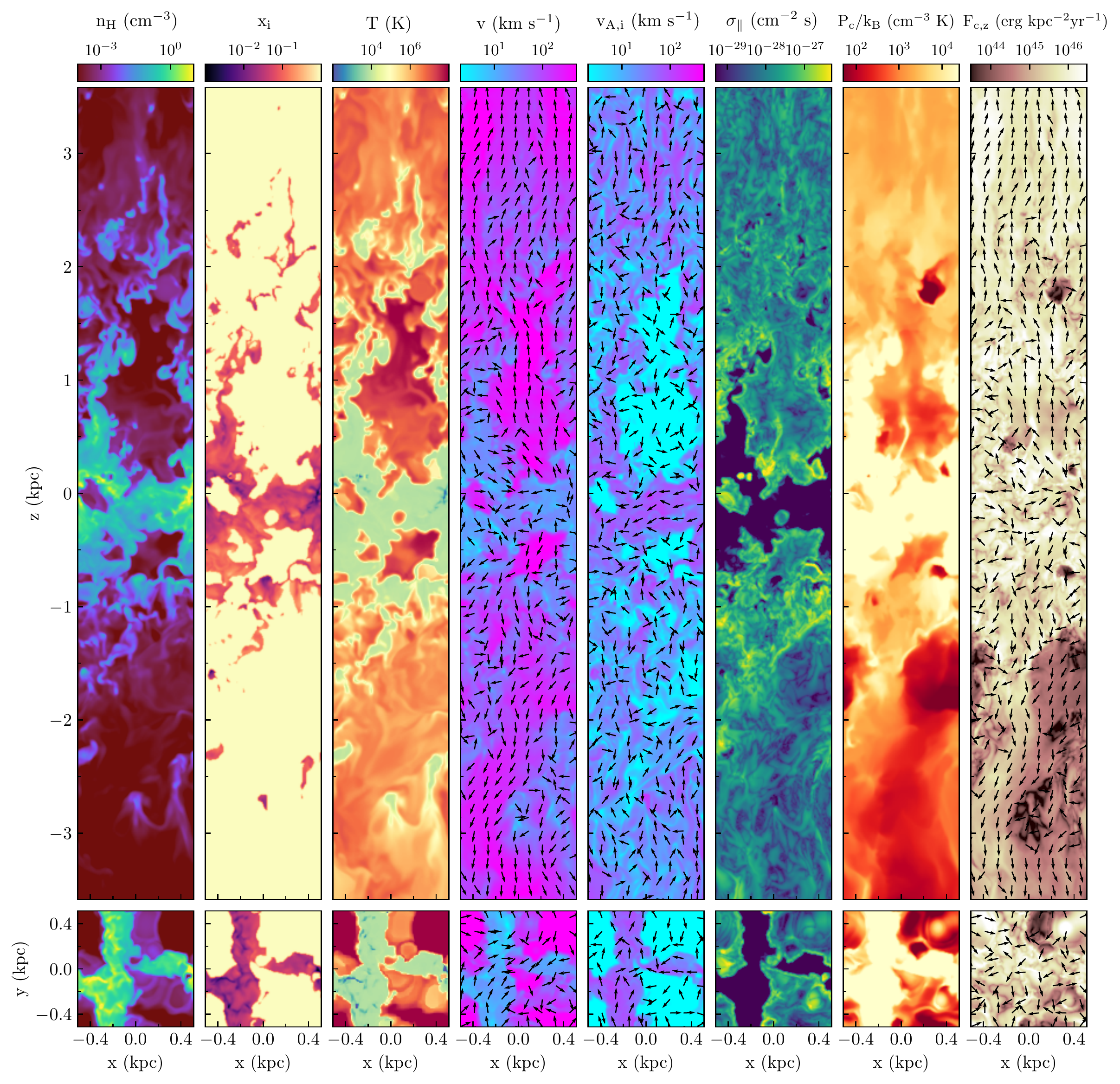}
\caption{Same as \autoref{fig:R2snapshot}, but for a sample snapshot from the R8 simulation.}
\label{fig:R8snapshot}
\end{figure*}

We use the algorithm presented in \autoref{sec:algorith} to post-process the snapshots selected from the three  TIGRESS simulations (\autoref{sec:Tigress}). 
In  post-processing, we freeze the MHD variables and evolve only the CR energy and flux density according to the methods of \autoref{sec:algorith} until the CR energy density has reached a steady state;  quantitatively we adopt the criterion
$(e_\mathrm{c,tot} (t) - e_\mathrm{c,tot} (t - 0.1 \,\mathrm{Myr}))/e_\mathrm{c,tot} (t) < 10^{-6}$, with $e_\mathrm{c,tot} = \int_\mathrm{Vol} e_\mathrm{c} dx^3$. The time required to reach equilibrium varies from a few tens to a few hundreds of Myr depending on the conditions of the background gas, tending to be shorter in systems with a faster outflow. In this section, we present the results of our post-processed runs. 

In \autoref{fig:R2snapshot}, \autoref{fig:R4snapshot}, and \autoref{fig:R8snapshot}, the first five columns from the left show the distribution on grid of some relevant MHD quantities in sample snapshots extracted from R2, R4, and R8, respectively. In particular, they display slices at $y=0$ (upper panels) and $z=0$ (lower panel) of hydrogen number density $n_\mathrm{H}$, ion fraction $x_\mathrm{i}$,  
gas temperature $T$,
magnitude of gas velocity $v$,
and magnitude of ion Alfv\'{e}n speed ${v_{\rm A,i}}$. 
The ion fraction is relevant for the calculation of both the ion Alfv\'{e}n speed ($v_\mathrm{A,i} \propto 1/\sqrt {n_\mathrm{i}}$) and the scattering coefficient (see \autoref{eq:sigmaNLL} and \autoref{eq:sigmaIN}). An accurate estimate of the ionization fraction is therefore important for a proper computation of CR transport. In particular, since most of the mass in the ISM is in neutral atomic and molecular gas that is too cool to be collisionally ionized and too shielded to be photoionized, the ionization is produced mainly by impact of low-energy CRs on atomic and molecular hydrogen \citep[e.g.][Ch. 16]{Draine11}. We refer to \citet{Armillotta+21} for details.

In all models, most of the computational volume is occupied by hot ($T > 10^6$~K) and rarefied gas, with a decrease in the hot-gas volume filling factor near the midplane.  Most of the mass resides near the midplane in the warm/cold ($T \lesssim 10^4$~K) ISM. As noted in \autoref{sec:Tigress}, the average ISM density decreases from R2 to R4 to R8 (see also \autoref{tab:tab1}). 
\autoref{fig:R2snapshot}-\autoref{fig:R8snapshot} show that the ion fraction is $x_i< 0.1$ in the higher-density, lower-temperature structures (both at the midplane and in the fountain region) where gas is mostly ionized by low-energy CRs, while  $x_i \approx 1.099$ in regions with temperature above $10^4$~K, where gas is assumed to be collisionally ionized. Therefore $n_\mathrm{i}\ll n_\mathrm{H}$ for $T\lesssim 10^4$~K, while $n_\mathrm{i}\approx n_\mathrm{H}$ for $T>10^4$~K. 

Regardless of the model, the gas velocity $v$ exceeds the ion Alfv\'{e}n speed $v_\mathrm{A,i}$ in the hot phase of the gas, while  $v_\mathrm{A,i}$ exceeds $v$ in the warm phase. A visual comparison between the three figures suggests that R2 is characterized by higher velocities in the hot gas and higher ion Alfv\'{e}n speeds in the warm gas; see \autoref{sec:streaming, diffusion and advection} for a more quantitative analysis of these quantities.

In the mostly-neutral warm and cold gas, the ionization fraction is low and the  ion-neutral collision frequency is small compared to the resonant frequencies relevant for CRs \citep[see e.g. Table 1 of][]{Plotnikov2021}.  As a result, for these high frequencies ions and neutrals are decoupled and Alfv\'en waves propagate only in the ions, at speed $v_\mathrm{A,i}\equiv {B}/\sqrt{4\pi\rho_i}$.  Since $\rho_i \ll \rho$ for the neutral gas, $v_\mathrm{A,i}$ exceeds the ideal Alfv\'{e}n speed ${v_{\rm A}} \equiv {B}/\sqrt{4\pi\rho}$ (which is commonly adopted in many models of CR transport).
In the three models analysed here, the average value of $x_\mathrm{i}$ in the warm gas is $\simeq 0.01-0.1$, which means $v_\mathrm{A,i}\simeq \sqrt{x_\mathrm{i}} v_\mathrm{A} \simeq (3-10) \,  v_\mathrm{A}$. For hot gas, the high ionization state implies $v_\mathrm{A,i} \approx v_\mathrm{A}$.  The distinction between $v_\mathrm{A,i}$ 
and $v_\mathrm{A}$ 
is important because the Alfv\'{e}n waves that interact with CRs propagate at $v_\mathrm{A,i}$, and this is reflected in the CR transport implementation of \citet{Jiang&Oh18}. Only at much higher density than we have in our simulations would the ion-neutral collision frequency be high enough for the well-coupled limit to apply, such that waves resonant with CRs propagate in the combined ion-neutral fluid at $v_\mathrm{A}$.

The three rightmost panels in \multiref{fig:R2snapshot}{fig:R4snapshot}{fig:R8snapshot} display some outputs of the CR transport algorithm: scattering coefficient $\sigma_\parallel$, CR pressure $P_\mathrm{c}/k_\mathrm{B}$, with $k_\mathrm{B}$ the Boltzmann constant, and magnitude of CR flux in the $z$-direction, $F_\mathrm{c,z}$. 
\footnote{We note that scale shown for the CR pressure is $\rm cm^{-3}\,K$ to enable straightforward comparison to MHD pressures; the CR energy density in $\rm eV\ cm^{-3}$ can be obtained by multiplying by a factor $8.6\times10^{-5}$.  Similarly, the CR flux, shown in units $\rm erg\ kpc^{-2}\ yr^{-1}$, can be converted to $\rm eV\ cm^{-3} km\ s^{-1}$ by multiplying by $2.1 \times 10^{-44}$.}  
The scattering coefficient distribution closely follows the distribution of the background MHD quantities. In particular, $\sigma_\parallel$ is relatively high (above $10^{-28}$~cm$^{-2}$~s) in hot, high-ionization regions and quite low (below $10^{-29}$~cm$^{-2}$~s) in cooler,  neutral regions. The highest values of $\sigma_\parallel$ are reached in intermediate-density regions at the interface between neutral and ionized gas. The main evidence that emerges from a visual comparison between the three figures is that the value of $\sigma_\parallel$ is overall higher in R2 and R4 than in R8. We refer to \autoref{sec:scattering rate} for a detailed analysis of $\sigma_\parallel$ as a function of gas density.

The qualitative distribution of CR pressure is overall similar in the three models: CRs accumulate in high-density regions, where the relatively-low gas velocities ($v<50\,\kms$) do not foster their removal, while CRs in regions with hot and fast-moving winds ($v \gg 100\,\kms$) are rapidly advected away from the mid-plane. One can note that the CR-flux streamlines mostly align with the velocity streamlines in regions with hot winds, meaning that CRs coupled to the hot gas escape the disk through these ``chimneys.''
Also, all models are characterized by extremely uniform CR pressure in high-density regions, where the very low scattering coefficient makes diffusion effective in smoothing out CR inhomogeneities. Although the three models share these qualitative features, the quantitative 
value of $P_\mathrm{c}$ and $F_\mathrm{c,z}$ increases from R8 to R4 to R2, as a consequence of the increasing SFR ($P_\mathrm{c} \propto Q \propto \dot{N}_\mathrm{SN} \propto \Sigma_\mathrm{SFR}$ -- see \autoref{eq:injectedenergy}). We shall come back to this point in the next section. 

\subsection{Cosmic-ray pressure}
\label{sec:pressure}

In all the TIGRESS simulations, the overall system reaches a quasi-steady state \citep[see \autoref{sec:Tigress} and also][]{WT_Kim2020,Vijayan+20}. Hereafter, we therefore focus on the analysis of CR properties averaged over time rather than at a single time, so that we can study mean trends. For each TIGRESS models, we use all the post-processed snapshots to construct temporally-averaged quantities.

\begin{figure*}
\centering
\includegraphics[width=\textwidth]{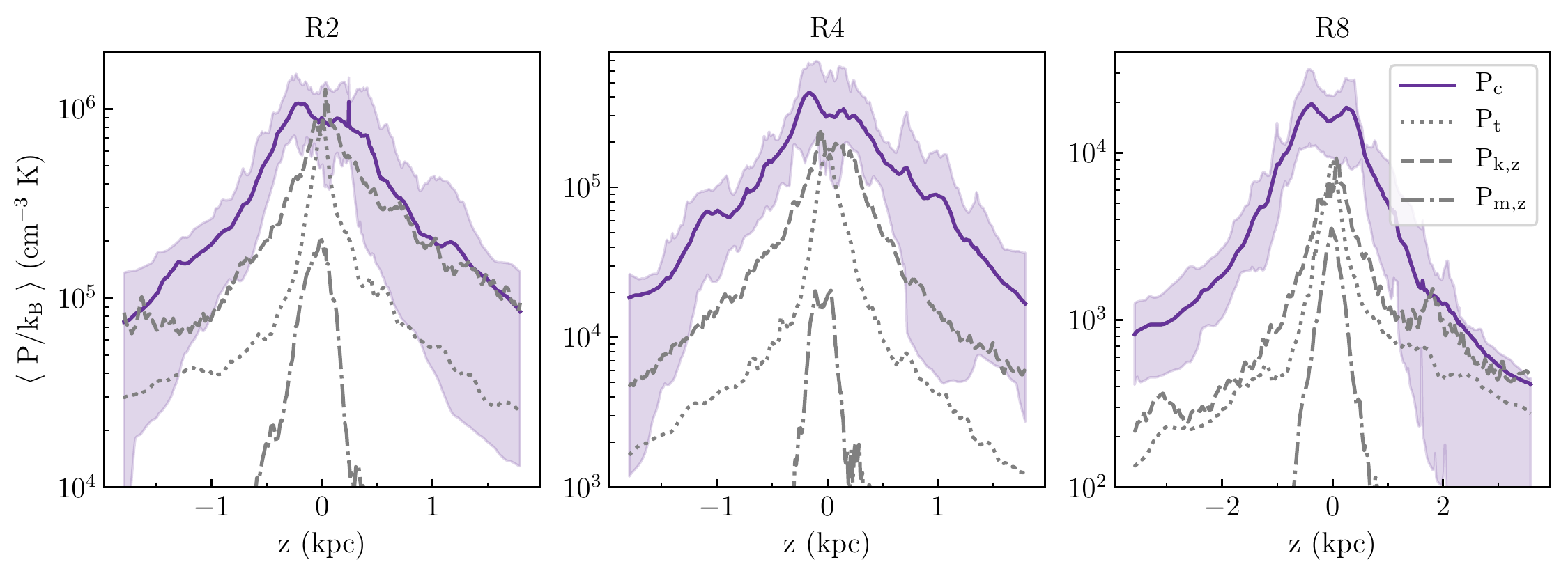}
\caption{Horizontally- and temporally-averaged vertical profiles of CR pressure $P_\mathrm{c}$ (purple), thermal pressure $P_\mathrm{t}$ (dotted gray), kinetic pressure $P_\mathrm{k,z}$ (dashed gray) and magnetic stress $P_\mathrm{m,z}$ (dot-dashed gray) for the R2 (\textit{left panel}), R4 (\textit{middle panel}) and R8 (\textit{right panel}) models. The shaded area covers the 16th and 84th percentiles from the temporal distribution of CR pressure.}
\label{fig:PressProfiles}
\end{figure*} 

In \autoref{fig:PressProfiles}, the purple lines show the horizontally- and temporally-averaged vertical profiles of CR pressure $P_\mathrm{c}$ in the three different galactic environments. In all models, the CR pressure peaks in the mid-plane, mostly occupied by slow-moving dense gas, and decreases at higher $z$. For comparison, the gray lines indicate the vertical profiles of thermal pressure $P_\mathrm{t}$, (averaged) vertical kinetic pressure $P_\mathrm{k,z} = \rho v_\mathrm{z}^2$, and vertical magnetic stress $P_\mathrm{m,z} = (B_\mathrm{x}^2 + B_\mathrm{y}^2 - B_\mathrm{z}^2)/8 \pi$, with $v_\mathrm{z}$ the gas velocity in the vertical direction,  and $B_\mathrm{x}$, $B_\mathrm{y}$, $B_\mathrm{z}$ the magnetic field components along the $x$-, $y$- and $z$-directions, respectively. Both CR and MHD pressures decrease going from R2 to R4 to R8. The overall reduction in pressure is due to the decrease in the feedback energy injection rate (per unit area)
from R2 to R4 to R8 as $\Sigma_{\rm SFR}$ decreases. However, the {\it ratio} between $P_\mathrm{c}$ and $P_\mathrm{k,z}$ (or $P_\mathrm{t}$) increases from R2 to R4 to R8. Near the midplane, thermal, kinetic and CR pressures are in equipartition in R2, while the CR pressure is more than a factor of two higher than the other pressures in R8. 

In steady state, thermal, kinetic, and magnetic pressure components in the ISM are set by balancing energy gains from star formation feedback and energy losses due to dissipative processes \citep{Ostriker2010, Ostriker2011, KimOstriker2015}. The efficiency of star formation feedback can be measured through so-called ``feedback yields'' $\Upsilon$, defined as the ratios between midplane pressure components and SFR surface density $\Sigma_\mathrm{SFR}$  \citep[note that][instead used the notation $\eta$ for the yield]{Kim+11,Kim2013}. The ratios among the individual pressure components therefore reflects the relative feedback yields. Analysis of the full set of TIGRESS models from  \citet{Kim+20} shows that the thermal yield in the warm/cold gas decreases at 
higher surface density $\Sigma$ (and $\Sigma_{\rm SFR}$) due to shielding, while the kinetic and total yield 
decrease only weakly at higher $\Sigma_\mathrm{SFR}$
\citep[Ostriker \& Kim 2021, in prep.; see also][]{Kim2013,KimOstriker2015}, meaning that total midplane MHD pressures are almost linearly proportional to $\Sigma_\mathrm{SFR}$. In the following, we investigate what sets the relation between CR pressure and $\Sigma_{\rm SFR}$ and how the CR pressure yield compares to other feedback yields.

\begin{deluxetable*}{ccccccccccccc}
\tablecaption{Comparison of properties related to CR distribution
\label{tab:tab2}}
\tablecolumns{13} 
\tablehead{
\colhead{Model} &
\colhead{$f_\mathrm{Coll.}$} &
\colhead{$f_\mathrm{Stream.}$} & 
\colhead{$f_\mathrm{Adiab.}$}  & 
\colhead{$f_\mathrm{tot.}$} & 
\colhead{$P_\mathrm{c}(0)/k_\mathrm{B}$}&  
\colhead{$F_\mathrm{c,z} (|z| = H_\mathrm{c,eff})$}&
\colhead{$H_\mathrm{c,eff}$}& 
\colhead{$\kappa_\mathrm{eff}$}&
\colhead{$|v|$}&
\colhead{$|v_\mathrm{A,i}|$}&
\colhead{$\Upsilon_\mathrm{c}$}&
\colhead{$\Upsilon_\mathrm{c}/\Upsilon_\mathrm{k}$}
\\
\colhead{} &
\colhead{} &
\colhead{} &
\colhead{} &
\colhead{} &
\colhead{\footnotesize(K/cm$^{3}$)} & 
\colhead{\footnotesize(erg/kpc$^{2}$/yr)}&
\colhead{\footnotesize(kpc)}& 
\colhead{\footnotesize(cm$^{2}$/s)}& 
\colhead{\footnotesize(km/s)}&
\colhead{\footnotesize(km/s)}&
\colhead{\footnotesize(km/s)}&
\colhead{}
}
\colnumbers
\startdata
R2 & $-0.42$ & $-1.40$  & 1.29 & $-0.53$ & $8.87 \times10^{5}$ & $5.20 \times 10^{47}$ & 0.61 & $2.75 \times 10^{28}$ & 320 & 110 & 185 & 1.16 \\
R4 & $-0.33$ & $-1.37$ & 1.12 & $-0.58$ & $3.01 \times10^{5}$ & $9.41 \times 10^{46}$ & 0.56 & $1.30 \times 10^{28}$ & 205 & 30& 346 & 1.86 \\
R8 & $-0.21$ & $-1.33$ &  1.05 & $-0.50$ & $1.57 \times10^{4}$ & $2.99 \times 10^{45}$ &0.81 &  $1.14 \times 10^{28}$ & 105 & 30 & 570 & 2.25 \\
\enddata
\tablecomments{
Columns: (1) Model name; (2) collisional loss relative to the injected energy; (3) streaming loss relative to the injected energy; (4) energy gained from the gas relative to the injected energy; (5) net loss relative to the injected energy; (6) CR pressure at the midplane; (7) horizontally-averaged vertical CR flux measured at $z = H_\mathrm{c,eff}$; (8) effective CR scale height; (9) effective diffusion coefficient calculated at $z = H_\mathrm{c,eff}$; (10) volume-weighted magnitude of gas-advection velocity; (11) volume-weighted magnitude of ion Alfv\'{e}n speed; (12) CR feedback yield; (13) ratio between CR and kinetic feedback yields (\autoref{eq:UpsilonCR}).
}
\end{deluxetable*}

In \autoref{tab:tab2}, we list the mean values of some quantities that are important in regulating the distribution of CRs. First, for each TIGRESS model, we calculate time-averaged sink/source energy terms. These consist of integrals over the whole simulation domain of the energy source terms, followed by averages over snapshots.  The total CR energy injected per unit time, 
$\dot{E}_{\rm c,inj}$, is 
the integral of $Q$ (\autoref{eq:injectedenergy}).  The
total rate of CR energy losses due to collisions is the integral 
of $-\Lambda_\mathrm{coll}(E) n_\mathrm{H} e_\mathrm{c}$ (\autoref{eq:lostenergy}).  From  the RHS of \autoref{eq:CRenergy}, the energy gain of CRs (or loss if negative) from adiabatic work done by the gas flow is the integral of 
$-\mathbf{v} \cdot  \tensor{\mathrm{\sigma}}_\mathrm{tot} \cdot ( \mathbf{F_\mathrm{c}} - 4/3 \mathbf{v} e_\mathrm{c}) $, while 
the CR energy loss due to CR steaming is 
$ -\mathbf{v_\mathrm{s}} \cdot  \tensor{\mathrm{\sigma}}_\mathrm{tot} \cdot   ( \mathbf{F_\mathrm{c}} - 4/3 \mathbf{v} e_\mathrm{c}) $ (streaming always drains energy from CRs based on the definition in \autoref{eq:vs}). The CR energy injected
per unit time per unit area $\dot{E}_\mathrm{inj}$ is $1.19 \times 10^{48}$~erg~kpc$^{-2}$~yr$^{-1}$ for R2, $2.19 \times 10^{47}$~kpc$^{-2}$~yr$^{-1}$ for R4 and $5.56 \times 10^{45}$~kpc$^{-2}$~yr$^{-1}$ for R8.

In \autoref{tab:tab2}, we report the fractional collisional loss $f_\mathrm{Coll.}$, the fractional streaming loss $f_\mathrm{Stream.}$, and the fractional gain from the gas work $f_\mathrm{Adiab.}$, where each is defined as a ratio of the term written above to the respective input energy.
In all cases, we find that the rate of work exchange is positive, meaning that on average the gas is doing work on the CR population, and the fractional exchange does not vary much for different models.
The fraction of energy lost to collisions decreases by a 
factor of two from model R2 to R8, the fractional streaming loss decreases by $5\%$, while the fractional work gain decreases by $18\%$. We note that the fraction of the original energy that escapes as a wind may be expressed as $f_{\rm wind}=1+f_\mathrm{Coll.}+f_\mathrm{Stream.}+f_\mathrm{Adiab.}$, which is in the range $\sim 0.4-0.5$.

Overall, the CR population is losing energy within the ISM in all models. This fractional loss relative to the input is roughly similar for the three models: $f_{\rm tot}=f_\mathrm{Coll} + f_\mathrm{Stream.} + f_\mathrm{Adiab.}$ is $-0.53$ for R2, $-0.58$ for R4, and $-0.50$ for R8. This result suggests that the different CR pressure relative to the MHD pressures cannot be explained by different fractional losses in the three environments. The difference must therefore owe to differences in CR transport.

To investigate the differences in transport for different environments, we start with an idealized ``average'' vertical diffusion equation that relates CR pressure to CR flux, 
\begin{equation}\label{eq:kappaeff}
 P_\mathrm{c}(0)\equiv H_\mathrm{c,eff} \frac{F_\mathrm{c,z}}{\kappa_\mathrm{eff}},   
\end{equation}
with $P_\mathrm{c}(0)$ the measured midplane pressure, $H_\mathrm{c,eff}= \langle |d\ln P_\mathrm{c}/dz| \rangle^{-1}$ an effective CR scale height (measured in the simulation through a linear fit of $\rm{ln} P_\mathrm{c}$ vs. $z$ within 1.5 kpc), 
 $F_\mathrm{c,z}$ the vertical CR flux measured at $|z|=H_\mathrm{c,eff}$, and $\kappa_\mathrm{eff} \equiv \sigma_\mathrm{eff}^{-1} $ an effective diffusion coefficient that is defined by this equation. All quantities for each model, using time-averaged CR profiles, are listed in \autoref{tab:tab2}. 
 
 In the case of negligible losses, the average vertical flux of CR energy above the SN input layer would be $0.5\,\epsilon_\mathrm{c}  E_\mathrm{SN}  \Sigma_\mathrm{SFR}  / m_\star$, where $m_\star$ is the total mass of new stars per supernova \citep[$95.5 M_\odot$  in][from a Kroupa IMF]{Kim&Ostriker17}. In our models, losses are on average not negligible (see above). Nevertheless, we note that the value of $F_\mathrm{c,z}$ computed at $|z|=H_\mathrm{c,eff}$ is not so different from the flux we would obtain in the absence of losses ($ \simeq 0.5\, \dot{E}_\mathrm{inj}/(L_x L_y) = 0.5 \, \epsilon_\mathrm{c}  E_\mathrm{SN}  \Sigma_\mathrm{SFR}  / m_\star $). This differs by a factor of 1.14, 1.16, 0.93 for model R2, R4, R8, respectively. From a detailed examination of the simulations, we find that the largest gain of energy from the gas comes from the disk region ($|z| < H_\mathrm{c,eff}$) at interfaces where hot gas is expanding at high velocity into warm/cold gas where CR densities are high. At the same time, most of the collisional losses and about 50\% of the streaming losses happen within $|z| < H_\mathrm{c,eff}$. Energy losses are therefore balanced by energy gains at low latitudes. In the coronal region, where the work term becomes negligible, streaming energy losses lead to a factor $\sim 2$ drop in the CR flux relative to the input value.

The values of $H_\mathrm{c,eff}$ and $\kappa_\mathrm{eff}$ are listed in \autoref{tab:tab2}. The effective scale heights differ by at most a factor of 1.5 for the three models, and no clear trend with $\Sigma_\mathrm{SFR}$ is present. On the other hand, $\kappa_\mathrm{eff}$ decreases from R2 to R4 to R8, meaning that the transport of CRs becomes less effective with decreasing $\Sigma_\mathrm{SFR}$. The larger CR pressure relative to the MHD pressure in model R8 can therefore be attributed primarily to its lower $\kappa_\mathrm{eff}$, and secondarily to its larger $H_\mathrm{eff}$.  Here it is important to note  that the effective diffusion coefficient defined in \autoref{eq:kappaeff} may be different from the actual diffusion coefficient ($\kappa_\parallel \equiv \sigma_\parallel^{-1}$), as $\kappa_\mathrm{eff}$ encodes the effects of advection and streaming, in addition to diffusion. In \autoref{sec:scattering rate} and \autoref{sec:streaming, diffusion and advection}, we shall analyse the individual contribution of advection, streaming and diffusion to the propagation of CRs.  There, we shall show that the main reason $\kappa_\mathrm{eff}$ is lower in model R8 is the lower advection speed in hot gas.  

Finally, we derive an expression for the CR feedback yield, $\Upsilon_\mathrm{c}$, as a function of $H_\mathrm{c,eff}$ and $\kappa_\mathrm{eff}$. Since $F_\mathrm{c,z} \approx 0.5\,\epsilon_\mathrm{c}  E_\mathrm{SN}  \Sigma_\mathrm{SFR}  / m_\star$ at $\vert z\vert = H_\mathrm{c,eff}$, we can write the CR feedback yield as
\begin{equation}
\Upsilon_\mathrm{c} \equiv \frac{P_\mathrm{c}(0)}{\Sigma_\mathrm{SFR}} \sim  \frac{1}{2}\epsilon_\mathrm{c}  \frac{H_\mathrm{c,eff}}{\kappa_\mathrm{eff}} \frac{E_\mathrm{SN}}{m_*}  \;.
\label{eq:UpsilonCR}
\end{equation}
The values of $\Upsilon_\mathrm{c}$, as well as the ratio between CR and kinetic feedback yields ($\Upsilon_\mathrm{c}/\Upsilon_\mathrm{k} = P_\mathrm{c}(0)/P_\mathrm{k}(0)$), are listed in \autoref{tab:tab2}. While both CR and kinetic feedback yields increase from R2 to R4 to R8, the former increase is larger ($\Upsilon_\mathrm{c} \propto \Sigma_\mathrm{SFR}^{-0.20}$, $\Upsilon_\mathrm{k} \propto \Sigma_\mathrm{SFR}^{-0.09}$).  It is worth recalling, however, that there is more than two orders of magnitude reduction in $\Sigma_\mathrm{SFR}$ from R2 to R8.

It is important to note that it is the difference $\Delta P$ between midplane pressures and pressures at the top of the atomic/molecular layer, rather than midplane pressure $P(0)$ itself, that contributes to the vertical support of the ISM against gravity. We compute the differences $\Delta P$ in \autoref{sec:VerticalSupport} and we show that $P_\mathrm{c}(0) \gtrsim P_\mathrm{k,z}(0)$ does not necessarily imply $\Delta P_\mathrm{c} \gtrsim \Delta P_\mathrm{k,z}$.

\subsection{Diffusion coefficient}
\label{sec:scattering rate}

\begin{figure*}
\centering
\includegraphics[width=\textwidth]{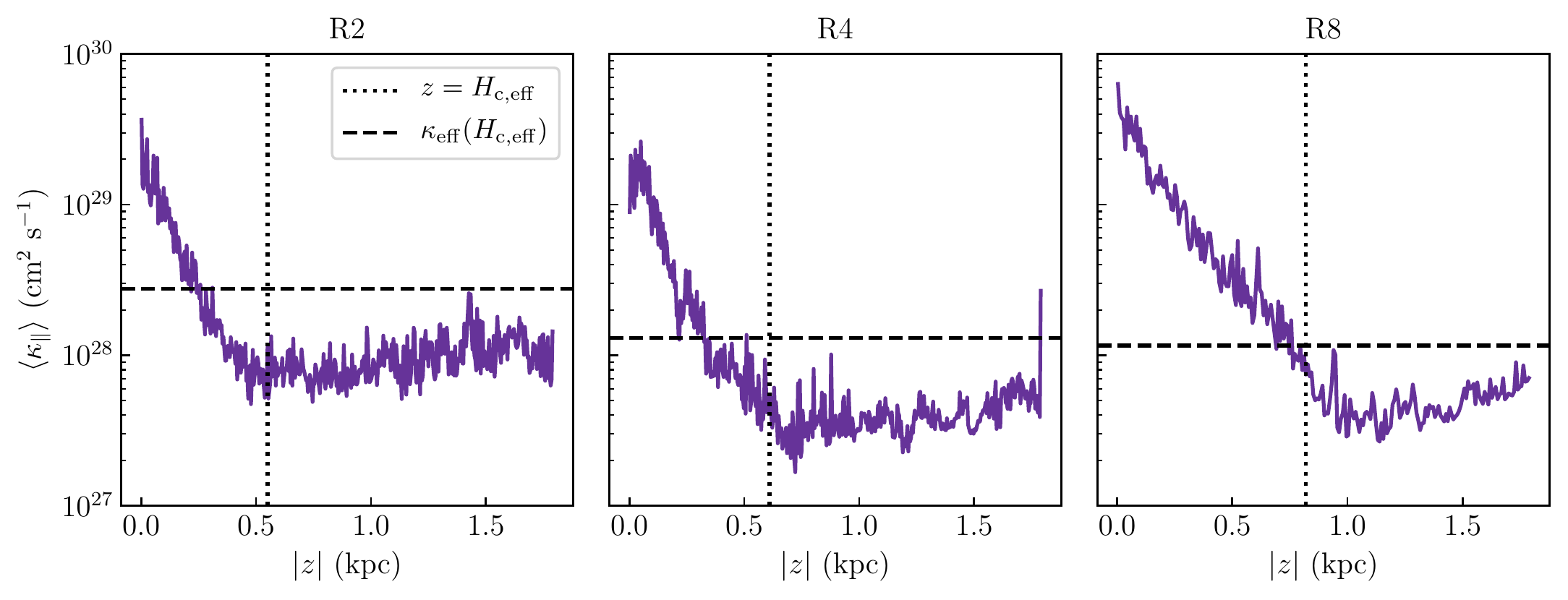}
\caption{Temporally-averaged weighted mean of the diffusion coefficient $\kappa_\parallel$ as a function of the distance from the midplane $\vert z \vert$ for the R2 (\textit{left panel}), R4 (\textit{middle panel}) and R8 (\textit{right panel}) models. The weighting factor applied in the average is the gradient of CR pressure along the magnetic field direction $\nabla P_\mathrm{c, \parallel}$. In each panel, the black dotted line indicates the average CR scale height $H_\mathrm{c,eff}$ for a given model (see also \autoref{tab:tab2}), while the black dashed line represents the value of the effective diffusion coefficient calculated at $z = H_\mathrm{c,eff}$.}
\label{fig:kappa}
\end{figure*}

In the previous section, we have seen that the effective diffusion coefficient increases with the SFR surface density. The effective diffusion coefficient must be understood as a measurement of the efficiency of CR propagation, including not only CR diffusion, but also advection and streaming. Hence, higher $\kappa_\mathrm{eff}$ does not necessarily mean stronger CR diffusivity.

\autoref{fig:kappa} shows the weighted mean of the actual diffusion coefficient $\kappa_\parallel$ as a function of $|z|$ averaged over time. At a given $z$, $\langle \kappa_\parallel \rangle$ is defined as:
\begin{equation}
\langle \kappa_\parallel \rangle (z) = \frac{\int \int \kappa_\parallel (x,y,z) | \nabla  P_\mathrm{c, \parallel} (x,y,z) | dx dy}{\int \int | \nabla P_\mathrm{c, \parallel} (x,y,z) | dx dy}\;.
\label{eq:averagekappa}
\end{equation}
where the weight $\nabla  P_\mathrm{c, \parallel} \equiv \vert \hat{\mathbf{B}} \cdot  \nabla {P_\mathrm{c}} \vert$ is the CR pressure gradient parallel to the magnetic field direction. In steady state, the RHS of \autoref{eq:averagekappa} can be written as the ratio between the moduli of the volume-weighted mean diffusive flux (in steady state $F_\mathrm{d, \parallel} = - \nabla P_\mathrm{c, \parallel}/ \sigma_\parallel = - \kappa_\parallel \nabla P_\mathrm{c, \parallel}$, see \autoref{eq:SteadyFlux}) and the volume-weighted mean CR pressure gradient along the magnetic field lines. \autoref{fig:kappa} shows that, in all cases, $\kappa_\parallel$ decreases with $|z|$ at low latitudes, while having a roughly constant value in the coronal region. As we shall see below, diffusion is particularly effective in the denser neutral gas, which is mostly located in the galactic disk (see also the distribution of $\sigma_\parallel$ in \multiref{fig:R2snapshot}{fig:R4snapshot}{fig:R8snapshot}). This explains why $\langle \kappa_\parallel \rangle$ is larger near the midplane, while it decreases with $|z|$ as the average gas density decreases. R8 exhibits the highest values of $\kappa_\parallel$ near the disk ($|z| \lesssim 0.5$~kpc), thus explaining the fact that the CR scale height is slightly larger for this model compared to the other two (see \autoref{tab:tab2}) -- the distribution of CRs is more extended due to stronger diffusion. 

In \autoref{fig:kappa}, the dotted vertical line and the dashed horizontal line respectively indicate the effective scale height and the value of the effective diffusion coefficient at $|z|\simeq H_\mathrm{c,eff}$ for a given model. The effective diffusion coefficient is always higher than the actual diffusion coefficient at $|z|\simeq H_\mathrm{c,eff}$, confirming that other mechanisms, in addition to diffusion, are at play to foster the transport of CRs out of the disk. The difference between effective and actual diffusion coefficient at $|z|\simeq H_\mathrm{c,eff}$ is smaller in R8 compared to R2 and R4, suggesting that diffusion plays a larger role in the former model.

We point out that, in all models, $\nabla  P_\mathrm{c,\parallel}$ is almost one order of magnitude lower than ${\vert \nabla  {P_\mathrm{c}} \vert}$, meaning that the magnetic field lines are mostly tangled or not aligned with the CR pressure gradients. If we neglected the real structure of the magnetic field and assumed open magnetic field lines parallel to the large-scale CR pressure gradient, $\kappa_\parallel$ ($\propto \nabla  P_\mathrm{c,\parallel}$, see \autoref{eq:sigmaNLL} and \autoref{eq:sigmaIN}) would be lower than what we found in this work.

\begin{figure*}
\centering
\includegraphics[width=\textwidth]{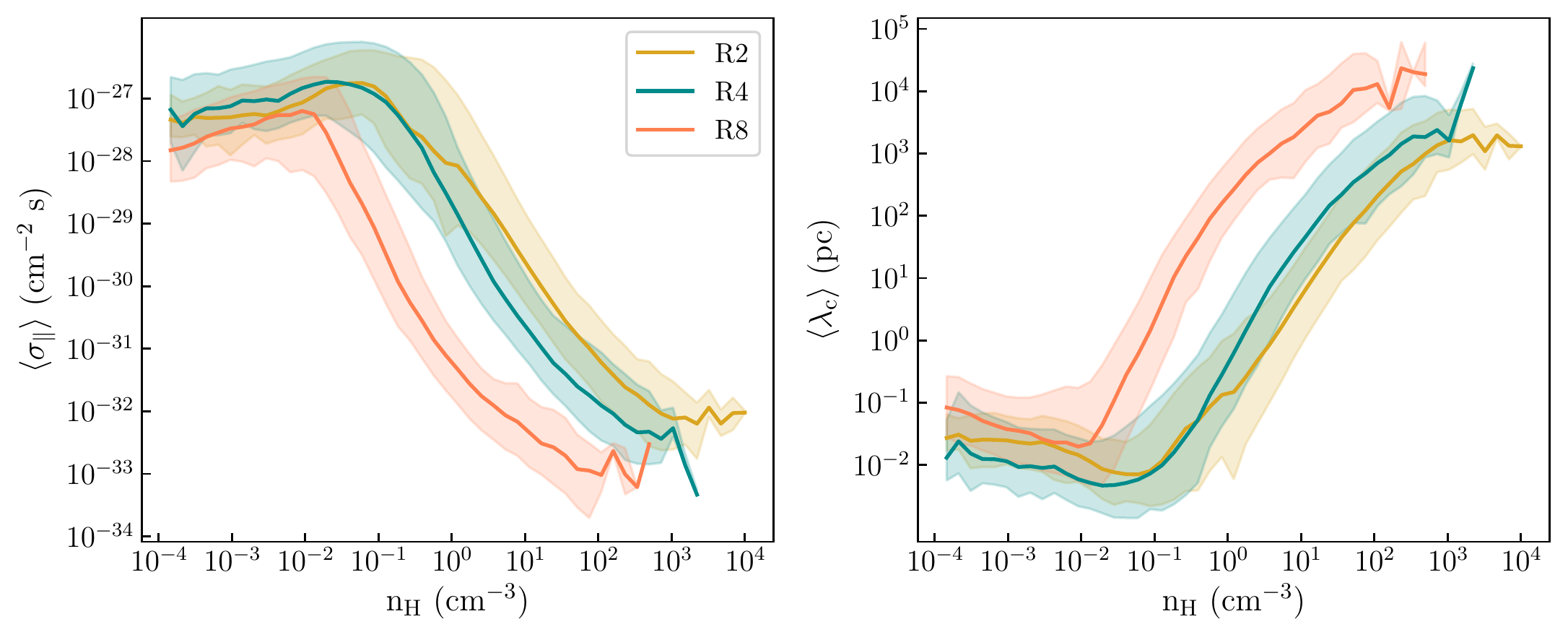}
\caption{Temporally-averaged median of the scattering coefficient $\sigma_\parallel$ (\textit{left panel}) and mean free path $\lambda_\mathrm{c}$ (\textit{right panel}). Different colors represent different models: gold for R2, dark cyan for R4 and coral for R2. The shaded areas cover the 16th to 84th percentiles of the temporally-averaged variations around the mean.}
\label{fig:sigmaprofiles}
\end{figure*}

For a better understanding of the importance of diffusion in the three different environments, in the left panel of \autoref{fig:sigmaprofiles}, we show the temporally-averaged median value of the scattering coefficient $\sigma_\parallel$ ($\equiv 1/\kappa_\parallel$) as a function of hydrogen density. The overall profiles are similar in the three models: $\sigma_\parallel$ slowly increases with $n_\mathrm{H}$ at low densities, where the gas is well ionized and nonlinear Landau damping dominates, while $\sigma_\parallel$ rapidly decreases at high densities, where the gas is mostly neutral and ion-neutral damping becomes stronger than nonlinear Landau damping (see \citealt{Armillotta+21} for a detailed explanation of the dependence of $\sigma_\parallel$ on $n_\mathrm{H}$). 

More specifically, in R2 $\sigma_\parallel$ goes from a few times $10^{-28}$~cm$^{-2}$~s at $n_\mathrm{H} \simeq 10^{-4}$~cm$^{-3}$ to $\simeq 10^{-27}$~cm$^{-2}$~s at $n_\mathrm{H} \simeq 10^{-1}$~cm$^{-3}$ and decreases at higher densities, becoming $\lesssim 10^{-31}$~cm$^{-2}$~s at $n_\mathrm{H} \simeq 10^{2}$~cm$^{-3}$; in R4 $\sigma_\parallel$ goes from a few times $10^{-28}$~cm$^{-2}$~s at $n_\mathrm{H} \simeq 10^{-4}$~cm$^{-3}$ to $\simeq 10^{-27}$~cm$^{-2}$~s at $n_\mathrm{H} \simeq 10^{-1}$~cm$^{-3}$ and then decreases down to $\simeq 10^{-32}$~cm$^{-2}$~s at $n_\mathrm{H} \simeq 10^{2}$~cm$^{-3}$; in R8 $\sigma_\parallel$ goes from $\simeq 10^{-28}$~cm$^{-2}$~s at $n_\mathrm{H} \simeq 10^{-4}$~cm$^{-3}$ to $\lesssim 10^{-27}$~cm$^{-2}$~s at $n_\mathrm{H} \simeq 10^{-2}$~cm$^{-3}$ and decreases at higher densities, assuming a value $\simeq 10^{-33}$~cm$^{-2}$~s at $n_\mathrm{H} \simeq 10^{2}$~cm$^{-3}$. At the average ISM density ($n_\mathrm{H} \simeq 7.7/1.4/0.86$~cm$^{-3}$ for R2/R4/R8, see \autoref{tab:tab1}), the average scattering coefficient is $\simeq 4-5 \times 10^{-30}$, $\simeq 4-5 \times 10^{-30}$, and $\simeq 10^{-31}$~cm$^{-2}$~s for R2, R4, and R8, respectively. 

Two main differences emerge from the comparison between the three models. First, the turnover happens at different densities: at $n_\mathrm{H} \simeq 10^{-2}$~cm$^{-3}$ for R8 and at $n_\mathrm{H} \simeq 10^{-1}$~cm$^{-3}$ for R2 and R4. 
Regardless of the model, gas becomes fully ionized at temperatures above a few times $10^4$~K \citep[][]{Sutherland&Dopita93}. As the average thermal pressure increases, the average density corresponding to the transition temperature between partially- versus fully-ionized regime increases from R8 to R4 to R2. The second difference in the scattering coefficient-density relation is that the value of $\sigma_\parallel$ at a given $n_\mathrm{H}$ increases going from R8 to R2, especially in the high-density regime. This difference can be mainly attributed to the different CR pressure gradients in the three models (see \autoref{eq:sigmaNLL} and \autoref{eq:sigmaIN}). The value of  $\sigma_\parallel$ is proportional to $(\vert \hat{\mathbf{B}} \cdot  \nabla {P_\mathrm{c}} \vert)^{1/2}$  at low densities and to ${\vert \hat{\mathbf{B}} \cdot  \nabla  {P_\mathrm{c}} \vert}$ at high densities. As a consequence of the increasing CR pressure (see \autoref{fig:PressProfiles}), the CR pressure gradient generally increases from R8 to R4 to R2. We note however that, even thought $P_\mathrm{c}$ is larger in R2 than in R4, the scattering coefficient is roughly similar in the two models. As we shall see in the next section, both advection and streaming are more effective in R2 than in R4, especially in low-density gas. The more effective transport makes the CR pressure gradients in the magnetic field direction smaller in R2 compared to R4.

From \autoref{fig:sigmaprofiles} we see that 
the environment of the R8 model, with lower density and star formation rate, is characterized by an overall higher physical  diffusivity (lower scattering rate and longer mean free path at a given density) than the other models. The differences are most pronounced at high density, and the midplane region in \autoref{fig:kappa} indeed shows the highest mean $\kappa_\parallel$ for model R8, with fairly similar high-latitude $\kappa_\parallel$ in all models.

Finally, in the right panel of \autoref{fig:sigmaprofiles}, we show the temporally-averaged CR mean free path as a function of hydrogen density. The mean free path $\lambda_\mathrm{c}$ is calculated as $(v_\mathrm{p} \sigma_\parallel)^{-1}$, where $v_\mathrm{p}$ is the CR velocity. The mean free path distribution reflects the scattering coefficient distribution. At low densities in ionized gas, where scattering is strong, the mean free path decreases from $ \lambda_\mathrm{c} \simeq 0.01-0.03$~pc ($\lambda_\mathrm{c} \simeq 0.07-0.08$~pc) at $n_\mathrm{H} = 10^{-4}$~cm$^{-3}$ to $\lambda_\mathrm{c} \simeq 0.005-0.006$~pc ($\lambda_\mathrm{c} \simeq 0.01-0.02$~pc) at $n_\mathrm{H} \simeq 10^{-1}$~cm$^{-3}$ ($n_\mathrm{H} \simeq 10^{-2}$~cm$^{-3}$) in R2 and R4 (R8). At higher densities, the mean free path quickly increases as the scattering coefficient decreases in denser, neutral gas. At $n_\mathrm{H} \simeq 10^{2}$~cm$^{-3}$ -- the characteristic density of cold atomic and diffuse molecular clouds -- $\lambda_\mathrm{c} \simeq 2 \times 10^2$~pc in R2, $\simeq 5-6 \times 10^2 $~pc in R4, and $\simeq 10^4$~pc in R8. With a mean free path in the cold dense gas comparable to or larger than the size of individual clouds ($\sim 10-10^2$~pc), CRs can freely stream across them. In \citet{Armillotta+21}, we found that, when scattering perpendicular to the magnetic field is neglected, the scattering coefficient increases by more than one order of magnitude at very high densities. Therefore, the actual value of $\lambda_\mathrm{c}$ at in cold, dense gas may be higher or lower than the one shown in \autoref{fig:sigmaprofiles} depending on whether the actual perpendicular scattering coefficient is lower or higher than the one assumed in this work ($\sigma_\perp = 10\, \sigma_\parallel$). Nevertheless, the conclusion that CRs would freely stream across dense, cold clouds is insensitive to the treatment of perpendicular diffusion since waves are strongly damped \citep[see also][]{Plotnikov2021} and magnetic fields are too strong to be tangled. 

\subsection {Role of streaming, diffusive and advective transport}
\label{sec:streaming, diffusion and advection}

\begin{figure*}
\centering
\includegraphics[width=\textwidth]{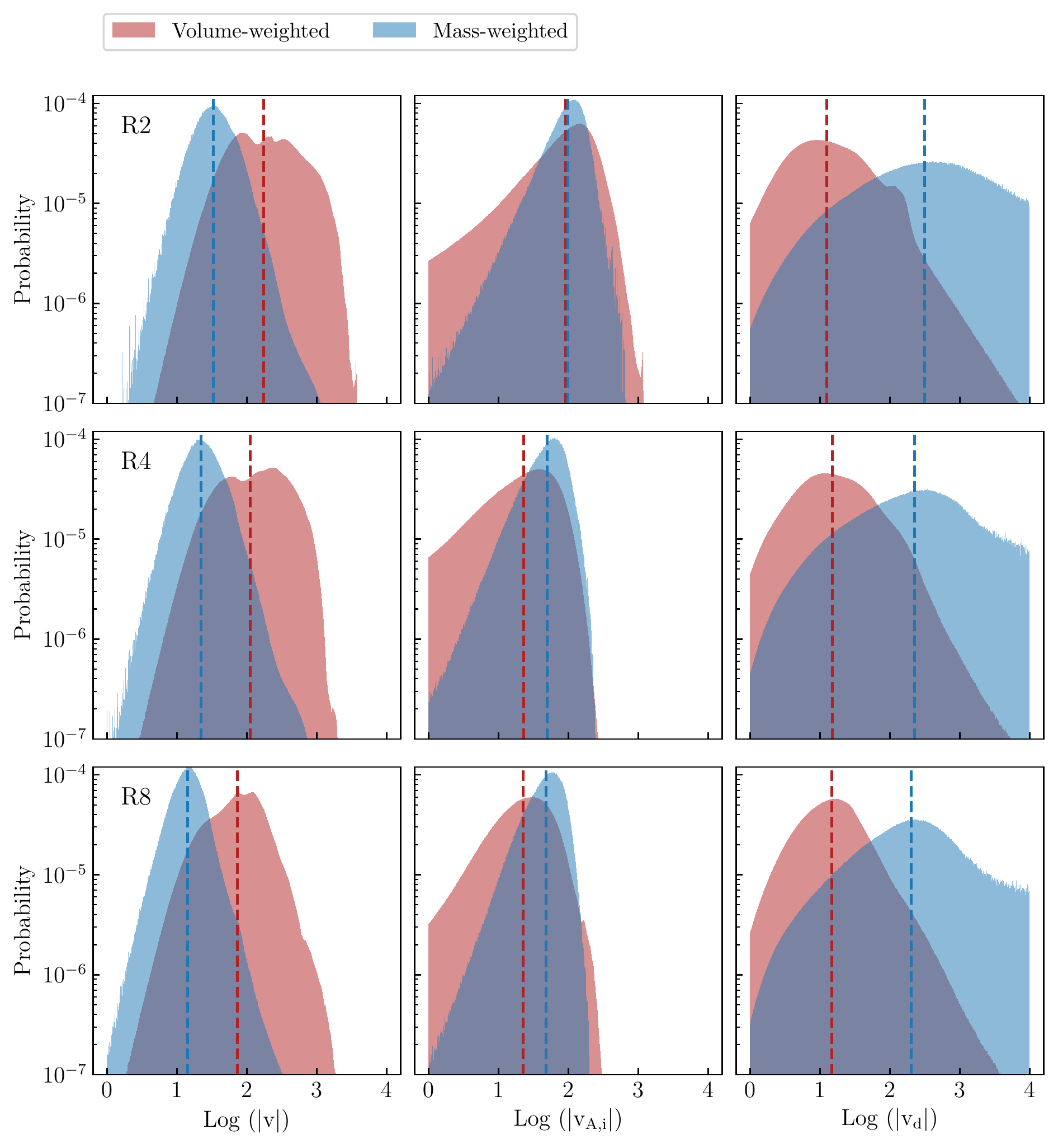}
\caption{Temporally-averaged volume-weighted (red histograms) and mass-weighted (blue histograms) probability distribution of gas velocity $\vert v \vert$ (\textit{left column}), ion Alfv\'{e}n speed $\vert v_\mathrm{A,i} \vert$ (\textit{middle column}), and diffusion velocity $\vert v_\mathrm{d} \vert$ (\textit{right column}) for the R2 (\textit{top row}), R4 (\textit{middle row}) and R8 (\textit{bottom row}) models. The red and blue dashed lines indicate the median values of the volume-weighted and mass-weighted distributions, respectively.}
\label{fig:histogram}
\end{figure*}

In the previous section, we have seen that diffusion is overall more effective in R8 than in the other two models, based on its lower scattering rates. This result at first may seem to conflict with our previous finding showing that the total propagation efficiency decreases going from R2 to R8 (as quantified by the decrease in $\kappa_{\rm eff}$ and increase in $\Upsilon_c$ shown in \autoref{tab:tab2}). To understand the different transport efficiencies in the three models, we now investigate the contribution of advection and streaming, in addition to diffusion. 

\autoref{fig:histogram} shows the volume-weighted (red histograms) and mass-weighted (blue histograms) probability distributions of the gas advection speed, the Alfv\'en speed, and the diffusion speed (see \autoref{eq:vd_def}) for the three models (see also volume-weighted mean values in \autoref{tab:tab2}). For all models, when volume-weighted the transport of CRs is mostly through advection, as the gas velocity dominates over the other relevant velocities in hot, low-density, well ionized regions which occupy most of the volume. In contrast, if we consider the mass-weighted distributions, both diffusion and streaming transport dominate over advection. In higher-density regions containing most of the gas mass, the ion Alfv\'{e}n speed is higher than the gas flow speed (see \multiref{fig:R2snapshot}{fig:R4snapshot}{fig:R8snapshot}). Moreover, the very low values of the scattering coefficient in poorly ionized gas (see \autoref{fig:sigmaprofiles}) makes CR diffusion quite strong. 

Comparing the distribution of individual propagation-velocity components, we can note some relevant differences among the three models. First, both the volume-weighted and the mass-weighted distribution shift towards higher velocity values going from R8 to R4 to R2: the median value of the volume-weighted distribution increases by a factor of $\sim 2$, while the median value of the mass-weighted distribution increases by a factor of $\sim 3$. This implies that on average the gas velocity increases with the SFR both in hot low-density regions and in warm/cold high-density regions. As advection is the dominant mechanism of CR propagation at least in the volume-filling low-density gas, this result explains why the efficiency of CR propagation increases from R8 to R4 to R2.

We can conclude that the propagation of CRs out of the galactic disk becomes more and more effective going from R8 to R2 mostly because the gas advection velocities become higher and higher, especially in hot gas. At the same time, the denser poorly-ionized gas that makes up most of the mass is dominated by diffusion.  Meanwhile, ion Alfv\'en speeds exceed advection speeds in the higher-density poorly-ionized gas and exceed diffusion speeds in the low-density well-ionized gas. 
Thus, in well-ionized hot gas, diffusion is always quite small and CRs are transported by a combination of advection (primary) and Alfv\'enic streaming (secondary), while in poorly-ionized dense gas the CRs are very strongly diffusive. The effect of all three transport mechanisms must therefore be considered to understand the relation between CR pressure in the disk and SFR surface density. 

\section{Predictions for the dynamical effects of cosmic rays}
\label{sec:results-IIpart}

Although the back-reaction of thermal gas and magnetic field to the CR pressure cannot be directly studied in this work, we can use the distribution of CR pressure inferred from our post-processed simulations to make predictions about the dynamical effect of CRs in galaxies. In the following, we investigate the potential impact of CRs on the dynamics of the ISM gas overall, as well as individual thermal phases. We define three different gas phases based on temperature: warm ($5050~\rm{K}<\rm{T}<2\times10^4~\rm{K}$), intermediate ($2\times10^4~\rm{K}<\rm{T}<5\times10^5~\rm{K}$), and hot ($T>5\times10^5~\rm{K}$) phase.

\subsection{Momentum Flux and Weight}
\label{sec:VerticalSupport}

In the presence of CRs, the gas-momentum equation becomes \citep[e.g.][]{Jiang&Oh18}:
\begin{equation}
\begin{split}
\frac{\partial (\rho \mathbf{v})}{\partial t}  & + \nabla \cdot (\rho \mathbf{v} \mathbf{v} + P_\mathrm{t} \tensor{\mathbf {I}} + \frac{B^2}{2} \tensor{\mathbf {I}} - \mathbf{B} \mathbf{B}) \\ & =  - \rho \nabla \Phi_\mathrm{tot} + \tensor{\mathrm{\sigma}}_\mathrm{tot} \cdot \left(  \mathbf{F_\mathrm{c}} - \frac{4}{3} e_\mathrm{c} \mathbf{v}\right)\;,
\end{split}
\label{eq:momeq}    
\end{equation}
where for our simulations $\Phi_\mathrm{tot}$ is given by the sum due to the ``external'' gravitational potential from the old stellar
disk and dark matter halo plus the gravitational potential of the gas obtained by solving Poisson's equation \citep[see][]{Kim&Ostriker17}.
The term $\tensor{\mathrm{\sigma}}_\mathrm{tot} \cdot ( \mathbf{F_\mathrm{c}} - 4/3 e_\mathrm{c} \mathbf{v})$ represents the force exerted from the CR population on the thermal gas. 

We now focus on the momentum equation in the $z$ direction, considering a shearing-periodic box and taking horizontal and temporal averages. We formally separate the terms from different thermal phases and sum over them, obtaining the following equation for the vertical momentum of gas:
\begin{equation}
\begin{split}
  \sum_\mathrm{ph} &\left <  \frac{\partial}{\partial t} ({\rho v_\mathrm{z}}) \right>_\mathrm{ph} 
+ \frac{d}{dz} \sum_\mathrm{ph} \left < P_\mathrm{k,z} + P_\mathrm{t} + P_\mathrm{m,z} \right >_\mathrm{ph} 
\\ & + \frac{d}{dz} \sum_\mathrm{ph} \left < P_\mathrm{c} \right >_\mathrm{ph} 
=  - \sum_\mathrm{ph} \left < \rho \frac{\partial}{\partial z} \Phi_\mathrm{tot}   \right>_\mathrm{ph} 
\;.
\end{split}
\label{eq:vertmomeq}    
\end{equation}
Here, $\left < {q} \right >_\mathrm{ph}$
is the average over time of $\bar{q}_\mathrm{ph}(z;t)$, the horizontal average of a quantity $q$ for a given thermal phase at height $z$, defined as
\begin{equation}
\bar{q}_\mathrm{ph}(z,t) =   \sum_\mathrm{x,y} \frac{q(x,y,z;t) \Theta_\mathrm{ph} (T) \Delta x \Delta y}{L_\mathrm{x}L_\mathrm{y}}\;,
\label{eq:average}
\end{equation}
with $\Theta_\mathrm{ph} (T)$ the top-hat function that returns 1 for gas at temperatures within the temperature range of each phase (ph = warm, intermediate, or hot) or 0 otherwise. In \autoref{eq:vertmomeq}, we have assumed that the time-dependent term in \autoref{eq:CRflux} is on average negligible, so that $\left < \tensor{\mathrm{\sigma}}_\mathrm{tot} \cdot ( \mathbf{F_\mathrm{c}} - 4/3 e_\mathrm{c} \mathbf{v}) \right >$ reduces to $- \left < \nabla P_\mathrm{c} \right >$.
From \autoref{eq:vertmomeq}, 
$\mathcal{F}_\mathrm{MHD,ph}(z) \equiv \langle P_\mathrm{k,z} + P_\mathrm{t} + P_\mathrm{m,z} \rangle_{\rm ph}$ is the contribution to the momentum flux from the MHD pressures of gas in a given phase,
while $\mathcal{F}_\mathrm{c,ph} (z)\equiv \langle P_\mathrm{c}\rangle_{\rm ph}$ is the contribution to 
the momentum flux from CRs co-located with that 
gas phase.  We note that the contribution to the momentum flux from 
each phase is equal to the area filling factor of that phase (at a given $z$) times the mean pressure of gas in that phase.

\autoref{eq:vertmomeq} is a function of $z$, and we may therefore integrate   from either the top or bottom of the simulation domain to an arbitrary height $z$.  In this way, we express the momentum equation in terms of momentum
flux differences across the ISM 
and the weight of gas \citep[see][]{KimOstriker2015,Vijayan+20}:
\begin{equation}
\begin{split}
-\sum_\mathrm{ph}  \langle \dot{p}_\mathrm{z} \rangle _\mathrm{ph}(z) &+ \sum_\mathrm{ph}  \Delta_\mathrm{z} \mathcal{F}_\mathrm{MHD,ph} (z) \\ &+ \sum_\mathrm{ph} \Delta_\mathrm{z} \mathcal{F}_\mathrm{c,ph} (z)  = \sum_\mathrm{ph} \mathcal{W}_\mathrm{ph}(z)\;.
\end{split}
\label{eq:vertmomeq2}    
\end{equation}
Here, $\langle \dot{p_\mathrm{z}} \rangle _\mathrm{ph}(z)$ is the volume-integrated rate of change in $z$-momentum normalized to the area of the horizontal plane,
\begin{equation}
\langle \dot{p}_\mathrm{z} \rangle _\mathrm{ph}(z) = \int_z^{\pm L_\mathrm{z}/2} \left < \frac{\partial}{\partial t}
(\rho v_\mathrm{z})
\right>_\mathrm{ph} 
d z'\;,
\end{equation}
$\mathcal{W}_\mathrm{ph}(z)$ is the gas weight in a given phase above $z$, and 
\begin{equation}
\mathcal{W}_\mathrm{ph}(z) = \int_z^{\pm L_\mathrm{z}/2} \left < \rho \frac{\partial \Phi_\mathrm{tot} }{\partial z}  \right>_\mathrm{ph} d z'\;.
\label{eq:weight}
\end{equation}
The differences 
\begin{equation}
  \Delta_\mathrm{z} \mathcal{F}_\mathrm{MHD,ph} (z) \equiv \mathcal{F}_\mathrm{MHD,ph} (z) - \mathcal{F}_\mathrm{MHD,ph} (\pm L_\mathrm{z}/2)  
\end{equation}
and 
\begin{equation}
\Delta_\mathrm{z} \mathcal{F}_\mathrm{c,ph} (z) \equiv \mathcal{F}_\mathrm{c,ph} (z) - \mathcal{F}_\mathrm{c,ph} (\pm L_\mathrm{z}/2)
\end{equation}
can be considered the MHD and CR vertical ``support against'' or ``counteraction of''  gravity. The former terminology is perhaps more appropriate for a quasi-hydrostatic region, while the latter may be more suited to a wind acceleration region. 

In the TIGRESS simulations, the system is in quasi-steady state, meaning that $\sum_\mathrm{ph} \langle \dot{p_z} \rangle _\mathrm{ph}(z) \approx 0$, and CRs are not included. Hence, \autoref{eq:vertmomeq2} reduces to:
\begin{equation}
\sum_\mathrm{ph}  \Delta_\mathrm{z} \mathcal{F}_\mathrm{MHD,ph} (z) = \sum_\mathrm{ph} \mathcal{W}_\mathrm{ph}(z)\;.
\label{eq:vertmomeq_TIGRESS}
\end{equation}

\begin{figure*}
\centering
\includegraphics[width=\textwidth]{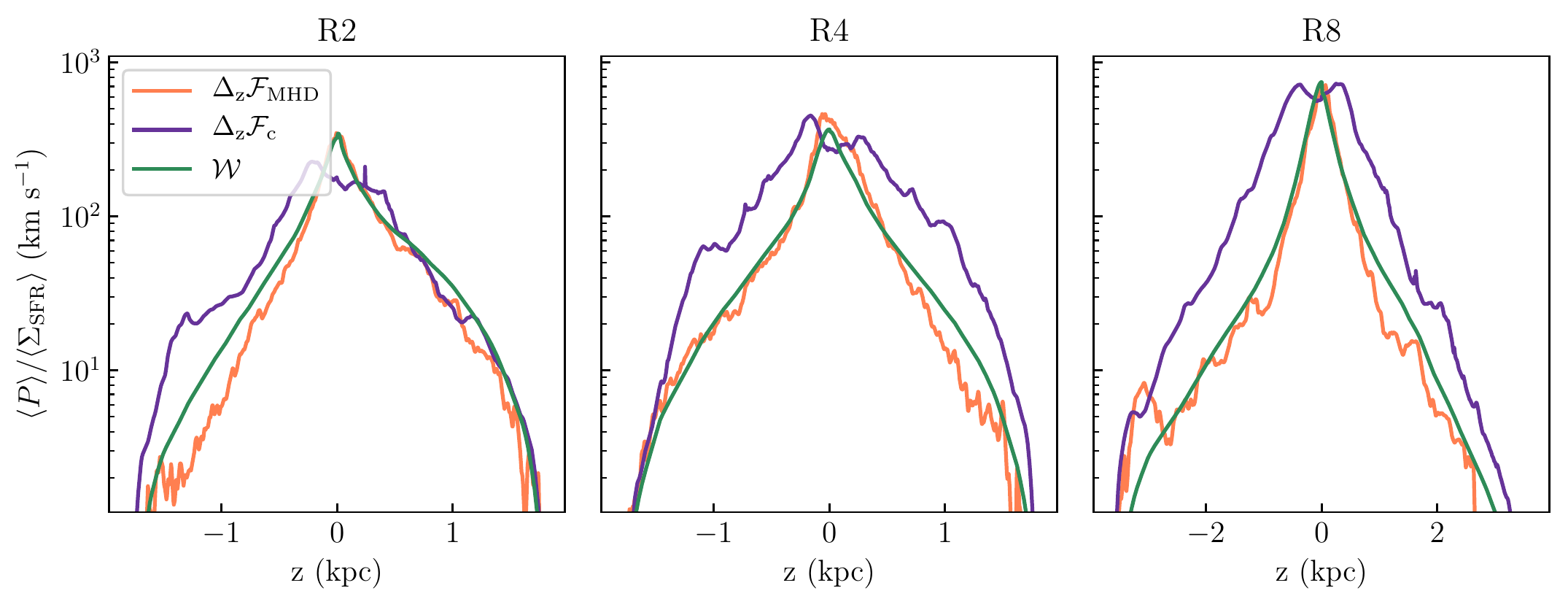}
\caption{Temporally-averaged MHD momentum flux difference (or vertical ``support'') $\Delta \mathcal{F}_\mathrm{MHD}$ (orange), CR momentum flux difference $\Delta \mathcal{F}_\mathrm{c}$ (purple) and weight $\mathcal{W}$ (green) of the total gas as a function of $z$ for the R2 (\textit{left panel}), R4 (\textit{middle panel}) and R8 (\textit{right panel}) models. The vertical profiles are divided by the temporally-averaged SFR surface density $\Sigma_\mathrm{SFR}$.}
\label{fig:supportandweight}
\end{figure*}

We compute the value of each term in \autoref{eq:vertmomeq2} using our post-processed simulations for the three galactic environments investigated in this paper. For each model, \autoref{fig:supportandweight} displays the MHD vertical support, the (potential) CR vertical support, and the weight of the total gas as a function of $z$; we do not show $\sum_\mathrm{ph} \langle \dot{p}_\mathrm{z} \rangle _\mathrm{ph}$ as its value is negligible. In each model, the MHD vertical support, $\Delta_\mathrm{z} \mathcal{F}_\mathrm{MHD} \equiv \sum_\mathrm{ph}  \Delta_\mathrm{z} \mathcal{F}_\mathrm{MHD,ph}$, fairly closely follows the gas weight, $\mathcal{W} \equiv \sum_\mathrm{ph} \mathcal{W}_\mathrm{ph}$, thus confirming that \autoref{eq:vertmomeq_TIGRESS} holds in the TIGRESS simulations (see \citealt{Vijayan+20} for a more detailed analysis of the solar neighborhood model). \autoref{fig:supportandweight} shows flux differences and weights normalized to $\langle \Sigma_{\rm SFR}\rangle$, so that the midplane value is equivalent to the total feedback yield $\Upsilon$; we note that this yield increases slightly from model R2 to R4 to R8 (see also Ostriker \& Kim 2021, in prep.).

\autoref{fig:supportandweight} also shows that the momentum flux difference due to CRs, $\Delta_\mathrm{z} \mathcal{F}_\mathrm{c} \equiv \sum_\mathrm{ph}  \Delta_\mathrm{z} \mathcal{F}_\mathrm{c,ph}$, is  larger than the gas weight at most $z$ away from the midplane  -- except in model R2 for the $z > 0$ region\footnote{Asymmetries in the CR distribution are due to the chaotic
nature of the turbulent ISM, which results in different injection of CR energy above and below the plane.}, where $\Delta_\mathrm{z} \mathcal{F}_\mathrm{c} \simeq \mathcal{W}$. 
Notably, the difference between $\Delta_\mathrm{z} \mathcal{F}_\mathrm{c}$ and $\mathcal{W}$ at a given height increase from R2 to R4 to R8.
This result suggests that the relative contribution of CRs to the vertical support against or counteraction of gravity might be more significant in environments with lower star formation. 

\begin{figure*}
\centering
\includegraphics[width=\textwidth]{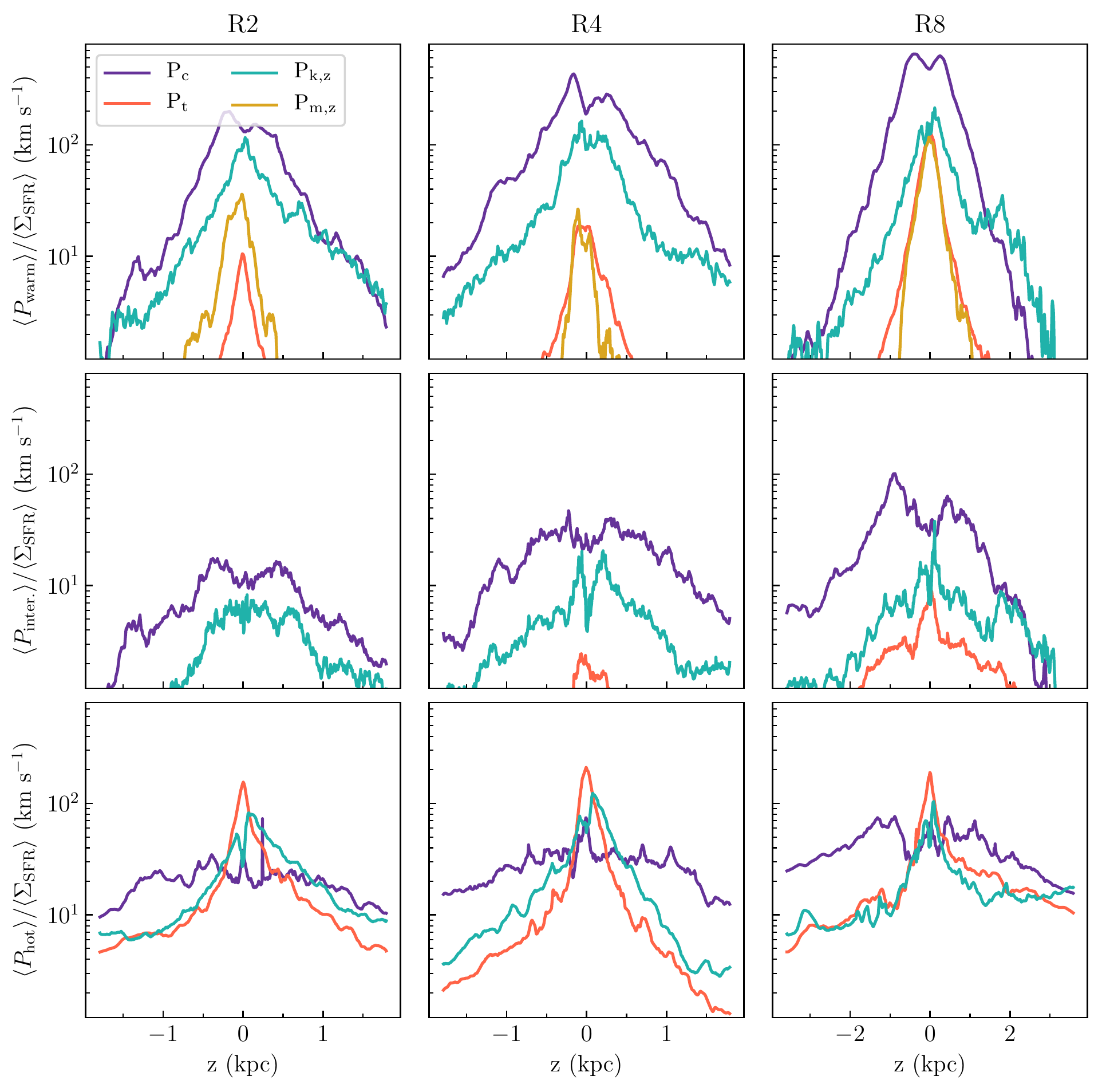}
\caption{Horizontally- and temporally-averaged vertical profiles of CR pressure $P_\mathrm{c}$ (purple), thermal pressure $P_\mathrm{t}$ (coral), kinetic pressure $P_\mathrm{k,z}$ (turquoise) and magnetic stress $P_\mathrm{m,z}$ (gold) divided by the temporally-averaged SFR surface density $\Sigma_\mathrm{SFR}$. Each column indicates a different model: R2 (\textit{left column}), R4 (\textit{middle column}) and R8 (\textit{right column}) models. Each row represents a different phase of the gas: from top to bottom warm ($T \leq 2 \times 10^4$~K), intermediate ($2 \times 10^4 < T \leq 5 \times 10^5$~K) and hot ($T > 5 \times 10^5$~K) phase. The magnetic pressure profiles of the intermediate and hot phases, as well as the thermal pressure profile of the intermediate phase in R2, are not shown as their values are much lower than the other pressure components ($\langle P_\mathrm{inter.}\rangle/\langle \Sigma_\mathrm{SFR} \rangle < 1\, \kms$).  }
\label{fig:diffphasepress}
\end{figure*}

In order to investigate the momentum flux further, in \autoref{fig:diffphasepress} we show the horizontally- and temporally-averaged vertical profiles of individual pressure contributions to the momentum flux for each thermal phase separately (\autoref{eq:average}). If we consider MHD pressures only, we can note that in the warm gas the vertical kinetic pressure is the largest contributor to the momentum flux at all $z$. In the hot gas, the thermal pressure is the largest momentum flux component near the midplane, while at higher latitude, the thermal and kinetic pressures are comparable in the hot gas, having been accelerated by pressure gradients. 
At the midplane, the contributions from warm and hot gas to the total momentum flux
are comparable.  
However, the contribution to total momentum flux from the warm gas drops more rapidly with $z$ (due to the turnaround of the warm fountain flow) than the contribution from the hot gas. 
Above $\sim 1$~kpc the hot gas is the largest contributor to the total momentum flux.  

If we now consider the CR momentum flux profile, we can see that the contribution from CRs associated with
warm gas exceeds the contribution from CRs associated with hot gas up to $|z| \sim 1-1.5$~kpc, 
while the latter dominates at higher latitudes. 
At high $z$, there is a relatively flat profile of 
$\langle P_c \rangle_{\rm hot}$.
This suggests that the contribution of CRs in the hot gas to 
offsetting gravity -- which is based on a  momentum flux difference -- is not more significant than the contribution of CRs in other phases of the gas. In the warm and intermediate-temperature gas, 
$\langle P_c \rangle$ is larger than $\langle P_{\rm k,z}\rangle$, and also 
$\langle P_c \rangle/\langle \Sigma_\mathrm{SFR} \rangle$ increases from R2 to R4 to R8, as advection of CRs becomes less effective (see \autoref{sec:pressure} and \autoref{sec:streaming, diffusion and advection}). This explains why $\Delta_z{\cal F}_{\rm c}/\langle \Sigma_\mathrm{SFR} \rangle$ increases with decreasing $\Sigma_\mathrm{SFR}$ (\autoref{fig:supportandweight}). 

Finally, we focus on the mass-containing disk region only\footnote{In the simulations, the regions at $|z| < 500$~pc contain almost $80\%$ of the total gas mass.} and integrate individual terms of \autoref{eq:vertmomeq} from $z = \pm 500$~pc\ to $z = 0$. In agreement with previous analysis of the TIGRESS simulations \citep[][Ostriker \& Kim 2021, in prep.]{Vijayan+20,WT_Kim2020}, we find that the disk is in vertical dynamical equilibrium, with the ISM weight balanced by the difference between the MHD momentum flux at the midplane and the MHD momentum flux at higher latitude (\autoref{eq:vertmomeq_TIGRESS}). Furthermore, the CR momentum flux difference across the midplane region is lower than the MHD momentum flux difference in all the galactic environments. $[\mathcal{F}_\mathrm{c}(z=0)-\mathcal{F}_\mathrm{c}(z=\pm500\,\rm{pc})]/k_\mathrm{B} \equiv [P_\mathrm{c}(z=0)-P_\mathrm{c}(z=\pm500\,\rm{pc})]/k_\mathrm{B}$ is $\simeq 3.8\times 10^5$, $1.3\times 10^5$ and $4.3\times 10^2$~cm$^{-3}$~K in R2, R4 and R8, respectively. For comparison, $[P_\mathrm{k,z}(z=0)-P_\mathrm{k,z}(z=\pm500\,\rm{pc})]/k_\mathrm{B}$ is $\simeq 5.3\times 10^5$, $1.8\times 10^5$ and $3.9\times 10^3$~cm$^{-3}$~K in R2, R4 and R8, respectively. Even though the CR pressure is higher than  the kinetic pressure near the midplane (based on the $\Upsilon_\mathrm{c}/\Upsilon_\mathrm{k}$ ratio in \autoref{tab:tab2}), the \textit{difference} between the CR pressure at the midplane and the CR pressure at $|z| = 0.5$~kpc is lower than the \textit{difference} between the kinetic pressure at the midplane and the kinetic pressure at $|z| = 0.5$~kpc. Of course, the exact ratio depends on the range of $\Delta z$, and for $\Delta z = \pm 0.35$~kpc (comparable to the MHD gas scale height) the ratio of $\Delta P_\mathrm{c}/\Delta P_\mathrm{k,z}$ is even smaller.  If we look at \autoref{fig:PressProfiles} or \autoref{fig:supportandweight}, we can indeed note that the vertical profile of CR pressure within $|z| < 500$~pc is flatter than the other pressure profiles. The ratio between the CR momentum flux difference and the kinetic momentum flux difference is especially small in R8. We also recall that this model is characterized by the highest diffusivity in the disk region (\autoref{fig:kappa} and \autoref{sec:scattering rate}), which makes the CR pressure profile even flatter than in the other models.

In conclusion, our analysis suggests that the contribution of CRs to offsetting gravity in disks is likely irrelevant in environments with low star formation, and subdominant even in environments with higher star formation rates. The main reason is that waves are strongly damped in neutral gas so the CR pressure is
highly uniform within the denser gas in the midplane region. By contrast, CRs could be more dynamically important at the interface between the mostly-neutral disk and the surrounding corona. In our present simulations, this region is characterized by very large CR pressure gradients, which in turn are a consequence of the primarily-horizontal magnetic field topology near the midplane that 
limits the propagation of CRs out of the disk \citep[see][]{Armillotta+21}. It is likely, however, that the magnetic field topology would be different if the back-reaction of the CR pressure on the gas were included. In particular, local  instabilities near the disk-corona interface, or possibly even global instabilities, could 
cause the magnetic field lines to bend and open up \citep{Parker69, Heintz+20}. 
Rearrangement of the magnetic field topology would enable CRs to stream and diffuse away from the midplane, leading to a significant decrease in the CR pressure gradients at the disk-corona interface. While we can intuitively expect that the self-consistent state is likely to have both lower CR pressure at the midplane and lower CR gradients at the disk-corona interface, testing this remains an important open question.

\subsection{Transfer of momentum to the warm extra-planar gas}
\label{sec:GainofMomentum}

In the previous subsection, we introduced a formalism to analyze the contributions of the various terms that appear in the momentum equation based on a horizontal and temporal averages.  We also quantitatively compared MHD and CR momentum flux terms to gravitational weight terms. 

In this section, we focus on dynamics of the warm extra-planar gas (here defined as gas at $|z| > 0.5$~kpc) only. Explaining the high observed velocities of the warm component of galactic outflows is a longstanding theoretical issue, and various mechanisms have been proposed \citep[see review by][]{Veilleux+20}. The hot gas accelerates under its own pressure gradients, and recent high resolution simulations of a starburst-driven wind have used passive scalars to show that momentum transfer from hot outflowing gas to cooler, denser clouds in a wind can be accomplished by mixing and subsequent cooling, enabling the cooler gas to reach velocities up to 600~$\kms$ \citep{Schneider+20}. The TIGRESS simulations
represent more normal star-forming disks rather than starbursts, and in this case previous analyses have also shown that the warm phase gains considerable momentum flux from the hot phase as gas flows away from the disk \citep[][]{Kim+20,Vijayan+20}.  In principle, additional momentum could be transferred from CRs to the warm gas, which would augment the momentum transfer from hot gas.  In the following, we estimate the potential gain of momentum flux from the CR population in comparison to the gain of momentum flux from the hot phase. 

To quantify the exchange of momentum flux between different gas phases and between gas and CRs outward along the vertical direction, we integrate \autoref{eq:vertmomeq} from an initial height $z_\mathrm{i}$ to an arbitrary height $z$ and separate the contribution of different phases:
\begin{equation}
\begin{split}
\Delta_\mathrm{z} \mathcal{F}_\mathrm{MHD,w} (z) - 
\Delta_\mathrm{z} \mathcal{W}_\mathrm{w}(z)  = 
& -  \Delta_\mathrm{z} \mathcal{F}_\mathrm{MHD,h} (z) \\
 & - \Delta_\mathrm{z}\mathcal{F}_\mathrm{c} (z) \;.
\end{split}
\label{eq:vertmomeq_phases}
\end{equation}
Here, the gas weight $\mathcal{W}$ is defined by \autoref{eq:weight}, the momentum flux and weight differences are defined as $\Delta_\mathrm{z} q \equiv q(z) - q(z_\mathrm{i})$, and $\mathcal{F}_\mathrm{c} (z)$ is a sum over CRs in all thermal phases. In \autoref{eq:vertmomeq_phases}, we retain only the weight term from warm gas as it dominates ($\sum_\mathrm{ph} \mathcal{W}_\mathrm{ph} \approx \mathcal{W}_\mathrm{w}$), and we have dropped 
contributions from the intermediate-temperature 
phase to the MHD momentum flux difference since these 
are small \citep[$|\Delta_\mathrm{z} \mathcal{F}_\mathrm{MHD,i}| \ll |\Delta_\mathrm{z} \mathcal{F}_\mathrm{MHD,w+h}| $; see \autoref{fig:diffphasepress} and also][]{Vijayan+20}. 

The LHS of \autoref{eq:vertmomeq_phases} can be understood as the ``net'' momentum flux difference of the warm gas that arises from interactions. The momentum flux in warm gas tends to decreases as the flow moves outward, $\Delta_\mathrm{z} \mathcal{F}_\mathrm{MHD,w} (z) < 0$, 
simply because it must climb out of the gravitational potential, with $\Delta_\mathrm{z} \mathcal{W}_\mathrm{w}(z)<0$ quantifying the corresponding gravitationally-induced loss of momentum flux. In the absence of CRs, the LHS and the RHS of \autoref{eq:vertmomeq_phases} would individually be equal to zero if there were no exchange of momentum between hot and warm phases. However, this is not the case in the TIGRESS simulations: the momentum flux of the hot gas decreases outward 
($\mathcal{F}_\mathrm{MHD,h} (z) < 0$) as a consequence of transferring momentum to the warm gas, with the LHS of \autoref{eq:vertmomeq_phases} positive. 

\begin{figure*}
\centering
\includegraphics[width=\textwidth]{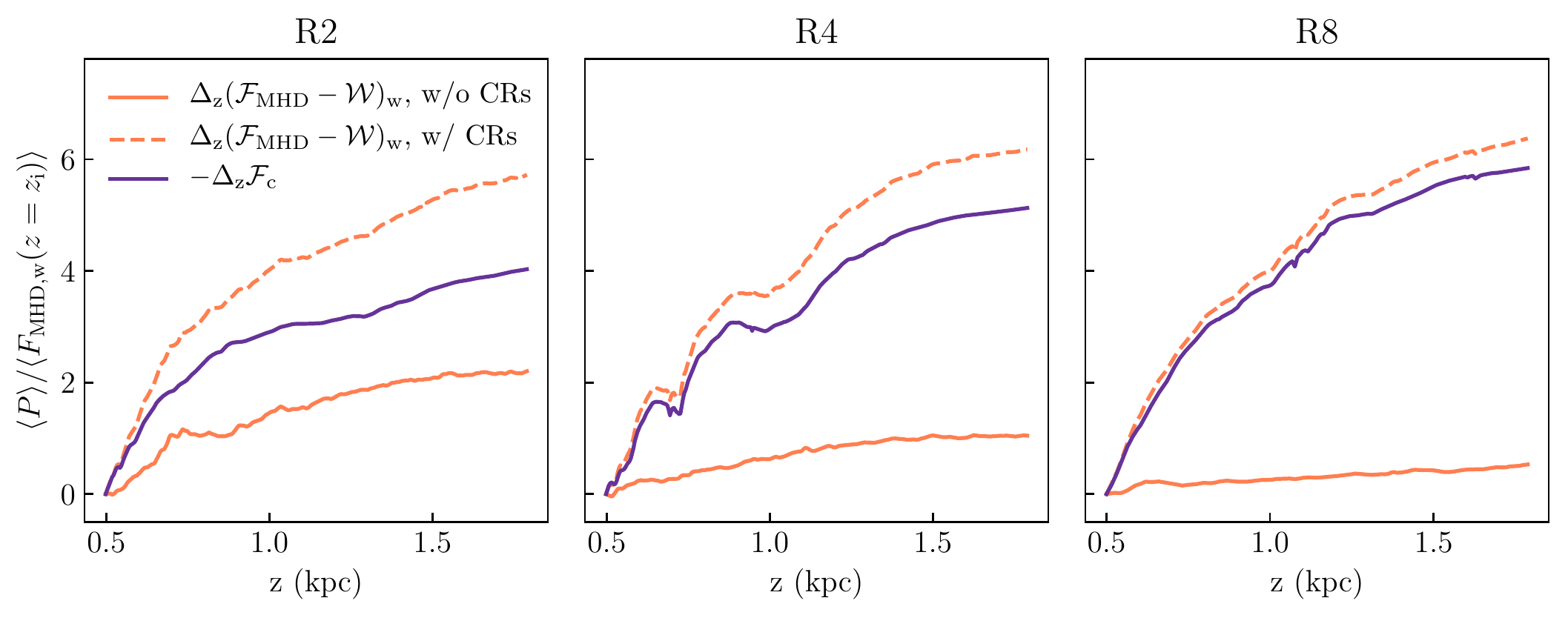}
\caption{Temporally-averaged vertical profiles of the ``net'' MHD momentum flux difference in the warm phase $\Delta_\mathrm{z} (\mathcal{F}_\mathrm{MHD,w} - \mathcal{W}_\mathrm{w})$ (solid coral lines) and of the CR momentum flux difference in the total gas $\Delta_\mathrm{z} \mathcal{F}_\mathrm{c}$ (purple lines). Different panels refer to different TIGRESS models: R2 (\textit{left panel}), R4 (\textit{middle panel}) and R8 (\textit{right panel}) models. The dashed lines show the profile that $\Delta_\mathrm{z} (\mathcal{F}_\mathrm{MHD,w} - \mathcal{W}_\mathrm{w})$ might assume if momentum were transferred from the CRs to the warm gas. The vertical profiles are divided by the MHD momentum flux in the warm phase at $z = z_\mathrm{i} = 500$~pc.}
\label{fig:momentumgain}
\end{figure*}

In \autoref{fig:momentumgain}, we display for each model
the vertical profile of $\Delta_\mathrm{z} (\mathcal{F}_\mathrm{MHD,w} - \mathcal{W}_\mathrm{w})$ as measured directly from the 
TIGRESS simulation in the absence of CRs, with  increase toward larger $z$ due to momentum transfer from hot gas.  We also show the profile $-\Delta_\mathrm{z}\mathcal{F}_\mathrm{c} (z)$ of the CR momentum flux change with height as measured from the post-processed simulations.   The latter corresponds to momentum flux that could in principle have been gained by the warm gas if the back-reaction were included, according to \autoref{eq:vertmomeq_phases}. Thus, by adding these two terms we obtain a ``virtual'' profile of the momentum flux in the warm medium, shown with a dashed curve
for each of the three TIGRESS models. To normalize, each profile is divided by the MHD momentum flux at $z=z_\mathrm{i}$.

\autoref{fig:momentumgain} shows that from the original 
TIGRESS MHD simulation (without CRs), the normalized momentum flux increase is enhanced from R8 to R4 to R2.  That is,  the fractional gain in momentum flux due to transfer from the hot  to
the warm phase is larger for higher $\Sigma_{\rm SFR}$. From $z = 1.0$~kpc to $z = 1.8$~kpc, the warm phase gains about 0.5, 1, and 2 times the original momentum flux in R8, R4, and R2, respectively. In contrast, the magnitude of momentum flux transfer from the CRs is higher at low $\Sigma_{\rm SFR}$, dropping from  R8 to R4 to R2 (see also \autoref{sec:VerticalSupport}).  Quantitatively, the change in $\mathcal{F}_\mathrm{c} $
from $z = 0.5$~kpc to $z = 1.8$~kpc is about 6, 5, and 4 times the original momentum flux in model R8, R4, and R2, respectively. The dashed lines in \autoref{fig:momentumgain} displaying ``virtual'' momentum flux profiles for warm gas show that the potential CR effect in model R8 is much more significant than in model R4.  Quantitatively, the momentum flux increase accounting for CRs could be as large as a factor of 
12, 6, or 3 compared to the increase due to the hot gas interaction alone in model R8, R4, or R2, respectively.  We note, however, that these values should be considered upper limits, since in fact not all of the CR momentum would be transferred to the warm gas.

In conclusion, this analysis shows that the impact of CRs on the dynamics of the warm extra-planar gas is potentially important in all the environments investigated in this paper. However, compared to the momentum flux gained from the hot and fast-moving gas, the potential 
enhancement from CRs is much greater in environments with 
relatively low $\Sigma_{\rm SFR}$. In other words,  CRs may have the most impact to launching of warm outflows in relatively quiescent environments. 

\subsection{Cosmic rays in extra-planar clouds}

In our TIGRESS simulation, warm gas structures are resolved in the extra-planar region, where they are surrounded by (faster-moving) low-density, hot gas (see \multiref{fig:R2snapshot}{fig:R4snapshot}{fig:R8snapshot}). It is interesting to use our post-processed CR distribution to estimate the acceleration that a given cloud might experience as 
a result of CR pressure forces. For this exercise, we focus on the R8 simulation modeling the solar neighborhood environment, where CRs are expected to give a more relevant contribution to the cloud dynamics (see \autoref{sec:GainofMomentum}). 

\begin{figure*}
\centering
\includegraphics[width=\textwidth]{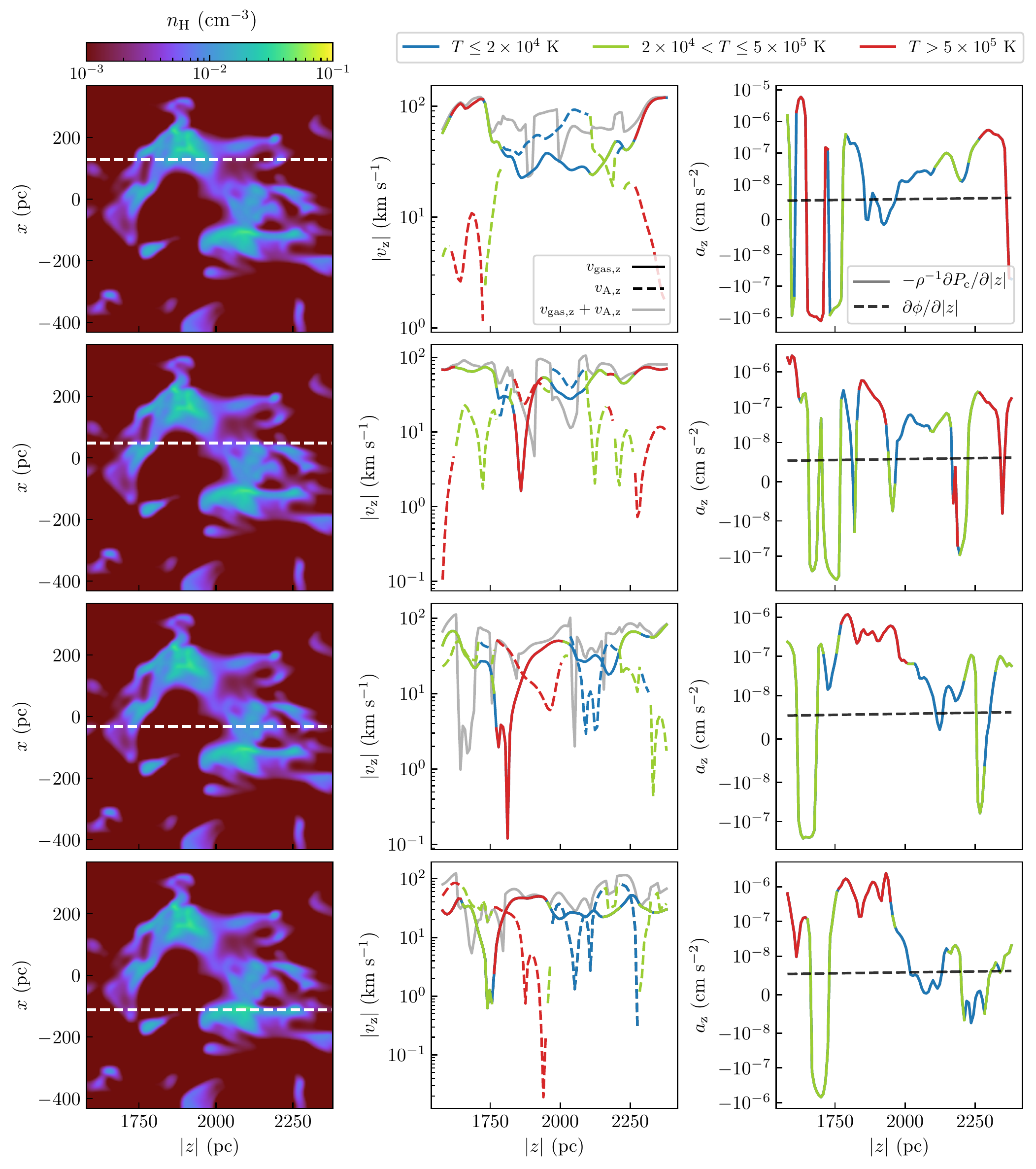}
\caption{Zoom-in on an extra-planar cloud in the simulation modeling the solar environment (R8 model). The cloud has been extracted from a snapshot representative of an outflow-dominated period ($t = 214$~Myr). In each row, the \textit{left panel} shows a slice through the cloud (in the $z-x$ plane) of hydrogen number density $n_\mathrm{H}$, the \textit{middle panel} shows the magnitude of gas velocity (solid line), ion Alfv\'{e}n speed (dashed line) and their sum (gray solid line) along the direction highlighted by the white dashed line in the left panel, while the \textit{right panel} shows the CR pressure-driven acceleration (solid line) and gravitational acceleration (dashed gray line) of the gas along the same direction. In the middle and right panels, colors represent different thermal phases of the gas: blue for warm ($T \leq 2 \times 10^4$~K), green for intermediate  ($2 \times 10^4 < T \leq 5 \times 10^5$~K) and red for hot ($T > 5 \times 10^5$~K). }
\label{fig:cloud1}
\end{figure*}

\begin{figure*}
\centering
\includegraphics[width=\textwidth]{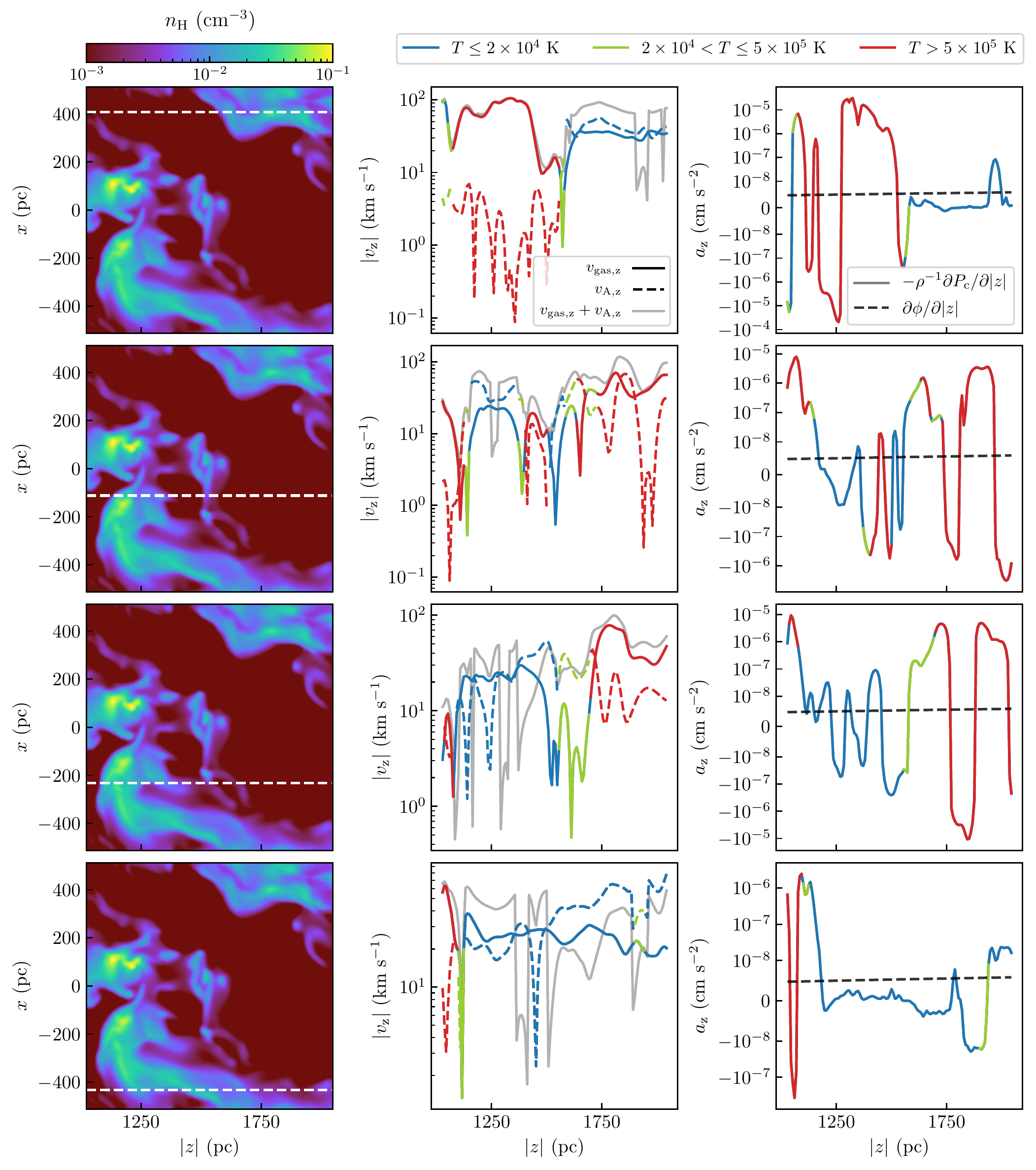}
\caption{Same as \autoref{fig:cloud1}, but showing a cloud extracted from a snapshot representative of an inflow-dominated period ($t = 250$~Myr).}
\label{fig:cloud2}
\end{figure*}

\autoref{fig:cloud1} and \autoref{fig:cloud2} show the results of this analysis for two clouds: the cloud of \autoref{fig:cloud1} is extracted from a snapshot representative of an outflow-dominated period, while the cloud of \autoref{fig:cloud2} is extracted from a snapshot representative of an inflow-dominated period. In both plots, the left-hand panels show a density slice through the cloud, and loci of selected pencil beam cuts along the $z$ direction through the cloud (shown with white dashed lines).  The middle panels show the profiles of gas speed, ion Alfv\'{e}n speed and their sum along the respective
pencil beam cuts.  The right panels shows the acceleration driven by CR pressure gradients $- \partial P_\mathrm{c}/\partial z/\rho$, as well as the absolute value of the gravitational acceleration, along the same lines. Colors (blue, green, red) indicate the temperature of the gas at a given point along each line.

In the cloud of \autoref{fig:cloud1}, representative of an outflow-dominated period, the CR-driven acceleration is mostly positive in the warm phase and larger than the absolute value of the gravitational acceleration, meaning that CRs pressure forces would push the cloud outward in the vertical direction. In contrast, for the cloud shown in \autoref{fig:cloud2}, representative of an inflow-dominated period, the CR-driven acceleration presents a less regular pattern in the warm gas, as it either oscillates between high positive and high negative values or it is lower than the absolute value of the gravitational acceleration. The net dynamical impact of CRs on this cloud would therefore be negligible. 
We conclude that, even though the CR pressure 
in the warm gas 
overall  decreases outward along the vertical direction (see top panels of \autoref{fig:diffphasepress})
the force arising from CR pressure gradients across individual clouds is not necessarily positive and significant compared to the other forces. The extent to which CRs impact the dynamics of individual clouds may vary with the local conditions of gas and magnetic field.

Previous idealized simulations of a CR front impinging on a warm cloud surrounded by a hotter and more tenuous medium have shown that CRs can accelerate the cloud through the so-called  ``bottleneck effect'' \citep[e.g.][]{Wiener+17, Wiener+19, Bruggen+20}. These simulations assume fully-ionized gas ($|v_\mathrm{s}|=|v_\mathrm{A}|$) and uniform magnetic field. As CRs stream down their pressure gradient, they encounter a decrease in the ideal Alfv´en speed on the upstream side of the cloud ($v_\mathrm{A}\propto\rho^{-1/2}$). In case of streaming-dominated transport ($v \ll v_\mathrm{A}$), this leads to a bottleneck in which CRs pile up at the cloud interface, build up their pressure, and exert a force on the cloud. In our simulations, the magnetic field structure and the ionization conditions are quite different form those adopted in idealised simulations. First, the magnetic field is not uniform across space. Second, we properly compute the ionization fraction of the gas and set $|v_\mathrm{s}|=|v_\mathrm{A,i}|$, with $v_\mathrm{A,i} \simeq 3-10 \, v_\mathrm{A}$ in partially-neutral warm gas (see \autoref{sec:results-Ipart}). As a consequence, the CR streaming velocity $v_\mathrm{A,i}$ does not necessarily decrease in the warm gas, as clearly visible in the top middle panels \autoref{fig:cloud1} and \autoref{fig:cloud2}. Moreover, we note that the gas velocity is comparable to the Alfv\'{e}n speed in the warm gas, meaning that advection and streaming are equally important for the transport of CRs (the diffusion velocity is slightly lower than the other components in these intermediate-density clouds). Even though the gas velocity generally decreases in the warm gas, the sum of gas velocity and ion Alfv\'{e}n speed is often comparable to the gas velocity in the hot gas. This may lead to a less effective (or even absent) bottleneck effect and reduced cloud acceleration \citep[e.g.][]{Bustard+21}.

\section{Summary and Discussion}
\label{sec:conclusions}

In this paper, we use the distribution of thermal gas, magnetic field, and supernova energy inputs computed in the TIGRESS MHD simulations \citep{Kim&Ostriker17, Kim+20} to study the propagation of GeV CRs in the multiphase, star-forming ISM. We employ the techniques developed in the previous work of \citet{Armillotta+21} and applied to a simulation with solar neighborhood conditions, now extending to simulations of additional galactic environments. Together, the three environments cover a wide range of galactic conditions typical of Milky Way-like star-forming galaxies in terms of gas surface density ($ \Sigma_\mathrm{gas} \sim 10-100 \, \mo$~pc$^{-2}$), SFR surface density ($\Sigma_\mathrm{SFR} \sim 0.005 - 1 \, \mo$~kpc$^{-2}$~yr$^{-1}$), and midplane total pressure ($P_\mathrm{mid}/k_\mathrm{B} \sim 10^4-10^6 $~cm$^{-3}$~K).

For this study, we extract a set of snapshots from the three TIGRESS simulations and post-process them using the algorithm for CR transport implemented in \textit{Athena}++ by \citet{Jiang&Oh18}. The propagation of CRs includes effects of advection by the background gas, streaming parallel to magnetic field lines down the CR pressure gradient at the local ion Alfv\'en speed, and diffusion relative to the Alfv\'en waves due to wave damping. We consider the realistic scenario in which Alfv\'{e}n waves excited by streaming are responsible for scattering, with a scattering coefficient that varies with the properties of both the background gas and the CRs.  We calculate the scattering coefficient assuming that 
the local wave amplitude is set by the balance of growth and damping, considering  both  ion-neutral damping and non-linear Landau damping.  

A key finding from our study is that the combined transport processes are 
more effective at removing CRs from galaxies in environments with higher SFR (see \autoref{sec:results-Ipart}). These environments are characterized by faster winds \citep[see also][]{Kim+20} that rapidly advect CRs away from the galactic midplane. 
As a result of more efficient transport, the CR feedback yield, defined as the ratio between midplane CR pressure and $\Sigma_\mathrm{SFR}$, decreases at higher $\Sigma_\mathrm{SFR}$. Since the CR pressure increases with $\Sigma_\mathrm{SFR}$ more slowly than other MHD pressures, the ratio between CR pressure and the kinetic/thermal/magnetic pressure decreases at higher $\Sigma_\mathrm{SFR}$. We find that the midplane CR pressure is in equipartition with the midplane thermal and kinetic pressures in the model with the highest $\Sigma_\mathrm{SFR}$, while it is more than a factor 2 larger than the other pressures in the solar neighborhood model.  

To our knowledge, numerical simulations systematically studying the propagation of CRs as a function of galaxy properties have not previously been conducted. Recently, however, there have been several analytic works modeling the transport of CRs in a broad range of galactic environments, from those typical of dwarf galaxies to those typical of extreme starbursts \citep[e.g.][]{Lacki11, Crocker+21a, Crocker+21b, Quataert+21a}. In agreement with our conclusion, these models find that the ratio of midplane CR pressure to midplane gas pressures decreases at higher $\Sigma_\mathrm{SFR}$. In these models, the midplane CR pressure is determined by the efficiency of CR transport and/or the fraction of hadronic losses relative to the energy input rate. In \autoref{sec:pressure}, we have seen that increased hadronic losses are not responsible for the fractional reduction in CR pressures at high $\Sigma_\mathrm{SFR}$ in our models; rather, we attribute the difference primarily to increased advection.  However, unlike the analytic works mentioned here, we do not consider highly star-forming environments representative of starburst galaxies ($\Sigma_\mathrm{SFR} > 10^3\, \mo$~kpc$^{-2}$~yr$^{-1}$). Due to their high gas densities, these environments would undergo significant CR energy losses that strongly reduce the ratio of CR to gas pressure.  

We point out that the above analytic works adopt simplified prescriptions for CR transport compared to our simulations. They assume CR diffusion only, with the diffusion coefficient either constrained through observations of non-thermal emission \citep{Lacki11, Quataert+21a} or estimated based on the average properties of the background gas \citep{Crocker+21a,Crocker+21b}. Despite this difference, they all find that the efficiency of CR diffusion increases with $\Sigma_\mathrm{SFR}$. Here, we also find that the ``effective'' diffusion coefficient -- which encodes the effects of advection and streaming, in addition to true diffusion -- increases with $\Sigma_\mathrm{SFR}$ (see \autoref{tab:tab2}). However, we find that the true diffusion coefficient (the inverse of the scattering rate) is in fact higher at given density in the environment with lower $\Sigma_\mathrm{SFR}$ 
(see \autoref{fig:sigmaprofiles}). This is because the CR pressure gradients are overall lower at lower $\Sigma_\mathrm{SFR}$ due to the lower CR pressure, which reduces scattering (cf. \autoref{eq:sigmaIN} and \autoref{eq:sigmaNLL}).
As we have seen in \autoref{sec:streaming, diffusion and advection}, diffusion dominates transport of CRs in the warm/cold neutral gas, while advection dominates in the hot, ionized gas that fills much of the volume. This renders advection the main mechanism responsible for the overall efficiency of CR transport in our models. Streaming at the ion Alfv\'en speed is secondary to diffusion when mass-weighted and secondary to advection when volume-weighted. 

The current work takes a post-processing approach for studying CR transport in realistic galactic ISM conditions, rather than self-consistently computing the MHD together with the CRs.  Nevertheless, we are able to use our results to investigate the potential dynamical impacts of CRs, and to make predictions for how these are likely to vary with galactic environmental conditions. Our analysis suggests that the CRs have only a minor contribution to disk dynamical equilibrium in the midplane regions, due to the high diffusion in the mostly-neutral gas there. In particular, our model R8 with the lowest $\Sigma_\mathrm{SFR}$ (representative of the solar neighborhood), the net force across the midplane region ($|z|<500$pc) from the vertical gradient of CR pressure is an order of magnitude smaller than the forces arising from the vertical gradient of thermal, kinetic, and magnetic pressures. By contrast, CR pressure gradient forces become much larger than the other pressure gradients in the extra-planar region ($|z|>0.5$~kpc) for model R8; for model R4 the extraplanar pressure forces still exceed MHD forces, while in model R2 they are comparable (see \autoref{fig:supportandweight}). 

Our predictions are in qualitative agreement with the results of recent simulations of Milky Way-like galaxies including CRs \citep{Chan+21}. Similar to our approach in \autoref{sec:VerticalSupport}, \citet{Chan+21} quantify gravitational weight and momentum flux differences (``vertical support'') from different pressure components as a function of the height from the disk. The momentum flux profiles of CRs in their simulations (see their Fig. 2) are flatter than in our simulations (\autoref{fig:supportandweight}). This is mainly because \citet{Chan+21} adopt a spatially-constant diffusion coefficient of $\kappa = 3 \times 10^{29}$ cm$^{2}$~s$^{-1}$, which is more than an order of magnitude larger than the average diffusion coefficient in the extra-planar regions of our models (\autoref{fig:kappa}). Nevertheless, they also find that CRs can become dynamically dominant beyond a few kpc from the midplane.  

Finally, our analysis of  extra-planar regions suggests that CRs may have important dynamical impacts on galactic fountains and/or winds. The contribution of CRs to the acceleration of warm clouds is of particular interest for understanding what drives observed fast outflows in gas at $T\lesssim 10^5$K. Based on  \autoref{fig:momentumgain}, transfer of momentum from CRs could significantly accelerate extra-planar warm gas in all our models, with an increasing impact at lower $\Sigma_\mathrm{SFR}$. In the R8 model, the momentum transfer from CRs to extra-planar warm gas could exceed the transfer from hot gas to warm gas at $|z|>0.5$~kpc by up to an order of magnitude.
Clearly, fully self-consistent simulations with time-dependent MHD and CRs are required to 
explore this intriguing possibility.

\section*{Acknowledgements}
We thank the anonymous referee for valuable comments and suggestions. This work was supported in part by grant 510940 from  the Simons Foundation to E.~C. Ostriker, and in part by Max-Planck/Princeton Center for Plasma Physics (NSF grant PHY-1804048). Computational resources were
provided by the Princeton Institute for Computational Science and Engineering
(PICSciE) and the Office of Information Technology's High Performance Computing
Center at Princeton University. The Center for Computational Astrophysics at the Flatiron Institute is supported by the Simons Foundation.

\bibliography{bib}{}

\begin{thebibliography}{}
\expandafter\ifx\csname natexlab\endcsname\relax\def\natexlab#1{#1}\fi
\providecommand{\url}[1]{\href{#1}{#1}}
\providecommand{\dodoi}[1]{doi:~\href{http://doi.org/#1}{\nolinkurl{#1}}}
\providecommand{\doeprint}[1]{\href{http://ascl.net/#1}{\nolinkurl{http://ascl.net/#1}}}
\providecommand{\doarXiv}[1]{\href{https://arxiv.org/abs/#1}{\nolinkurl{https://arxiv.org/abs/#1}}}

\bibitem[{{Abdo} {et~al.}(2010{\natexlab{a}}){Abdo}, {Ackermann}, {Ajello},
  {Allafort}, {Atwood}, {Baldini}, {Ballet}, {Barbiellini}, {Bastieri},
  {Bechtol}, {Bellazzini}, {Berenji}, {Blandford}, {Bloom}, {Bonamente},
  {Borgland}, {Bouvier}, {Brandt}, {Bregeon}, {Brigida}, {Bruel}, {Buehler},
  {Burnett}, {Buson}, {Caliandro}, {Cameron}, {Cannon}, {Caraveo},
  {Casandjian}, {Cecchi}, {{\c{C}}elik}, {Charles}, {Chekhtman}, {Chiang},
  {Ciprini}, {Claus}, {Cohen-Tanugi}, {Conrad}, {Dermer}, {de Angelis}, {de
  Palma}, {Digel}, {Silva}, {Drell}, {Drlica-Wagner}, {Dubois}, {Favuzzi},
  {Fegan}, {Fortin}, {Frailis}, {Fukazawa}, {Funk}, {Fusco}, {Gargano},
  {Germani}, {Giglietto}, {Giordano}, {Giroletti}, {Glanzman}, {Godfrey},
  {Grenier}, {Grondin}, {Guiriec}, {Gustafsson}, {Hadasch}, {Harding},
  {Hayashi}, {Hayashida}, {Hays}, {Healey}, {Jean}, {J{\'o}hannesson},
  {Johnson}, {Johnson}, {Johnson}, {Kamae}, {Katagiri}, {Kataoka}, {Kerr},
  {Kn{\"o}dlseder}, {Kuss}, {Lande}, {Latronico}, {Lee}, {Lemoine-Goumard},
  {Longo}, {Loparco}, {Lott}, {Lovellette}, {Lubrano}, {Madejski}, {Makeev},
  {Martin}, {Mazziotta}, {Mehault}, {Michelson}, {Mitthumsiri}, {Mizuno},
  {Moiseev}, {Monte}, {Monzani}, {Morselli}, {Moskalenko}, {Murgia},
  {Naumann-Godo}, {Nolan}, {Norris}, {Nuss}, {Ohsugi}, {Okumura}, {Omodei},
  {Orlando}, {Ormes}, {Ozaki}, {Paneque}, {Panetta}, {Parent}, {Pepe},
  {Persic}, {Pesce-Rollins}, {Piron}, {Porter}, {Rain{\`o}}, {Rando},
  {Razzano}, {Reimer}, {Reimer}, {Ritz}, {Romani}, {Sadrozinski}, {Saz
  Parkinson}, {Sgr{\`o}}, {Siskind}, {Smith}, {Smith}, {Spandre}, {Spinelli},
  {Strickman}, {Strigari}, {Strong}, {Suson}, {Takahashi}, {Takahashi},
  {Tanaka}, {Thayer}, {Thompson}, {Tibaldo}, {Torres}, {Tosti}, {Tramacere},
  {Uchiyama}, {Usher}, {Vandenbroucke}, {Vianello}, {Vilchez}, {Vitale},
  {Waite}, {Wang}, {Winer}, {Wood}, {Yang}, \& {Ziegler}}]{Abdo+10a}
{Abdo}, A.~A., {Ackermann}, M., {Ajello}, M., {et~al.} 2010{\natexlab{a}},
  \aap, 523, L2, \dodoi{10.1051/0004-6361/201015759}

\bibitem[{{Abdo} {et~al.}(2010{\natexlab{b}}){Abdo}, {Ackermann}, {Ajello},
  {Baldini}, {Ballet}, {Barbiellini}, {Bastieri}, {Bechtol}, {Bellazzini},
  {Berenji}, {Blandford}, {Bloom}, {Bonamente}, {Borgland}, {Bouvier},
  {Brandt}, {Bregeon}, {Brez}, {Brigida}, {Bruel}, {Buehler}, {Buson},
  {Caliandro}, {Cameron}, {Caraveo}, {Carrigan}, {Casandjian}, {Cecchi},
  {{\c{C}}elik}, {Charles}, {Chekhtman}, {Cheung}, {Chiang}, {Ciprini},
  {Claus}, {Cohen-Tanugi}, {Conrad}, {Dermer}, {de Palma}, {Digel}, {Silva},
  {Drell}, {Dubois}, {Dumora}, {Favuzzi}, {Fegan}, {Fukazawa}, {Funk}, {Fusco},
  {Gargano}, {Gasparrini}, {Gehrels}, {Germani}, {Giglietto}, {Giordano},
  {Giroletti}, {Glanzman}, {Godfrey}, {Grenier}, {Grondin}, {Grove}, {Guiriec},
  {Hadasch}, {Harding}, {Hayashida}, {Hays}, {Horan}, {Hughes}, {Jean},
  {J{\'o}hannesson}, {Johnson}, {Johnson}, {Kamae}, {Katagiri}, {Kataoka},
  {Kerr}, {Kn{\"o}dlseder}, {Kuss}, {Lande}, {Latronico}, {Lee},
  {Lemoine-Goumard}, {Llena Garde}, {Longo}, {Loparco}, {Lovellette},
  {Lubrano}, {Makeev}, {Martin}, {Mazziotta}, {McEnery}, {Michelson},
  {Mitthumsiri}, {Mizuno}, {Monte}, {Monzani}, {Morselli}, {Moskalenko},
  {Murgia}, {Nakamori}, {Naumann-Godo}, {Nolan}, {Norris}, {Nuss}, {Ohsugi},
  {Okumura}, {Omodei}, {Orlando}, {Ormes}, {Panetta}, {Parent}, {Pelassa},
  {Pepe}, {Pesce-Rollins}, {Piron}, {Porter}, {Rain{\`o}}, {Rando}, {Razzano},
  {Reimer}, {Reimer}, {Reposeur}, {Ripken}, {Ritz}, {Romani}, {Sadrozinski},
  {Sander}, {Saz Parkinson}, {Scargle}, {Sgr{\`o}}, {Siskind}, {Smith},
  {Smith}, {Spandre}, {Spinelli}, {Strickman}, {Strong}, {Suson}, {Takahashi},
  {Takahashi}, {Tanaka}, {Thayer}, {Thayer}, {Thompson}, {Tibaldo}, {Torres},
  {Tosti}, {Tramacere}, {Uchiyama}, {Usher}, {Vandenbroucke}, {Vasileiou},
  {Vilchez}, {Vitale}, {Waite}, {Wang}, {Winer}, {Wood}, {Yang}, {Ylinen}, \&
  {Ziegler}}]{Abdo+10b}
---. 2010{\natexlab{b}}, \aap, 523, A46, \dodoi{10.1051/0004-6361/201014855}

\bibitem[{{Abdo} {et~al.}(2010{\natexlab{c}}){Abdo}, {Ackermann}, {Ajello},
  {Atwood}, {Baldini}, {Ballet}, {Barbiellini}, {Bastieri}, {Baughman},
  {Bechtol}, {Bellazzini}, {Berenji}, {Blandford}, {Bloom}, {Bonamente},
  {Borgland}, {Bregeon}, {Brez}, {Brigida}, {Bruel}, {Burnett}, {Buson},
  {Caliandro}, {Cameron}, {Caraveo}, {Casandjian}, {Cecchi}, {{\c{C}}elik},
  {Chekhtman}, {Cheung}, {Chiang}, {Ciprini}, {Claus}, {Cohen-Tanugi},
  {Cominsky}, {Conrad}, {Cutini}, {Dermer}, {de Angelis}, {de Palma}, {Digel},
  {Silva}, {Drell}, {Dubois}, {Dumora}, {Farnier}, {Favuzzi}, {Fegan}, {Focke},
  {Fortin}, {Frailis}, {Fukazawa}, {Fusco}, {Gargano}, {Gasparrini}, {Gehrels},
  {Germani}, {Giavitto}, {Giebels}, {Giglietto}, {Giordano}, {Glanzman},
  {Godfrey}, {Gotthelf}, {Grenier}, {Grondin}, {Grove}, {Guillemot}, {Guiriec},
  {Hanabata}, {Harding}, {Hayashida}, {Hays}, {Horan}, {Hughes}, {Jackson},
  {Jean}, {J{\'o}hannesson}, {Johnson}, {Johnson}, {Johnson}, {Johnson},
  {Kamae}, {Katagiri}, {Kataoka}, {Kawai}, {Kerr}, {Kn{\"o}dlseder}, {Kocian},
  {Kuss}, {Lande}, {Latronico}, {Lemoine-Goumard}, {Longo}, {Loparco}, {Lott},
  {Lovellette}, {Lubrano}, {Madejski}, {Makeev}, {Marshall}, {Martin},
  {Mazziotta}, {McConville}, {McEnery}, {Meurer}, {Michelson}, {Mitthumsiri},
  {Mizuno}, {Moiseev}, {Monte}, {Monzani}, {Morselli}, {Moskalenko}, {Murgia},
  {Nolan}, {Norris}, {Nuss}, {Ohsugi}, {Omodei}, {Orlando}, {Ormes}, {Paneque},
  {Parent}, {Pelassa}, {Pepe}, {Pesce-Rollins}, {Piron}, {Porter}, {Rain{\`o}},
  {Rando}, {Razzano}, {Reimer}, {Reimer}, {Reposeur}, {Ritz}, {Rodriguez},
  {Romani}, {Roth}, {Ryde}, {Sadrozinski}, {Sanchez}, {Sander}, {Saz
  Parkinson}, {Scargle}, {Sellerholm}, {Sgr{\`o}}, {Siskind}, {Smith}, {Smith},
  {Spandre}, {Spinelli}, {Starck}, {Strickman}, {Strong}, {Suson}, {Tajima},
  {Takahashi}, {Tanaka}, {Thayer}, {Thayer}, {Thompson}, {Tibaldo}, {Torres},
  {Tosti}, {Tramacere}, {Uchiyama}, {Usher}, {Vasileiou}, {Venter}, {Vilchez},
  {Vitale}, {Waite}, {Wang}, {Weltevrede}, {Winer}, {Wood}, {Ylinen}, \&
  {Ziegler}}]{Abdo+10c}
---. 2010{\natexlab{c}}, \aap, 512, A7, \dodoi{10.1051/0004-6361/200913474}

\bibitem[{{Acero} {et~al.}(2016){Acero}, {Ackermann}, {Ajello}, {Albert},
  {Baldini}, {Ballet}, {Barbiellini}, {Bastieri}, {Bellazzini}, {Bissaldi},
  {Bloom}, {Bonino}, {Bottacini}, {Brandt}, {Bregeon}, {Bruel}, {Buehler},
  {Buson}, {Caliandro}, {Cameron}, {Caragiulo}, {Caraveo}, {Casandjian},
  {Cavazzuti}, {Cecchi}, {Charles}, {Chekhtman}, {Chiang}, {Chiaro}, {Ciprini},
  {Claus}, {Cohen-Tanugi}, {Conrad}, {Cuoco}, {Cutini}, {D'Ammando}, {de
  Angelis}, {de Palma}, {Desiante}, {Digel}, {Di Venere}, {Drell}, {Favuzzi},
  {Fegan}, {Ferrara}, {Focke}, {Franckowiak}, {Funk}, {Fusco}, {Gargano},
  {Gasparrini}, {Giglietto}, {Giordano}, {Giroletti}, {Glanzman}, {Godfrey},
  {Grenier}, {Guiriec}, {Hadasch}, {Harding}, {Hayashi}, {Hays}, {Hewitt},
  {Hill}, {Horan}, {Hou}, {Jogler}, {J{\'o}hannesson}, {Kamae}, {Kuss},
  {Landriu}, {Larsson}, {Latronico}, {Li}, {Li}, {Longo}, {Loparco},
  {Lovellette}, {Lubrano}, {Maldera}, {Malyshev}, {Manfreda}, {Martin},
  {Mayer}, {Mazziotta}, {McEnery}, {Michelson}, {Mirabal}, {Mizuno}, {Monzani},
  {Morselli}, {Nuss}, {Ohsugi}, {Omodei}, {Orienti}, {Orlando}, {Ormes},
  {Paneque}, {Pesce-Rollins}, {Piron}, {Pivato}, {Rain{\`o}}, {Rando},
  {Razzano}, {Razzaque}, {Reimer}, {Reimer}, {Remy}, {Renault},
  {S{\'a}nchez-Conde}, {Schaal}, {Schulz}, {Sgr{\`o}}, {Siskind}, {Spada},
  {Spandre}, {Spinelli}, {Strong}, {Suson}, {Tajima}, {Takahashi}, {Thayer},
  {Thompson}, {Tibaldo}, {Tinivella}, {Torres}, {Tosti}, {Troja}, {Vianello},
  {Werner}, {Wood}, {Wood}, {Zaharijas}, \& {Zimmer}}]{Acero+16}
{Acero}, F., {Ackermann}, M., {Ajello}, M., {et~al.} 2016, \apjs, 223, 26,
  \dodoi{10.3847/0067-0049/223/2/26}

\bibitem[{{Ackermann} {et~al.}(2012){Ackermann}, {Ajello}, {Allafort},
  {Baldini}, {Ballet}, {Bastieri}, {Bechtol}, {Bellazzini}, {Berenji}, {Bloom},
  {Bonamente}, {Borgland}, {Bouvier}, {Bregeon}, {Brigida}, {Bruel}, {Buehler},
  {Buson}, {Caliandro}, {Cameron}, {Caraveo}, {Casandjian}, {Cecchi},
  {Charles}, {Chekhtman}, {Cheung}, {Chiang}, {Cillis}, {Ciprini}, {Claus},
  {Cohen-Tanugi}, {Conrad}, {Cutini}, {de Palma}, {Dermer}, {Digel}, {Silva},
  {Drell}, {Drlica-Wagner}, {Favuzzi}, {Fegan}, {Fortin}, {Fukazawa}, {Funk},
  {Fusco}, {Gargano}, {Gasparrini}, {Germani}, {Giglietto}, {Giordano},
  {Glanzman}, {Godfrey}, {Grenier}, {Guiriec}, {Gustafsson}, {Hadasch},
  {Hayashida}, {Hays}, {Hughes}, {J{\'o}hannesson}, {Johnson}, {Kamae},
  {Katagiri}, {Kataoka}, {Kn{\"o}dlseder}, {Kuss}, {Lande}, {Longo}, {Loparco},
  {Lott}, {Lovellette}, {Lubrano}, {Madejski}, {Martin}, {Mazziotta},
  {McEnery}, {Michelson}, {Mizuno}, {Monte}, {Monzani}, {Morselli},
  {Moskalenko}, {Murgia}, {Nishino}, {Norris}, {Nuss}, {Ohno}, {Ohsugi},
  {Okumura}, {Omodei}, {Orlando}, {Ozaki}, {Parent}, {Persic}, {Pesce-Rollins},
  {Petrosian}, {Pierbattista}, {Piron}, {Pivato}, {Porter}, {Rain{\`o}},
  {Rando}, {Razzano}, {Reimer}, {Reimer}, {Ritz}, {Roth}, {Sbarra}, {Sgr{\`o}},
  {Siskind}, {Spandre}, {Spinelli}, {Stawarz}, {Strong}, {Takahashi}, {Tanaka},
  {Thayer}, {Tibaldo}, {Tinivella}, {Torres}, {Tosti}, {Troja}, {Uchiyama},
  {Vandenbroucke}, {Vianello}, {Vitale}, {Waite}, {Wood}, \&
  {Yang}}]{Ackermann+12}
{Ackermann}, M., {Ajello}, M., {Allafort}, A., {et~al.} 2012, \apj, 755, 164,
  \dodoi{10.1088/0004-637X/755/2/164}

\bibitem[{{Ackermann} {et~al.}(2014){Ackermann}, {Ajello}, {Albert}, {Baldini},
  {Ballet}, {Barbiellini}, {Bastieri}, {Bellazzini}, {Bissaldi}, {Blandford},
  {Bloom}, {Bottacini}, {Brandt}, {Bregeon}, {Bruel}, {Buehler}, {Buson},
  {Caliandro}, {Cameron}, {Caragiulo}, {Caraveo}, {Cavazzuti}, {Charles},
  {Chekhtman}, {Cheung}, {Chiang}, {Chiaro}, {Ciprini}, {Claus},
  {Cohen-Tanugi}, {Conrad}, {Corbel}, {D'Ammando}, {de Angelis}, {den Hartog},
  {de Palma}, {Dermer}, {Desiante}, {Digel}, {Di Venere}, {do Couto e Silva},
  {Donato}, {Drell}, {Drlica-Wagner}, {Favuzzi}, {Ferrara}, {Focke},
  {Franckowiak}, {Fuhrmann}, {Fukazawa}, {Fusco}, {Gargano}, {Gasparrini},
  {Germani}, {Giglietto}, {Giordano}, {Giroletti}, {Glanzman}, {Godfrey},
  {Grenier}, {Grove}, {Guiriec}, {Hadasch}, {Harding}, {Hayashida}, {Hays},
  {Hewitt}, {Hill}, {Hou}, {Jean}, {Jogler}, {J{\'o}hannesson}, {Johnson},
  {Johnson}, {Kerr}, {Kn{\"o}dlseder}, {Kuss}, {Larsson}, {Latronico},
  {Lemoine-Goumard}, {Longo}, {Loparco}, {Lott}, {Lovellette}, {Lubrano},
  {Manfreda}, {Martin}, {Massaro}, {Mayer}, {Mazziotta}, {McEnery},
  {Michelson}, {Mitthumsiri}, {Mizuno}, {Monzani}, {Morselli}, {Moskalenko},
  {Murgia}, {Nemmen}, {Nuss}, {Ohsugi}, {Omodei}, {Orienti}, {Orlando},
  {Ormes}, {Paneque}, {Panetta}, {Perkins}, {Pesce-Rollins}, {Piron}, {Pivato},
  {Porter}, {Rain{\`o}}, {Rando}, {Razzano}, {Razzaque}, {Reimer}, {Reimer},
  {Reposeur}, {Saz Parkinson}, {Schaal}, {Schulz}, {Sgr{\`o}}, {Siskind},
  {Spandre}, {Spinelli}, {Stawarz}, {Suson}, {Takahashi}, {Tanaka}, {Thayer},
  {Thayer}, {Thompson}, {Tibaldo}, {Tinivella}, {Torres}, {Tosti}, {Troja},
  {Uchiyama}, {Vianello}, {Winer}, {Wolff}, {Wood}, {Wood}, {Wood},
  {Charbonnel}, {Corbet}, {De Gennaro Aquino}, {Edlin}, {Mason}, {Schwarz},
  {Shore}, {Starrfield}, {Teyssier}, \& {Fermi-LAT
  Collaboration}}]{Ackermann+13}
{Ackermann}, M., {Ajello}, M., {Albert}, A., {et~al.} 2014, Science, 345, 554,
  \dodoi{10.1126/science.1253947}

\bibitem[{{Aguilar} {et~al.}(2014){Aguilar}, {Aisa}, {Alvino}, {Ambrosi},
  {Andeen}, {Arruda}, {Attig}, {Azzarello}, {Bachlechner}, {Barao}, {Barrau},
  {Barrin}, {Bartoloni}, {Basara}, {Battarbee}, {Battiston}, {Bazo}, {Becker},
  {Behlmann}, {Beischer}, {Berdugo}, {Bertucci}, {Bigongiari}, {Bindi},
  {Bizzaglia}, {Bizzarri}, {Boella}, {de Boer}, {Bollweg}, {Bonnivard},
  {Borgia}, {Borsini}, {Boschini}, {Bourquin}, {Burger}, {Cadoux}, {Cai},
  {Capell}, {Caroff}, {Casaus}, {Cascioli}, {Castellini}, {Cernuda},
  {Cervelli}, {Chae}, {Chang}, {Chen}, {Chen}, {Cheng}, {Chen}, {Cheng},
  {Chikanian}, {Chou}, {Choumilov}, {Choutko}, {Chung}, {Clark}, {Clavero},
  {Coignet}, {Consolandi}, {Contin}, {Corti}, {Coste}, {Cui}, {Dai}, {Delgado},
  {Della Torre}, {Demirk{\"o}z}, {Derome}, {Di Falco}, {Di Masso}, {Dimiccoli},
  {D{\'\i}az}, {von Doetinchem}, {Du}, {Duranti}, {D'Urso}, {Eline}, {Eppling},
  {Eronen}, {Fan}, {Farnesini}, {Feng}, {Fiandrini}, {Fiasson}, {Finch},
  {Fisher}, {Galaktionov}, {Gallucci}, {Garc{\'\i}a}, {Garc{\'\i}a-L{\'o}pez},
  {Gast}, {Gebauer}, {Gervasi}, {Ghelfi}, {Gillard}, {Giovacchini}, {Goglov},
  {Gong}, {Goy}, {Grabski}, {Grand i}, {Graziani}, {Guandalini}, {Guerri},
  {Guo}, {Habiby}, {Haino}, {Han}, {He}, {Heil}, {Hoffman}, {Hsieh}, {Huang},
  {Huh}, {Incagli}, {Ionica}, {Jang}, {Jinchi}, {Kanishev}, {Kim}, {Kim},
  {Kirn}, {Kossakowski}, {Kounina}, {Kounine}, {Koutsenko}, {Krafczyk}, {Kunz},
  {La Vacca}, {Laudi}, {Laurenti}, {Lazzizzera}, {Lebedev}, {Lee}, {Lee},
  {Leluc}, {Li}, {Li}, {Li}, {Li}, {Li}, {Li}, {Li}, {Li}, {Li}, {Lim}, {Lin},
  {Lipari}, {Lippert}, {Liu}, {Liu}, {Lomtadze}, {Lu}, {Lu}, {Luebelsmeyer},
  {Luo}, {Luo}, {Lv}, {Majka}, {Malinin}, {Ma{\~n}{\'a}}, {Mar{\'\i}n},
  {Martin}, {Mart{\'\i}nez}, {Masi}, {Maurin}, {Menchaca-Rocha}, {Meng}, {Mo},
  {Morescalchi}, {Mott}, {M{\"u}ller}, {Ni}, {Nikonov}, {Nozzoli}, {Nunes},
  {Obermeier}, {Oliva}, {Orcinha}, {Palmonari}, {Palomares}, {Paniccia},
  {Papi}, {Pedreschi}, {Pensotti}, {Pereira}, {Pilo}, {Piluso}, {Pizzolotto},
  {Plyaskin}, {Pohl}, {Poireau}, {Postaci}, {Putze}, {Quadrani}, {Qi},
  {Rancoita}, {Rapin}, {Ricol}, {Rodr{\'\i}guez}, {Rosier-Lees}, {Rozhkov},
  {Rozza}, {Sagdeev}, {Sandweiss}, {Saouter}, {Sbarra}, {Schael}, {Schmidt},
  {Schuckardt}, {von Dratzig}, {Schwering}, {Scolieri}, {Seo}, {Shan}, {Shan},
  {Shi}, {Shi}, {Shi}, {Siedenburg}, {Son}, {Spada}, {Spinella}, {Sun}, {Sun},
  {Tacconi}, {Tang}, {Tang}, {Tang}, {Tao}, {Tescaro}, {Ting}, {Ting},
  {Tomassetti}, {Torsti}, {T{\"u}rko{\v{g}}lu}, {Urban}, {Vagelli}, {Valente},
  {Vannini}, {Valtonen}, {Vaurynovich}, {Vecchi}, {Velasco}, {Vialle}, {Wang},
  {Wang}, {Wang}, {Wang}, {Wang}, {Weng}, {Whitman}, {Wienkenh{\"o}ver}, {Wu},
  {Xia}, {Xie}, {Xie}, {Xiong}, {Xin}, {Xu}, {Xu}, {Yan}, {Yang}, {Yang}, {Ye},
  {Yi}, {Yu}, {Yu}, {Zeissler}, {Zhang}, {Zhang}, {Zhang}, {Zhang}, {Zheng},
  {Zhuang}, {Zhukov}, {Zichichi}, {Zimmermann}, {Zuccon}, {Zurbach}, \& {AMS
  Collaboration}}]{Aguilar+14}
{Aguilar}, M., {Aisa}, D., {Alvino}, A., {et~al.} 2014, \prl, 113, 121102,
  \dodoi{10.1103/PhysRevLett.113.121102}

\bibitem[{{Aguilar} {et~al.}(2015){Aguilar}, {Aisa}, {Alpat}, {Alvino},
  {Ambrosi}, {Andeen}, {Arruda}, {Attig}, {Azzarello}, {Bachlechner}, {Barao},
  {Barrau}, {Barrin}, {Bartoloni}, {Basara}, {Battarbee}, {Battiston}, {Bazo},
  {Becker}, {Behlmann}, {Beischer}, {Berdugo}, {Bertucci}, {Bindi},
  {Bizzaglia}, {Bizzarri}, {Boella}, {de Boer}, {Bollweg}, {Bonnivard},
  {Borgia}, {Borsini}, {Boschini}, {Bourquin}, {Burger}, {Cadoux}, {Cai},
  {Capell}, {Caroff}, {Casaus}, {Castellini}, {Cernuda}, {Cerreta}, {Cervelli},
  {Chae}, {Chang}, {Chen}, {Chen}, {Chen}, {Chen}, {Cheng}, {Chou},
  {Choumilov}, {Choutko}, {Chung}, {Clark}, {Clavero}, {Coignet}, {Consolandi},
  {Contin}, {Corti}, {Gil}, {Coste}, {Creus}, {Crispoltoni}, {Cui}, {Dai},
  {Delgado}, {Della Torre}, {Demirk{\"o}z}, {Derome}, {Di Falco}, {Di Masso},
  {Dimiccoli}, {D{\'\i}az}, {von Doetinchem}, {Donnini}, {Duranti}, {D'Urso},
  {Egorov}, {Eline}, {Eppling}, {Eronen}, {Fan}, {Farnesini}, {Feng},
  {Fiandrini}, {Fiasson}, {Finch}, {Fisher}, {Formato}, {Galaktionov},
  {Gallucci}, {Garc{\'\i}a}, {Garc{\'\i}a-L{\'o}pez}, {Gargiulo}, {Gast},
  {Gebauer}, {Gervasi}, {Ghelfi}, {Giovacchini}, {Goglov}, {Gong}, {Goy},
  {Grabski}, {Grandi}, {Graziani}, {Guand alini}, {Guerri}, {Guo}, {Haas},
  {Habiby}, {Haino}, {Han}, {He}, {Heil}, {Hoffman}, {Hsieh}, {Huang}, {Huh},
  {Incagli}, {Ionica}, {Jang}, {Jinchi}, {Kanishev}, {Kim}, {Kim}, {Kirn},
  {Korkmaz}, {Kossakowski}, {Kounina}, {Kounine}, {Koutsenko}, {Krafczyk}, {La
  Vacca}, {Laudi}, {Laurenti}, {Lazzizzera}, {Lebedev}, {Lee}, {Lee}, {Leluc},
  {Li}, {Li}, {Li}, {Li}, {Li}, {Li}, {Li}, {Li}, {Li}, {Li}, {Lim}, {Lin},
  {Lipari}, {Lippert}, {Liu}, {Liu}, {Liu}, {Lolli}, {Lomtadze}, {Lu}, {Lu},
  {Lu}, {Luebelsmeyer}, {Luo}, {Luo}, {Lv}, {Majka}, {Ma{\~n}{\'a}},
  {Mar{\'\i}n}, {Martin}, {Mart{\'\i}nez}, {Masi}, {Maurin}, {Menchaca-Rocha},
  {Meng}, {Mo}, {Morescalchi}, {Mott}, {M{\"u}ller}, {Nelson}, {Ni}, {Nikonov},
  {Nozzoli}, {Nunes}, {Obermeier}, {Oliva}, {Orcinha}, {Palmonari},
  {Palomares}, {Paniccia}, {Papi}, {Pauluzzi}, {Pedreschi}, {Pensotti},
  {Pereira}, {Picot-Clemente}, {Pilo}, {Piluso}, {Pizzolotto}, {Plyaskin},
  {Pohl}, {Poireau}, {Putze}, {Quadrani}, {Qi}, {Qin}, {Qu}, {R{\"a}ih{\"a}},
  {Rancoita}, {Rapin}, {Ricol}, {Rodr{\'\i}guez}, {Rosier-Lees}, {Rozhkov},
  {Rozza}, {Sagdeev}, {Sandweiss}, {Saouter}, {Schael}, {Schmidt}, {von
  Dratzig}, {Schwering}, {Scolieri}, {Seo}, {Shan}, {Shan}, {Shi}, {Shi},
  {Shi}, {Siedenburg}, {Son}, {Song}, {Spada}, {Spinella}, {Sun}, {Sun},
  {Tacconi}, {Tang}, {Tang}, {Tang}, {Tao}, {Tescaro}, {Ting}, {Ting},
  {Tomassetti}, {Torsti}, {T{\"u}rko{\v{g}}lu}, {Urban}, {Vagelli}, {Valente},
  {Vannini}, {Valtonen}, {Vaurynovich}, {Vecchi}, {Velasco}, {Vialle},
  {Vitale}, {Vitillo}, {Wang}, {Wang}, {Wang}, {Wang}, {Wang}, {Wang}, {Weng},
  {Whitman}, {Wienkenh{\"o}ver}, {Willenbrock}, {Wu}, {Wu}, {Xia}, {Xie},
  {Xie}, {Xiong}, {Xu}, {Xu}, {Yan}, {Yang}, {Yang}, {Yang}, {Ye}, {Yi}, {Yu},
  {Yu}, {Zeissler}, {Zhang}, {Zhang}, {Zhang}, {Zhang}, {Zhang}, {Zhang},
  {Zhang}, {Zheng}, {Zhuang}, {Zhukov}, {Zichichi}, {Zimmermann}, {Zuccon}, \&
  {AMS Collaboration}}]{Aguilar+15}
{Aguilar}, M., {Aisa}, D., {Alpat}, B., {et~al.} 2015, \prl, 115, 211101,
  \dodoi{10.1103/PhysRevLett.115.211101}

\bibitem[{{Aharonian} {et~al.}(2020){Aharonian}, {Peron}, {Yang}, {Casanova},
  \& {Zanin}}]{Ahronian2020}
{Aharonian}, F., {Peron}, G., {Yang}, R., {Casanova}, S., \& {Zanin}, R. 2020,
  \prd, 101, 083018, \dodoi{10.1103/PhysRevD.101.083018}

\bibitem[{{Amato} \& {Blasi}(2018)}]{Amato&Blasi18}
{Amato}, E., \& {Blasi}, P. 2018, Advances in Space Research, 62, 2731,
  \dodoi{10.1016/j.asr.2017.04.019}

\bibitem[{{Armillotta} {et~al.}(2021){Armillotta}, {Ostriker}, \&
  {Jiang}}]{Armillotta+21}
{Armillotta}, L., {Ostriker}, E.~C., \& {Jiang}, Y.-F. 2021, \apj, 922, 11,
  \dodoi{10.3847/1538-4357/ac1db2}

\bibitem[{{Bai} {et~al.}(2019){Bai}, {Ostriker}, {Plotnikov}, \&
  {Stone}}]{Bai2019}
{Bai}, X.-N., {Ostriker}, E.~C., {Plotnikov}, I., \& {Stone}, J.~M. 2019, \apj,
  876, 60, \dodoi{10.3847/1538-4357/ab1648}

\bibitem[{{Bambic} {et~al.}(2021){Bambic}, {Bai}, \& {Ostriker}}]{Bambic2021}
{Bambic}, C.~J., {Bai}, X.-N., \& {Ostriker}, E.~C. 2021, \apj, 920, 141,
  \dodoi{10.3847/1538-4357/ac0ce7}

\bibitem[{{Beck}(2001)}]{Beck01}
{Beck}, R. 2001, \ssr, 99, 243.
\newblock \doarXiv{astro-ph/0012402}

\bibitem[{{Bell}(2004)}]{Bell04}
{Bell}, A.~R. 2004, \mnras, 353, 550, \dodoi{10.1111/j.1365-2966.2004.08097.x}

\bibitem[{{Blasi} {et~al.}(2012){Blasi}, {Amato}, \& {Serpico}}]{Blasi2012}
{Blasi}, P., {Amato}, E., \& {Serpico}, P.~D. 2012, \prl, 109, 061101,
  \dodoi{10.1103/PhysRevLett.109.061101}

\bibitem[{{Booth} {et~al.}(2013){Booth}, {Agertz}, {Kravtsov}, \&
  {Gnedin}}]{Booth+13}
{Booth}, C.~M., {Agertz}, O., {Kravtsov}, A.~V., \& {Gnedin}, N.~Y. 2013,
  \apjl, 777, L16, \dodoi{10.1088/2041-8205/777/1/L16}

\bibitem[{{Boulares} \& {Cox}(1990)}]{Boulares&Cox90}
{Boulares}, A., \& {Cox}, D.~P. 1990, \apj, 365, 544, \dodoi{10.1086/169509}

\bibitem[{{Breitschwerdt} {et~al.}(1991){Breitschwerdt}, {McKenzie}, \&
  {Voelk}}]{Breitschwerdt+91}
{Breitschwerdt}, D., {McKenzie}, J.~F., \& {Voelk}, H.~J. 1991, \aap, 245, 79

\bibitem[{{Br{\"u}ggen} \& {Scannapieco}(2020)}]{Bruggen+20}
{Br{\"u}ggen}, M., \& {Scannapieco}, E. 2020, \apj, 905, 19,
  \dodoi{10.3847/1538-4357/abc00f}

\bibitem[{{Bustard} \& {Zweibel}(2021)}]{Bustard+21}
{Bustard}, C., \& {Zweibel}, E.~G. 2021, \apj, 913, 106,
  \dodoi{10.3847/1538-4357/abf64c}

\bibitem[{{Butsky} {et~al.}(2020){Butsky}, {Fielding}, {Hayward}, {Hummels},
  {Quinn}, \& {Werk}}]{Butsky+20}
{Butsky}, I.~S., {Fielding}, D.~B., {Hayward}, C.~C., {et~al.} 2020, \apj, 903,
  77, \dodoi{10.3847/1538-4357/abbad2}

\bibitem[{{Chan} {et~al.}(2021){Chan}, {Keres}, {Gurvich}, {Hopkins}, {Trapp},
  {Ji}, \& {Faucher-Giguere}}]{Chan+21}
{Chan}, T.~K., {Keres}, D., {Gurvich}, A.~B., {et~al.} 2021, arXiv e-prints,
  arXiv:2110.06231.
\newblock \doarXiv{2110.06231}

\bibitem[{{Chan} {et~al.}(2019){Chan}, {Kere{\v{s}}}, {Hopkins}, {Quataert},
  {Su}, {Hayward}, \& {Faucher-Gigu{\`e}re}}]{Chan+19}
{Chan}, T.~K., {Kere{\v{s}}}, D., {Hopkins}, P.~F., {et~al.} 2019, \mnras, 488,
  3716, \dodoi{10.1093/mnras/stz1895}

\bibitem[{{Chandran}(2000)}]{Chandran00}
{Chandran}, B. D.~G. 2000, \apj, 529, 513, \dodoi{10.1086/308232}

\bibitem[{{Crocker} {et~al.}(2021{\natexlab{a}}){Crocker}, {Krumholz}, \&
  {Thompson}}]{Crocker+21a}
{Crocker}, R.~M., {Krumholz}, M.~R., \& {Thompson}, T.~A. 2021{\natexlab{a}},
  \mnras, 502, 1312, \dodoi{10.1093/mnras/stab148}

\bibitem[{{Crocker} {et~al.}(2021{\natexlab{b}}){Crocker}, {Krumholz}, \&
  {Thompson}}]{Crocker+21b}
---. 2021{\natexlab{b}}, \mnras, 503, 2651, \dodoi{10.1093/mnras/stab502}

\bibitem[{{Cummings} {et~al.}(2016){Cummings}, {Stone}, {Heikkila}, {Lal},
  {Webber}, {J{\'o}hannesson}, {Moskalenko}, {Orlando}, \&
  {Porter}}]{Cummings+16}
{Cummings}, A.~C., {Stone}, E.~C., {Heikkila}, B.~C., {et~al.} 2016, \apj, 831,
  18, \dodoi{10.3847/0004-637X/831/1/18}

\bibitem[{{Dashyan} \& {Dubois}(2020)}]{Dashyan+20}
{Dashyan}, G., \& {Dubois}, Y. 2020, \aap, 638, A123,
  \dodoi{10.1051/0004-6361/201936339}

\bibitem[{{Dorfi} \& {Breitschwerdt}(2012)}]{Dorfi&Breitschwerdt12}
{Dorfi}, E.~A., \& {Breitschwerdt}, D. 2012, \aap, 540, A77,
  \dodoi{10.1051/0004-6361/201118082}

\bibitem[{{Draine}(2011)}]{Draine11}
{Draine}, B.~T. 2011, {Physics of the Interstellar and Intergalactic Medium}

\bibitem[{{Everett} {et~al.}(2008){Everett}, {Zweibel}, {Benjamin}, {McCammon},
  {Rocks}, \& {Gallagher}}]{Everett+08}
{Everett}, J.~E., {Zweibel}, E.~G., {Benjamin}, R.~A., {et~al.} 2008, \apj,
  674, 258, \dodoi{10.1086/524766}

\bibitem[{{Evoli} {et~al.}(2018){Evoli}, {Blasi}, {Morlino}, \&
  {Aloisio}}]{Evoli+18}
{Evoli}, C., {Blasi}, P., {Morlino}, G., \& {Aloisio}, R. 2018, \prl, 121,
  021102, \dodoi{10.1103/PhysRevLett.121.021102}

\bibitem[{{Farber} {et~al.}(2018){Farber}, {Ruszkowski}, {Yang}, \&
  {Zweibel}}]{Farber+18}
{Farber}, R., {Ruszkowski}, M., {Yang}, H. Y.~K., \& {Zweibel}, E.~G. 2018,
  \apj, 856, 112, \dodoi{10.3847/1538-4357/aab26d}

\bibitem[{{Girichidis} {et~al.}(2018){Girichidis}, {Naab}, {Hanasz}, \&
  {Walch}}]{Girichidis+18}
{Girichidis}, P., {Naab}, T., {Hanasz}, M., \& {Walch}, S. 2018, \mnras, 479,
  3042, \dodoi{10.1093/mnras/sty1653}

\bibitem[{{Girichidis} {et~al.}(2021){Girichidis}, {Pfrommer}, {Pakmor}, \&
  {Springel}}]{Girichidis+21}
{Girichidis}, P., {Pfrommer}, C., {Pakmor}, R., \& {Springel}, V. 2021, arXiv
  e-prints, arXiv:2109.13250.
\newblock \doarXiv{2109.13250}

\bibitem[{{Girichidis} {et~al.}(2016){Girichidis}, {Naab}, {Walch}, {Hanasz},
  {Mac Low}, {Ostriker}, {Gatto}, {Peters}, {W{\"u}nsch}, {Glover}, {Klessen},
  {Clark}, \& {Baczynski}}]{Girichidis+16}
{Girichidis}, P., {Naab}, T., {Walch}, S., {et~al.} 2016, \apjl, 816, L19,
  \dodoi{10.3847/2041-8205/816/2/L19}

\bibitem[{{Grenier} {et~al.}(2015){Grenier}, {Black}, \& {Strong}}]{Grenier+15}
{Grenier}, I.~A., {Black}, J.~H., \& {Strong}, A.~W. 2015, \araa, 53, 199,
  \dodoi{10.1146/annurev-astro-082214-122457}

\bibitem[{{Hanasz} {et~al.}(2013){Hanasz}, {Lesch}, {Naab}, {Gawryszczak},
  {Kowalik}, \& {W{\'o}lta{\'n}ski}}]{Hanasz2013}
{Hanasz}, M., {Lesch}, H., {Naab}, T., {et~al.} 2013, \apjl, 777, L38,
  \dodoi{10.1088/2041-8205/777/2/L38}

\bibitem[{{Hanasz} {et~al.}(2021){Hanasz}, {Strong}, \&
  {Girichidis}}]{Hanasz+21}
{Hanasz}, M., {Strong}, A., \& {Girichidis}, P. 2021, arXiv e-prints,
  arXiv:2106.08426.
\newblock \doarXiv{2106.08426}

\bibitem[{{Heintz} {et~al.}(2020){Heintz}, {Bustard}, \& {Zweibel}}]{Heintz+20}
{Heintz}, E., {Bustard}, C., \& {Zweibel}, E.~G. 2020, \apj, 891, 157,
  \dodoi{10.3847/1538-4357/ab7453}

\bibitem[{{Hopkins} {et~al.}(2020){Hopkins}, {Squire}, {Chan}, {Quataert},
  {Ji}, {Keres}, \& {Faucher-Giguere}}]{Hopkins+20}
{Hopkins}, P.~F., {Squire}, J., {Chan}, T.~K., {et~al.} 2020, arXiv e-prints,
  arXiv:2002.06211.
\newblock \doarXiv{2002.06211}

\bibitem[{{Ipavich}(1975)}]{Ipavich75}
{Ipavich}, F.~M. 1975, \apj, 196, 107, \dodoi{10.1086/153397}

\bibitem[{{Ji} {et~al.}(2020){Ji}, {Chan}, {Hummels}, {Hopkins}, {Stern},
  {Kere{\v{s}}}, {Quataert}, {Faucher-Gigu{\`e}re}, \& {Murray}}]{Ji+20}
{Ji}, S., {Chan}, T.~K., {Hummels}, C.~B., {et~al.} 2020, \mnras, 496, 4221,
  \dodoi{10.1093/mnras/staa1849}

\bibitem[{{Jiang} \& {Oh}(2018)}]{Jiang&Oh18}
{Jiang}, Y.-F., \& {Oh}, S.~P. 2018, \apj, 854, 5,
  \dodoi{10.3847/1538-4357/aaa6ce}

\bibitem[{{J{\'o}hannesson} {et~al.}(2016){J{\'o}hannesson}, {Ruiz de Austri},
  {Vincent}, {Moskalenko}, {Orlando}, {Porter}, {Strong}, {Trotta}, {Feroz},
  {Graff}, \& {Hobson}}]{Johannesson+16}
{J{\'o}hannesson}, G., {Ruiz de Austri}, R., {Vincent}, A.~C., {et~al.} 2016,
  \apj, 824, 16, \dodoi{10.3847/0004-637X/824/1/16}

\bibitem[{{Kempski} \& {Quataert}(2020)}]{Kempski+20}
{Kempski}, P., \& {Quataert}, E. 2020, \mnras, 493, 1801,
  \dodoi{10.1093/mnras/staa385}

\bibitem[{{Kim} {et~al.}(2011){Kim}, {Kim}, \& {Ostriker}}]{Kim+11}
{Kim}, C.-G., {Kim}, W.-T., \& {Ostriker}, E.~C. 2011, \apj, 743, 25,
  \dodoi{10.1088/0004-637X/743/1/25}

\bibitem[{{Kim} \& {Ostriker}(2015)}]{KimOstriker2015}
{Kim}, C.-G., \& {Ostriker}, E.~C. 2015, \apj, 815, 67,
  \dodoi{10.1088/0004-637X/815/1/67}

\bibitem[{{Kim} \& {Ostriker}(2017)}]{Kim&Ostriker17}
---. 2017, \apj, 846, 133, \dodoi{10.3847/1538-4357/aa8599}

\bibitem[{{Kim} \& {Ostriker}(2018)}]{Kim&Ostriker18}
---. 2018, \apj, 853, 173, \dodoi{10.3847/1538-4357/aaa5ff}

\bibitem[{{Kim} {et~al.}(2013){Kim}, {Ostriker}, \& {Kim}}]{Kim2013}
{Kim}, C.-G., {Ostriker}, E.~C., \& {Kim}, W.-T. 2013, \apj, 776, 1,
  \dodoi{10.1088/0004-637X/776/1/1}

\bibitem[{{Kim} {et~al.}(2020{\natexlab{a}}){Kim}, {Ostriker}, {Somerville},
  {Bryan}, {Fielding}, {Forbes}, {Hayward}, {Hernquist}, \& {Pandya}}]{Kim+20}
{Kim}, C.-G., {Ostriker}, E.~C., {Somerville}, R.~S., {et~al.}
  2020{\natexlab{a}}, arXiv e-prints, arXiv:2006.16315.
\newblock \doarXiv{2006.16315}

\bibitem[{{Kim} {et~al.}(2020{\natexlab{b}}){Kim}, {Ostriker}, {Fielding},
  {Smith}, {Bryan}, {Somerville}, {Forbes}, {Genel}, \& {Hernquist}}]{Kim+20b}
{Kim}, C.-G., {Ostriker}, E.~C., {Fielding}, D.~B., {et~al.}
  2020{\natexlab{b}}, \apjl, 903, L34, \dodoi{10.3847/2041-8213/abc252}

\bibitem[{{Kim} {et~al.}(2020{\natexlab{c}}){Kim}, {Kim}, \&
  {Ostriker}}]{WT_Kim2020}
{Kim}, W.-T., {Kim}, C.-G., \& {Ostriker}, E.~C. 2020{\natexlab{c}}, \apj, 898,
  35, \dodoi{10.3847/1538-4357/ab9b87}

\bibitem[{{Kroupa}(2001)}]{Kroupa01}
{Kroupa}, P. 2001, \mnras, 322, 231, \dodoi{10.1046/j.1365-8711.2001.04022.x}

\bibitem[{{Kulsrud} \& {Pearce}(1969)}]{Kulsrud&Pearce69}
{Kulsrud}, R., \& {Pearce}, W.~P. 1969, \apj, 156, 445, \dodoi{10.1086/149981}

\bibitem[{{Kulsrud}(2005)}]{Kulsrud05}
{Kulsrud}, R.~M. 2005, {Plasma physics for astrophysics}

\bibitem[{{Kulsrud} \& {Cesarsky}(1971)}]{Kulsrud&Cesarsky1971}
{Kulsrud}, R.~M., \& {Cesarsky}, C.~J. 1971, \aplett, 8, 189

\bibitem[{{Lacki} {et~al.}(2011){Lacki}, {Thompson}, {Quataert}, {Loeb}, \&
  {Waxman}}]{Lacki11}
{Lacki}, B.~C., {Thompson}, T.~A., {Quataert}, E., {Loeb}, A., \& {Waxman}, E.
  2011, \apj, 734, 107, \dodoi{10.1088/0004-637X/734/2/107}

\bibitem[{{Leitherer} {et~al.}(1999){Leitherer}, {Schaerer}, {Goldader},
  {Delgado}, {Robert}, {Kune}, {de Mello}, {Devost}, \&
  {Heckman}}]{Leitherer+99}
{Leitherer}, C., {Schaerer}, D., {Goldader}, J.~D., {et~al.} 1999, \apjs, 123,
  3, \dodoi{10.1086/313233}

\bibitem[{{Mao} \& {Ostriker}(2018)}]{Mao&Ostriker18}
{Mao}, S.~A., \& {Ostriker}, E.~C. 2018, \apj, 854, 89,
  \dodoi{10.3847/1538-4357/aaa88e}

\bibitem[{{Morlino} \& {Caprioli}(2012)}]{Morlino&Caprioli12}
{Morlino}, G., \& {Caprioli}, D. 2012, \aap, 538, A81,
  \dodoi{10.1051/0004-6361/201117855}

\bibitem[{{Ostriker} {et~al.}(2010){Ostriker}, {McKee}, \&
  {Leroy}}]{Ostriker2010}
{Ostriker}, E.~C., {McKee}, C.~F., \& {Leroy}, A.~K. 2010, \apj, 721, 975,
  \dodoi{10.1088/0004-637X/721/2/975}

\bibitem[{{Ostriker} \& {Shetty}(2011)}]{Ostriker2011}
{Ostriker}, E.~C., \& {Shetty}, R. 2011, \apj, 731, 41,
  \dodoi{10.1088/0004-637X/731/1/41}

\bibitem[{{Padovani} {et~al.}(2018){Padovani}, {Ivlev}, {Galli}, \&
  {Caselli}}]{Padovani+18}
{Padovani}, M., {Ivlev}, A.~V., {Galli}, D., \& {Caselli}, P. 2018, \aap, 614,
  A111, \dodoi{10.1051/0004-6361/201732202}

\bibitem[{{Padovani} {et~al.}(2020){Padovani}, {Ivlev}, {Galli}, {Offner},
  {Indriolo}, {Rodgers-Lee}, {Marcowith}, {Girichidis}, {Bykov}, \&
  {Kruijssen}}]{Padovani+20}
{Padovani}, M., {Ivlev}, A.~V., {Galli}, D., {et~al.} 2020, \ssr, 216, 29,
  \dodoi{10.1007/s11214-020-00654-1}

\bibitem[{{Pakmor} {et~al.}(2016){Pakmor}, {Pfrommer}, {Simpson}, \&
  {Springel}}]{Pakmor+16}
{Pakmor}, R., {Pfrommer}, C., {Simpson}, C.~M., \& {Springel}, V. 2016, \apjl,
  824, L30, \dodoi{10.3847/2041-8205/824/2/L30}

\bibitem[{{Parker}(1969)}]{Parker69}
{Parker}, E.~N. 1969, \ssr, 9, 651, \dodoi{10.1007/BF00174032}

\bibitem[{{Peron} {et~al.}(2021){Peron}, {Aharonian}, {Casanova}, {Yang}, \&
  {Zanin}}]{Peron2021}
{Peron}, G., {Aharonian}, F., {Casanova}, S., {Yang}, R., \& {Zanin}, R. 2021,
  \apjl, 907, L11, \dodoi{10.3847/2041-8213/abcaa9}

\bibitem[{{Plotnikov} {et~al.}(2021){Plotnikov}, {Ostriker}, \&
  {Bai}}]{Plotnikov2021}
{Plotnikov}, I., {Ostriker}, E.~C., \& {Bai}, X.-N. 2021, \apj, 914, 3,
  \dodoi{10.3847/1538-4357/abf7b3}

\bibitem[{{Quataert} {et~al.}(2021{\natexlab{a}}){Quataert}, {Thompson}, \&
  {Jiang}}]{Quataert+21a}
{Quataert}, E., {Thompson}, T.~A., \& {Jiang}, Y.-F. 2021{\natexlab{a}},
  \mnras, \dodoi{10.1093/mnras/stab3273}

\bibitem[{{Quataert} {et~al.}(2021{\natexlab{b}}){Quataert}, {Thompson}, \&
  {Jiang}}]{Quataert+21b}
---. 2021{\natexlab{b}}, \mnras, \dodoi{10.1093/mnras/stab3273}

\bibitem[{{Recchia}(2021)}]{Recchia21}
{Recchia}, S. 2021, arXiv e-prints, arXiv:2101.02052.
\newblock \doarXiv{2101.02052}

\bibitem[{{Ruszkowski} {et~al.}(2017){Ruszkowski}, {Yang}, \&
  {Zweibel}}]{Ruszkowski+17}
{Ruszkowski}, M., {Yang}, H. Y.~K., \& {Zweibel}, E. 2017, \apj, 834, 208,
  \dodoi{10.3847/1538-4357/834/2/208}

\bibitem[{{Salem} \& {Bryan}(2014)}]{Salem&Bryan13}
{Salem}, M., \& {Bryan}, G.~L. 2014, \mnras, 437, 3312,
  \dodoi{10.1093/mnras/stt2121}

\bibitem[{{Schneider} {et~al.}(2020){Schneider}, {Ostriker}, {Robertson}, \&
  {Thompson}}]{Schneider+20}
{Schneider}, E.~E., {Ostriker}, E.~C., {Robertson}, B.~E., \& {Thompson}, T.~A.
  2020, \apj, 895, 43, \dodoi{10.3847/1538-4357/ab8ae8}

\bibitem[{{Shalchi}(2019)}]{Shalchi2019}
{Shalchi}, A. 2019, \apjl, 881, L27, \dodoi{10.3847/2041-8213/ab379d}

\bibitem[{{Shalchi}(2020)}]{Shalchi2020}
---. 2020, \ssr, 216, 23, \dodoi{10.1007/s11214-020-0644-4}

\bibitem[{{Simpson} {et~al.}(2016){Simpson}, {Pakmor}, {Marinacci}, {Pfrommer},
  {Springel}, {Glover}, {Clark}, \& {Smith}}]{Simpson+16}
{Simpson}, C.~M., {Pakmor}, R., {Marinacci}, F., {et~al.} 2016, \apjl, 827,
  L29, \dodoi{10.3847/2041-8205/827/2/L29}

\bibitem[{{Stone} \& {Gardiner}(2010)}]{Stone&Gardiner10}
{Stone}, J.~M., \& {Gardiner}, T.~A. 2010, \apjs, 189, 142,
  \dodoi{10.1088/0067-0049/189/1/142}

\bibitem[{{Stone} {et~al.}(2008){Stone}, {Gardiner}, {Teuben}, {Hawley}, \&
  {Simon}}]{Stone+08}
{Stone}, J.~M., {Gardiner}, T.~A., {Teuben}, P., {Hawley}, J.~F., \& {Simon},
  J.~B. 2008, \apjs, 178, 137, \dodoi{10.1086/588755}

\bibitem[{{Stone} {et~al.}(2020){Stone}, {Tomida}, {White}, \&
  {Felker}}]{Stone+20}
{Stone}, J.~M., {Tomida}, K., {White}, C.~J., \& {Felker}, K.~G. 2020, \apjs,
  249, 4, \dodoi{10.3847/1538-4365/ab929b}

\bibitem[{{Strong} {et~al.}(2007){Strong}, {Moskalenko}, \&
  {Ptuskin}}]{Strong+07}
{Strong}, A.~W., {Moskalenko}, I.~V., \& {Ptuskin}, V.~S. 2007, Annual Review
  of Nuclear and Particle Science, 57, 285,
  \dodoi{10.1146/annurev.nucl.57.090506.123011}

\bibitem[{{Sutherland} \& {Dopita}(1993)}]{Sutherland&Dopita93}
{Sutherland}, R.~S., \& {Dopita}, M.~A. 1993, \apjs, 88, 253,
  \dodoi{10.1086/191823}

\bibitem[{{Trotta} {et~al.}(2011){Trotta}, {J{\'o}hannesson}, {Moskalenko},
  {Porter}, {Ruiz de Austri}, \& {Strong}}]{Trotta+11}
{Trotta}, R., {J{\'o}hannesson}, G., {Moskalenko}, I.~V., {et~al.} 2011, \apj,
  729, 106, \dodoi{10.1088/0004-637X/729/2/106}

\bibitem[{{Veilleux} {et~al.}(2020){Veilleux}, {Maiolino}, {Bolatto}, \&
  {Aalto}}]{Veilleux+20}
{Veilleux}, S., {Maiolino}, R., {Bolatto}, A.~D., \& {Aalto}, S. 2020, \aapr,
  28, 2, \dodoi{10.1007/s00159-019-0121-9}

\bibitem[{{Vijayan} {et~al.}(2020){Vijayan}, {Kim}, {Armillotta}, {Ostriker},
  \& {Li}}]{Vijayan+20}
{Vijayan}, A., {Kim}, C.-G., {Armillotta}, L., {Ostriker}, E.~C., \& {Li}, M.
  2020, \apj, 894, 12, \dodoi{10.3847/1538-4357/ab8474}

\bibitem[{{Wentzel}(1974)}]{Wentzel74}
{Wentzel}, D.~G. 1974, \araa, 12, 71,
  \dodoi{10.1146/annurev.aa.12.090174.000443}

\bibitem[{{Wiener} {et~al.}(2017){Wiener}, {Oh}, \& {Zweibel}}]{Wiener+17}
{Wiener}, J., {Oh}, S.~P., \& {Zweibel}, E.~G. 2017, \mnras, 467, 646,
  \dodoi{10.1093/mnras/stx109}

\bibitem[{{Wiener} {et~al.}(2019){Wiener}, {Zweibel}, \&
  {Ruszkowski}}]{Wiener+19}
{Wiener}, J., {Zweibel}, E.~G., \& {Ruszkowski}, M. 2019, \mnras, 489, 205,
  \dodoi{10.1093/mnras/stz2007}

\bibitem[{{Yan} \& {Lazarian}(2002)}]{Yan&Lazarian02}
{Yan}, H., \& {Lazarian}, A. 2002, \prl, 89, 281102,
  \dodoi{10.1103/PhysRevLett.89.281102}

\bibitem[{{Yoast-Hull} {et~al.}(2013){Yoast-Hull}, {Everett}, {Gallagher}, \&
  {Zweibel}}]{Yoast-Hull+13}
{Yoast-Hull}, T.~M., {Everett}, J.~E., {Gallagher}, J.~S., I., \& {Zweibel},
  E.~G. 2013, \apj, 768, 53, \dodoi{10.1088/0004-637X/768/1/53}

\bibitem[{{Yoast-Hull} {et~al.}(2016){Yoast-Hull}, {Gallagher}, \&
  {Zweibel}}]{Yoast-Hull+16}
{Yoast-Hull}, T.~M., {Gallagher}, J.~S., \& {Zweibel}, E.~G. 2016, \mnras, 457,
  L29, \dodoi{10.1093/mnrasl/slv195}

\bibitem[{{Zweibel}(2013)}]{Zweibel13}
{Zweibel}, E.~G. 2013, Physics of Plasmas, 20, 055501,
  \dodoi{10.1063/1.4807033}

\bibitem[{{Zweibel}(2017)}]{Zweibel17}
---. 2017, Physics of Plasmas, 24, 055402, \dodoi{10.1063/1.4984017}

\end{thebibliography}
\bibliographystyle{aasjournal}

\end{CJK*}
\end{document}